\documentclass{article}

\usepackage{arxiv}

\newcommand{\uu}{\mathbf{u}}

\newcommand{\vv}{\mathbf{v}}

\newcommand{\mm}{\mathbf{m}}

\newcommand{\CC}{\mathbf{C}}

\newcommand{\xx}{\mathbf{x}}

\newcommand{\GG}{\mathbf{G}}

\newcommand{\cV}{\mathcal{V}}

\newcommand{\vsigma}{\bm{\sigma}}

\newcommand{\vbeta}{\bm{\beta}}

\newcommand{\veta}{\bm{\eta}}

\newcommand{\cref}[1]{Chapter~\ref{chapt:#1}}

\newcommand{\cmatrixb}{\left\{ \begin{matrix}}
\newcommand{\cmatrixe}{\end{matrix} \right\}}

\newcommand{\bc}{\begin{center}}
\newcommand{\ec}{\end{center}}
\newcommand{\bitem}{\begin{itemize}}
\newcommand{\eitem}{\end{itemize}}

\newcommand{\beq}{\begin{equation}}
\newcommand{\eeq}{\end{equation}}
\newcommand{\beqa}{\begin{eqnarray}}
\newcommand{\eeqa}{\end{eqnarray}}

\newcommand{\bv}{\begin{verbatim}}
\newcommand{\V}{\verb}                  % EX: \V=-d{#@~}= Expr must

\newcommand{\testpix}[1]{\fbox{\begin{minipage}[c]{\textwidth}
                      #1 \end{minipage} }}

\newcommand{\putpstex}[1]{\includegraphics{#1.pstex_t}}

\usepackage{graphicx}
\usepackage{graphics}
\usepackage{color}
\usepackage{epsfig}
\usepackage{amsfonts}
\usepackage{amsmath}
\usepackage{amssymb}
\usepackage{amsthm}
\usepackage{amsopn}
\usepackage{lineno}
\usepackage{times}
\usepackage{xspace}
\usepackage{array}
\usepackage{multirow}
\usepackage{subcaption}
\usepackage{caption}
\usepackage[Symbol]{upgreek}
\usepackage{longtable}
\usepackage{mathptmx}
\usepackage[toc]{appendix}
\usepackage{bm}

\newcommand{\Frefs}[1]{Figs.~\ref{#1}}
\newcommand{\Trefs}[1]{Tables~\ref{#1}}
\newcommand{\Erefs}[1]{Eqs.~(\ref{#1})}

\newcommand{\osigma}{\overline {\sigma}}
\newcommand{\otau}{\overline{\tau}}
\newcommand{\ow}{\overline{w}}

\newcommand{\bfi}{\begin{figure}}
\newcommand{\efi}{\end{figure}}
\newcommand{\change}[1]{\textcolor{red}{#1}}  

\usepackage[dvipsnames]{xcolor}
\newcommand{\red}[1]{\textcolor{red}{#1}}
\newcommand{\blue}[1]{\textcolor{blue}{#1}}

\usepackage{hyperref}
\usepackage{standalone}

\usepackage[utf8]{inputenc} 
\usepackage[T1]{fontenc}    
\usepackage{url}            
\usepackage{booktabs}       
\usepackage{amsfonts}       
\usepackage{nicefrac}       
\usepackage{microtype}      
\usepackage{lipsum}		
\usepackage{graphicx}
\usepackage[numbers]{natbib}
\usepackage{doi}

\title{Modelling of functionally graded triply periodic minimal surface (FG-TPMS) plates}

\author{ \href{https://orcid.org/0000-0002-1746-8297}{\includegraphics[scale=0.06]{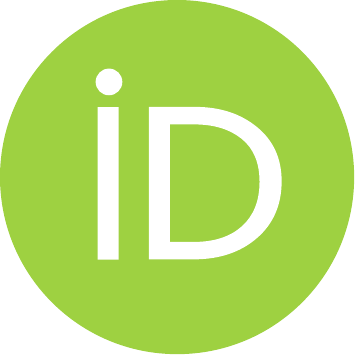}\hspace{1mm}H. Nguyen-Xuan}\thanks{Corresponding author.} \\
	CIRTECH Institute\\
	HUTECH University\\
	Ho Chi Minh City, Viet Nam \\
	\texttt{ngx.hung@hutech.edu.vn} \\
	\And
	\href{https://orcid.org/0000-0003-0924-9235}{\includegraphics[scale=0.06]{orcid.pdf}\hspace{1mm}
    Kim Q. Tran} \\
	CIRTECH Institute\\
	HUTECH University\\
	Ho Chi Minh City, Viet Nam \\
	\texttt{tqkim.work@gmail.com} \\
	\And
	\href{https://orcid.org/0000-0003-1019-4286}{\includegraphics[scale=0.06]{orcid.pdf}\hspace{1mm}
    Chien H. Thai} \\
    Division of Computational Mechanics \\
    Institute for Computational Sciences \\
    Ton Duc Thang University \\
	Ho Chi Minh City, Viet Nam \\
	\And
	\href{https://orcid.org/0000-0002-5056-829X}{\includegraphics[scale=0.06]{orcid.pdf}\hspace{1mm}
    Jaehong Lee} \\
    Deep Learning Architecture Research Center \\ 
    Sejong University \\
    209 Neungdong Ro, Gwangjin Ku, Seoul 05006, Korea \\
	\texttt{jhlee@sejong.ac.kr} \\
}

\renewcommand{\shorttitle}{Modelling of FG-TPMS plates}

\begin{document}
\maketitle

\begin{abstract}
Functionally graded porous plates have been validated as remarkable lightweight structures with excellent mechanical characteristics and numerous applications. With inspiration from the high strength-to-volume ratio of triply periodic minimal surface (TPMS) structures, a new model of porous plates, which is called a functionally graded TPMS (FG-TPMS) plate, is investigated in this paper. Three TPMS architectures including Primitive (P), Gyroid (G), and wrapped package-graph (IWP) with different graded functions are presented. To predict the mechanical responses, a new fitting technique based on a two-phase piece-wise function is employed to evaluate the effective moduli of TPMS structures, including elastic modulus, shear modulus, and bulk modulus. In addition, this function corresponds to the cellular structure formulation in the context of relative density. The separated phases of the function are divided by the different deformation behaviors. Furthermore, another crucial mechanical property of porous structure, i.e, Poisson's ratio, is also achieved by a similar fitting technique. To verify the mechanical characteristics of the FG-TPMS plate, the generalized displacement field is modeled by a seventh-order shear deformation theory (SeSDT). A NURBS-based isogeometric analysis (IGA) is then employed to capture the $C^1$ continuity in approximations. Numerical examples regarding static, buckling, and free vibration analyses of FG-TPMS plates are illustrated to confirm the reliability and accuracy of the proposed approach. 
Consequently, these FG-TPMS structures can provide much higher stiffness than the same-weight isotropic plate. The greater stiffness-to-weight ratio of these porous plates compared to the full-weight isotropic ones should be considered the most remarkable feature. Thus, these complex porous structures have numerous practical applications because of these high ratios and their fabrication ability through additive manufacturing (AM) technology.
\end{abstract}


\section{Introduction}
\label{Section 1}
Biomimicry structure has been known as a robust group of nature-inspired structures. Its applications could be found in various engineering fields such as civil construction, biomedical engineering, crash-worthiness and aerospace material, etc \cite{Ha2020}. Thanks to the development of additive manufacturing (AM) or 3D printing, numerous sophisticated structures could be fabricated precisely. For instance, a group of nature-inspired structures which consists of zero mean curvature surfaces so-called TPMS structure has been manufactured by 3D printing technology. Recently, several solid-type cellular structures have been developed to apply these complex geometries including skeletal-based, network-based, thicken-based, and sheet-based structures. The latter is suggested to have higher mechanical performances than other solid types in view of thermal, electrical conductivity, and mechanical properties \cite{Abu16}. Studies on this structure type have been conducted using the finite element method (FEM) and experiments. An overview of several related studies can be found in \cite{al-ketan19_multifunctional}.

Furthermore, wide-range applications of sheet-based TPMS structures have been reported in recent studies. From the prospect of energy absorption, investigations on protection system strategies might be noticed. For instance, plastic TPMSs have been discovered in the spatial transportation field \citep{Sychov18}. Stainless steel materials have also been utilized to fabricate and assess the energy-absorbing capacity of sheet-based Primitive, Gyroid, and Diamond structures \citep{zhang18_energy}. Another report by Liang \textit{et al.} \citep{liang20_energy} has conducted biaxial and triaxial compression tests on Gyroid blocks to investigate the energy aspects of porous materials. In addition, the sound absorption properties of these structures were also determined in the study by Yang \textit{et al.} \citep{Yang20}. Recently, the sandwich panels with TPMS cores have been conducted to specify their sound insulation effect with the thin plate model theory \citep{lin22_revealing}. In a biomechanics view, the low relative density of TPMS should be used to mimic the human bone joints \citep{Zhu19} or even the bone replacements based on the great strength-to-weight ratio and the controllable pore size characteristic \citep{poltue21_design}. Several applications of TPMS as sandwich plates were reviewed, in which the Gyroid structure exhibited a strong influence on both impact and explosive resistances \citep{tran20_triply}. The semi-theoretical approach was adopted to verify the numerical results of Primitive, Neovius, and IWP sandwich beams \citep{peng21_3d-printed}. Moreover, TPMS reinforcement strategies have been recently found to be an interesting research area. Researchers have verified the enhancement in bending behaviors of cementitious beams reinforced with plastic 3D printed Primitive TPMS cores subjected to both static \citep{Vuong22} and dynamic loads \citep{nguyen-van22_dynamic}. Dang-Bao \textit{et al.} \citep{Dang22} proposed a new cement breakwater solution for the improved hydrodynamic responses of Gyroid-TPMS structures.

In another aspect of understanding TPMS structures, tunable properties (i.e. pore size, anisotropic index, Poisson's ratio, plastic deformation, etc) can be accomplished by modifying the implicit geometry function. The functional grading and hybridizing relative density might also change these properties. Karuna \textit{et al.} \citep{karuna22_mechanical} studied the fluid properties of various modified geometries of IWP structures. However, diverge schemes of TPMS anisotropy are probably difficult to be applied. For that reason, an isotropic design based on the optimization of hollow structures has been operated \citep{fu22_isotropic}. Furthermore, negative Poisson's ratios of sheet-based Primitive and Gyroid composites were discovered by Chawla \textit{et al.} \citep{chawla22_numerical}. With different functionally grading strategies, the mechanical properties and the deformation mechanism of Primitive TPMS can be simply controlled for specific purposes \citep{al-ketan21_programmed}.

Despite a large amount of researches on these TPMS structures, there are possibly only a few reports on the functionally graded TPMS structures. For a specific case, buckling and free vibration behaviors of FG-TPMS beams were pointed out in \citep{Viet21}. In the following study, the flexural properties of several FG-TPMS and hybrid TPMS beams were experimentally and numerically explored by Ejeh \textit{et al.} \citep{ejeh22_flexural}. However, the FG-TPMS plate should be considered an improved structure in the aspects of mechanical properties and fabrication ability. In a relevant work, a bio-inspired functionally graded Gyroid sandwich panel \citep{peng20_bioinspired} exhibited its effectiveness in blast resistance. Furthermore, to address the potential application of FG-TPMS plate structures, popular plate theories consisting of the classical thin plate theory (CPT), the first-order shear deformation theory (FSDT), and the higher-order shear deformation theory (HSDT) can be utilized for modelling of FG-TPMS plates. It was shown that all these theories could be unified by a common polynomial form \citep{Tuan2016}. In addition, a high-order shear deformation theory \citep{Tuan17_3DOF} with three variables (THSDT) is suitable for modelling FG-TPMS plates. Practically, computational approaches based on higher-order shear deformation theories with five variables have been widely used because they can capture the nonlinear distribution of shear terms through the thickness of the plate and satisfy the zero-shear strains/stresses without taking into account shear correction factors (SCF). Numerous recent research on HSDT considers the justified distribution function \citep{Tuan2016}. It was seen that better results could be obtained as indicated in many reports including third-order shear deformation theory (TSDT) \citep{Reddy_HSDT, Shimpi02}, fifth-order shear deformation theory (FiSDT) \citep{Hung13}, various trigonometric shear deformation theories \citep{Mantari2012,Chien14a} and so on. However, the $C^1$ continuity in HSDT models causes difficulties in implementing itself into the standard finite elements. To address this issue, isogeometric analysis based on NURBS is adopted to fulfill naturally the high-order derivatives in the weak form.

Isogeometric analysis (IGA) coined by Hughes \textit{et al.} \citep{Hughes05} allows us to fill the gap between finite element analysis (FEA) and computer-aided design (CAD). IGA used the same NURBS-based basis functions to model the exact geometry and approximate the FE solution. When IGA is used, accessing the CAD system during the FEA process is unnecessary and the exact geometry is unchanged at all discretization levels. IGA has been successfully applied to laminated composite plate/shell structures  \citep{ChienIJNME,Guo2015,Tuan17_3DOF,Faroughi2020CMAME,Huang2022COS}, functionally graded material (FGM) structure \citep{Valizadeh13,Chien14b,Hung17,TanNguyen2019}. Several advanced IGA techniques \citep{Zhang2021,Patton2021,Borjesson2022} have also been developed to enhance the approximation of the laminated composite structure.

In the current work, the mechanical behaviors of FG-TPMS plates are inspected according to the change in the relative density, or the porous characteristic of materials. A new fitting method is proposed based on a two-phase piece-wise function to fit and predict the effective features of FG-TPM plates. The higher-order shear deformation theories (HSDTs) are then employed to approximate the displacement field. After deriving the weak form, the NURBS-based isogeometric approach is then used to analyze static, buckling, and free vibration behaviors of FG-TPMS plates. In this study, three types of TPMS are adopted in the plate structures, whose properties depend on the relative density varying gradually through the plate thickness. Various plates with different porosity are utilized to describe the efficiency of these new porous plates. 

\section{Mechanical properties of FG-TPMS materials}
\label{Section 2}
In this section, A novel fitting approach is proposed to investigate the mechanical characteristics of the FG-TPMS foam structure. This study focuses on the FG-TPMS plate model based on three different types of sheet-based architecture including Primitive, Gyroid, and I-graph and wrapped package-graph (IWP). They can be described by Eqs. (\ref{eq.TPMS}), and (\ref{eq. Sheet-based}), which provide the sheet-based solid type constructed from TPMS geometry.

\begin{equation} \label{eq.TPMS}
    \begin{array}{*{20}{l}}
        \textrm {Primitive} & \phi = \textrm{cos}(w_1 x_1) + \textrm{cos}(w_2 x_2) +\textrm{cos}(w_3 x_3),  \\[2mm]
        \textrm {Gyroid} & \phi = \textrm{sin}(w_1 x_1)\textrm{cos}(w_2 x_2) + \textrm{sin}(w_2 x_2)\textrm{cos}(w_3 x_3) + \textrm{sin}(w_3 x_3) \textrm{cos}(w_1 x_1), \\[2mm]
        \textrm {IWP} & \phi = 2 \left(\textrm{cos}(w_1 x_1)\textrm{cos}(w_2 x_2) + \textrm{cos}(w_2 x_2)\textrm{cos}(w_3 x_3) + \textrm{cos}(w_3 x_3)\textrm{cos}(w_1 x_1) \right)\\
        & - \left(\textrm{cos}(2w_1 x_1) + \textrm{cos}(2w_2 x_2) + \textrm{cos}(2w_3 x_3) \right).
    \end{array}
\end{equation}
where
\begin{equation} \label{eq.Primitive}
    w_i = \dfrac{2\pi n_i}{l_i}, i= 1,2,3
\end{equation}
where $\xx =(x_1, x_2, x_3)^T$ denotes an arbitrary point in a three-dimensional domain, $n_i$ and $l_i$ indicate the unit cell quantity and its lengths, respectively. In this study, the same unit properties are applied to all three perpendicular directions.

\begin{equation} \label{eq. Sheet-based}
    -t \leq \phi \leq t
\end{equation}
where $t$ is the TPMS control parameter. The relationships of $t$ and the relative density  ($\rho$) of TPMS types are different from each other. Fig. \ref{fig:t-RD} presents the fitting curves of the sheet-based TPMS relationship. In addition, $\rho$ is defined by
\begin{equation} \label{eq. rho}
    \rho = \dfrac{V}{V_s}
\end{equation}
in which $V$ and $V_s$ are the total volumes of the TPMS cell and the surrounding cube, respectively.
\begin{figure}
    \centering
    \includegraphics[trim=1cm 6cm 1cm 6cm, clip=true, width=0.6\textwidth]{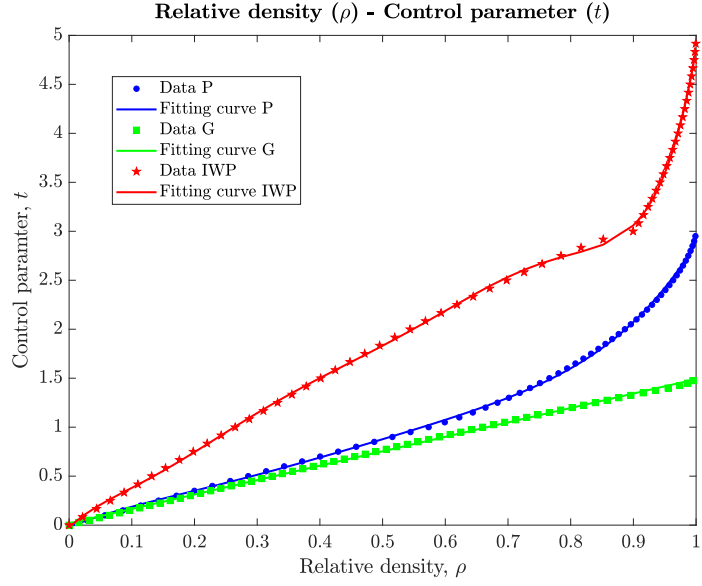}
    \caption {Relationship of control parameter $t$ and relative density $\rho$ of sheet-based TPMS}
    \label{fig:t-RD}
\end{figure}

Fig. \ref{fig.por.dis} demonstrates two groups of density distribution patterns through the plate thickness. These functions can be described in Eq. (\ref{eq.Distribution}): 
\begin{equation} \label{eq.Distribution}
\rho = \left\{ 
{\begin{array}{*{20}{l}}
    \rho_{min} + (\rho_{max}-\rho_{min}) \times \left( \dfrac{x_3}{h} + \dfrac{1}{2} \right) ^n, & \textrm{Pattern A} \\
    \rho_{min} + (\rho_{max}-\rho_{min}) \times \left( 1 - \textrm{cos}(\dfrac{\pi x_3}{h}) \right) ^n. & \textrm{Pattern B} \\
\end{array}} \right.
\end{equation}

These distribution groups consist of non-uniform functions whose ranges vary from $\rho_{min}$ to $\rho_{max}$. Using the power $n$, various relative density functions can be produced. In fact, numerical investigations on the influence of these parameters on the plate behaviors are implemented in Section \ref{Section 4}. Fig. \ref{fig:FGplate} illustrates three types of TPMS structures which are P, G, and IWP with density distribution pattern A and the power value of $3$.
\begin{figure}[ht!]
  \centering
  \begin{subfigure}[c]{0.48\textwidth}
    \centering
    \includegraphics[trim=1.5cm 7.4cm 0.3cm 6cm,clip=true,width=1\textwidth]{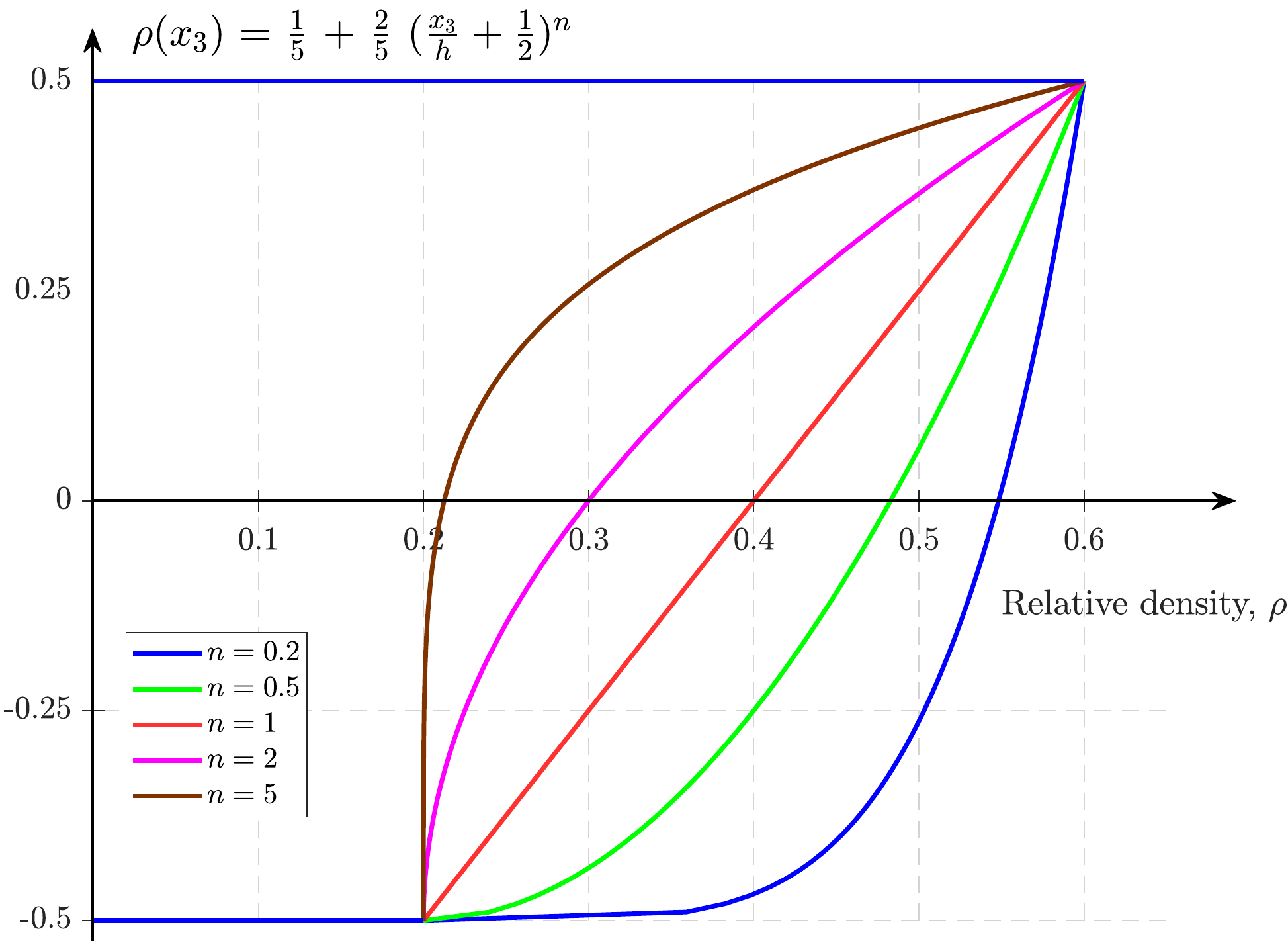}
    \caption {Pattern A}
  \end{subfigure}
  \hfill
  \begin{subfigure}[c]{0.48\textwidth}
    \centering
    \includegraphics[trim=1.5cm 7.4cm 0.3cm 6cm,clip=true,width=1\textwidth]{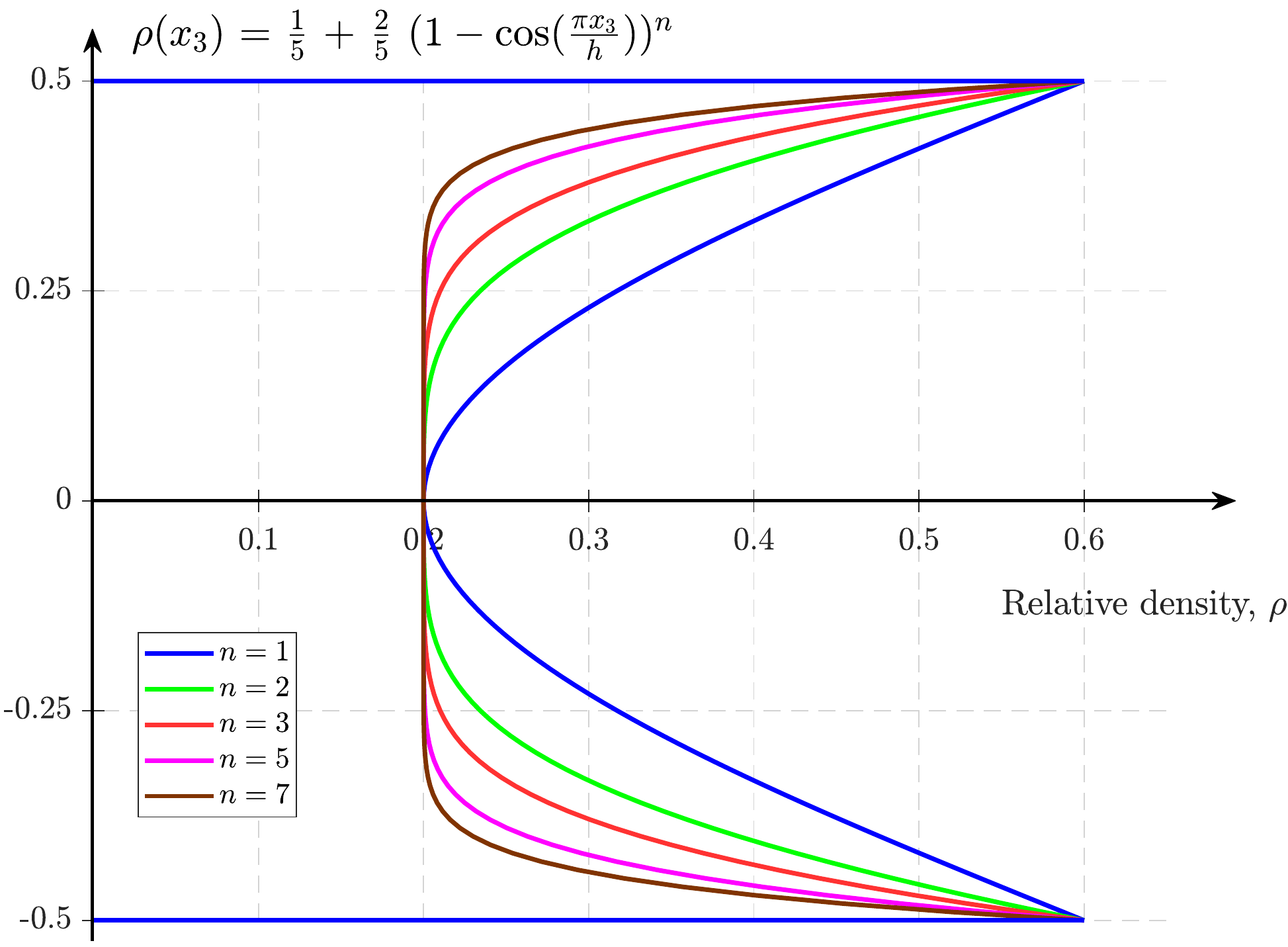}
    \caption {Pattern B}
  \end{subfigure}
\caption {Relative density distribution of FG-TPMS}
\label{fig.por.dis}
\end{figure}

\begin{figure}[!ht]
    \centering
    \begin{subfigure}[c]{0.32\textwidth}
        \centering
        \includegraphics[trim=4.2cm 7cm 3.5cm 6.5cm,clip=true,width=\textwidth]{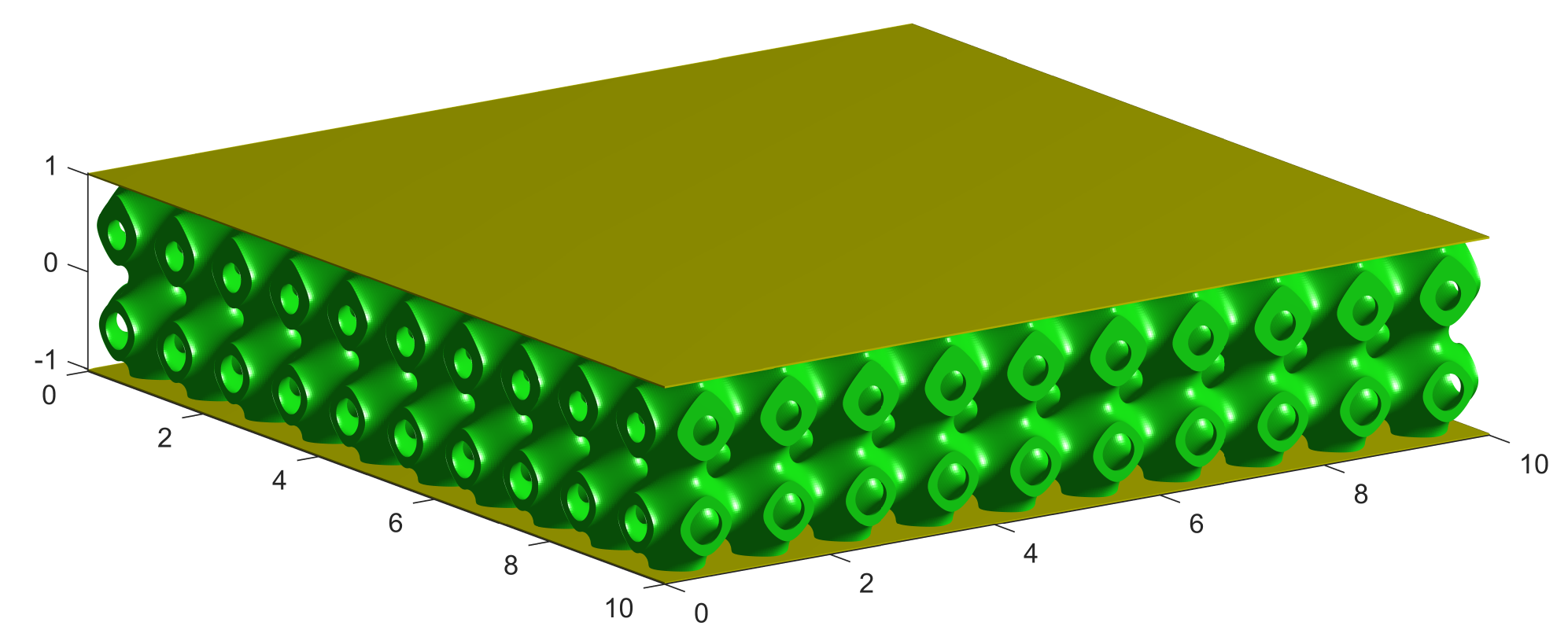}
        \caption {Primitive (P)}
    \end{subfigure}
    \begin{subfigure}[c]{0.32\textwidth}
        \centering
        \includegraphics[trim=4.2cm 7cm 3.5cm 6.5cm,clip=true,width=\textwidth]{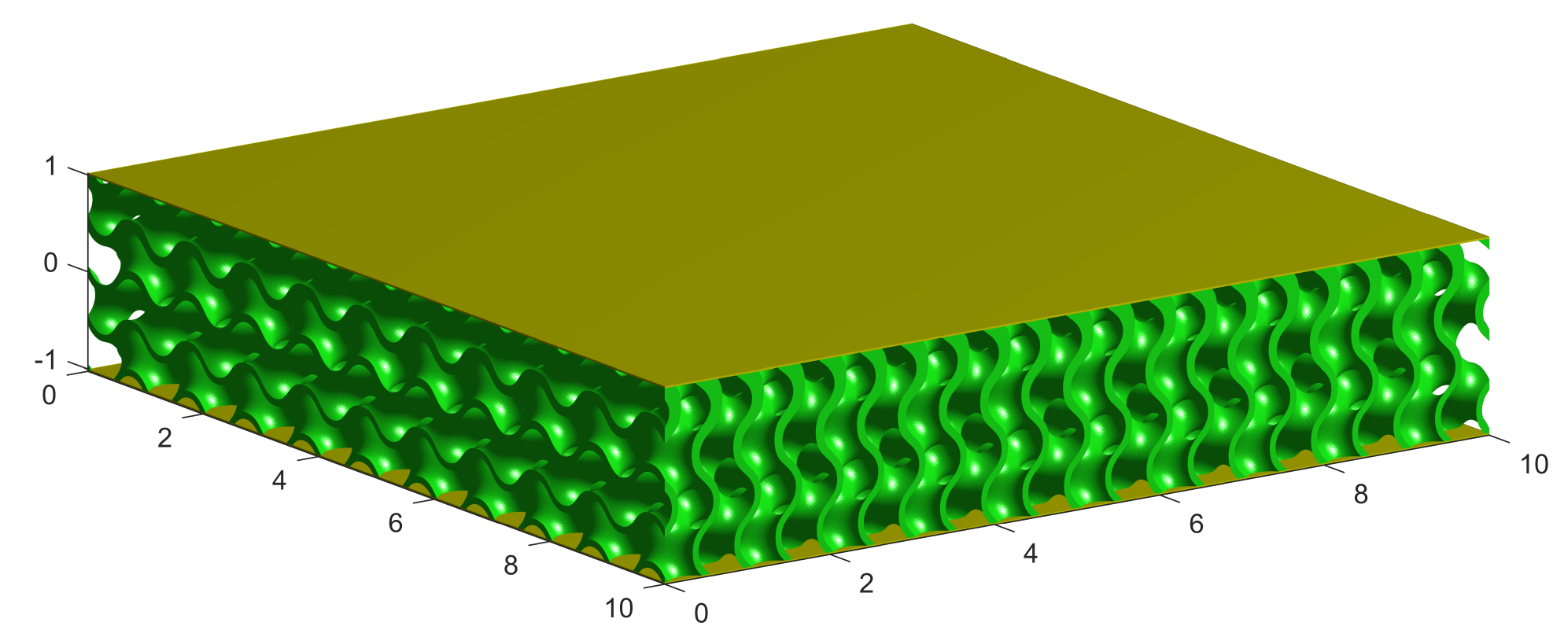}
        \caption {Gyroid (G)}
    \end{subfigure}
    \begin{subfigure}[c]{0.32\textwidth}
        \centering
        \includegraphics[trim=4.2cm 7cm 3.5cm 6.5cm,clip=true,width=\textwidth]{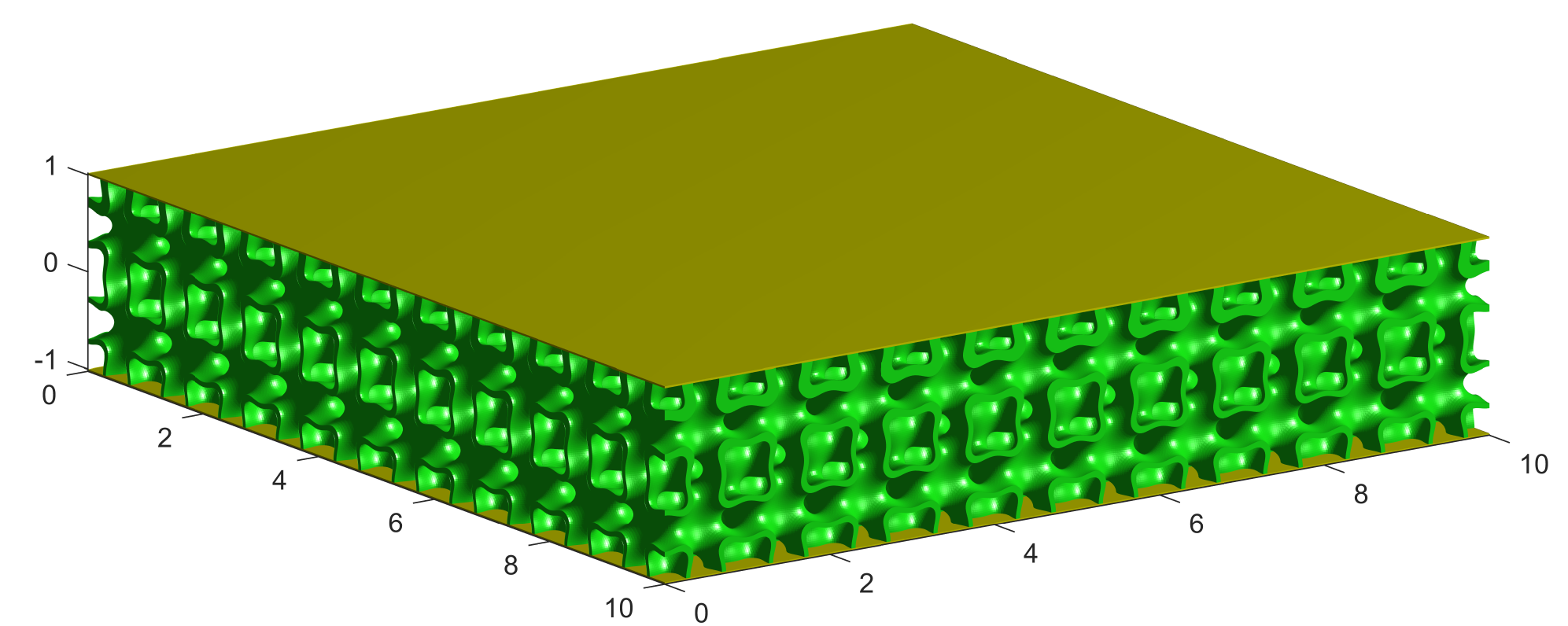}
        \caption {I wrapped package-graph (IWP)}
    \end{subfigure}
    \caption {An illustration for FG-TPMS plate model with density pattern A and the power $n=3$.}
    \label{fig:FGplate}
\end{figure}

Regarding the elastic properties, the parent (base) material is assumed to be isotropic. In addition, the representative volume elements (RVE) of TPMS structures are indicated as cubic symmetric materials \citep{Zener1949}. Their elasticity material coefficients are expressed as an "isotropic-like form" as follows
\begin{equation}
    \mathbf{C} = \begin{bmatrix}
        \lambda+2\mu & \lambda & \lambda & 0 & 0 & 0 \\
        & \lambda+2\mu & \lambda & 0 & 0 & 0 \\
        &  & \lambda+2\mu & 0 & 0 & 0 \\
        &  &  & G & 0 & 0 \\
        &  sym & & & G & 0 \\
        &  &  & & & G \\
    \end{bmatrix} 
    \label{eq: TPMS_constitutive}
\end{equation}
where $\lambda=\dfrac{E\nu}{(1+\nu)(1-2\nu)}$, and $\mu=\dfrac{E}{2(1+\nu)}$ describe the elastic Lam\'{e} coefficients, and a material is isotropic if $G=\mu$. In other words, $G$ depends on the relative density $\rho$.  


To apply the TPMS geometry in practice, all three independent variables including uniaxial elastic modulus ($E$), shear modulus ($G$), and Poisson's ratio ($\nu$) of the TPMS cell need to be defined explicitly. Furthermore, Poisson's ratio is derived from the bulk modulus ($K$) using Eq. (\ref{eq. nu-E-K_Relationship}). These mechanical moduli of porous structures can be written in a scalar form based on the base material as described in Eq. (\ref{eq. effective_properties}).
\begin{equation} \label{eq. nu-E-K_Relationship}
    \nu = \dfrac{3K - E}{6K}
\end{equation}
\begin{equation} \label{eq. effective_properties}
    E^* = \dfrac{E}{E_s}; \hphantom{000} G^* = \dfrac{G}{G_s}; \hphantom{000} K^* = \dfrac{K}{K_s}.
\end{equation}
where $E$, $G$, and $K$ are the elastic modulus, shear modulus, and bulk modulus of the TPMS cell, respectively; the corresponding properties of the parent (base) or isotropic elastic material are denoted as $E_s$, $G_s$, and $K_s$, respectively.

In recent studies, the effective mechanical properties of sheet-based TPMS structures were established using the power scaling law of cellular structure in the context of the relative density $\rho$ \citep{gibson03_cellular}. However, the value of relative density might not be studied in the full range of $0$ to $1$, and the differences in mechanical responses of structures with various porous states were also ignored. In the current work, we propose a fitting method that can be divided into three continuous processes. At first, the elastic moduli of the structures are approximated by a two-phase piece-wise function based on their deformation modes. Next, the step values in the previous results are adopted to fit the shear and bulk moduli with the same function. Finally, the Poison's ratios of these structures are generated by the relationship among the mechanical properties and estimated by another piece-wise function. This fitting process is shown in the flowchart in Fig. \ref{fig: fit_process}
\begin{figure}[ht!]
    \centering
    \includegraphics[width=\textwidth]{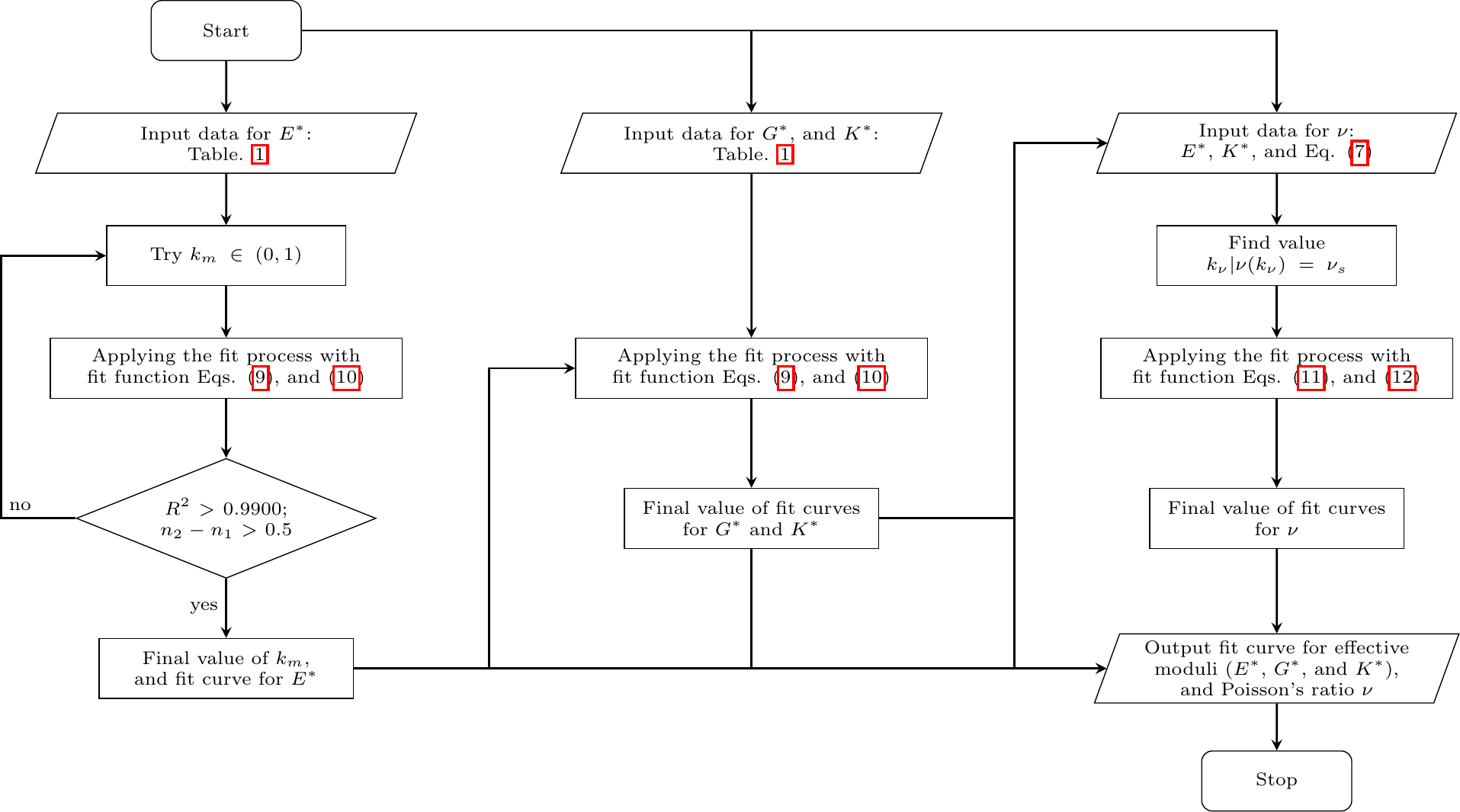} 
    \caption{A flowchart for computing the effective mechanical properties using the fitting method.}
    \label{fig: fit_process}
\end{figure}

The mechanical moduli of the TPMSs, namely elastic modulus, shear modulus, and bulk modulus are fitted by a two-phase model in Eq. (\ref{eq. fit.function}). In this function, the coefficients can be minimized, bringing the value from $6$ to $4$, due to two conditions. They are the continuous condition at $\rho = k_m$ and the boundary condition at $\rho = 1$ which leads to $E^* = 1$. These conditions are described in Eq. (\ref{eq. fit_E.BC}):
\begin{equation} \label{eq. fit.function}
    E^* = 
    \begin{cases}
		C_1 \rho^{n_1}, & \rho \leq k_m \\
		C_2 \rho^{n_2} + C_3, & \rho > k_m
	\end{cases}
\end{equation}
\begin{equation} \label{eq. fit_E.BC}
    \begin{array}{*{20}{l}}
        \textrm {Continuous Condition} & C_1 k_m^{n_1} =  C_2 k_m^{n_2} + C_3 \\
        \textrm {Boundary Condition} & C_2 + C_3 = 1.
	\end{array}
\end{equation}

Consequently, the two coefficients, $C_2 = (C_1 k_m^{n_1} - 1)/(k_m^{n_2} - 1)$ and $C_3 = 1 - C_2$, can be calculated. In addition, the power coefficients ($n_1$ and $n_2$) depend on the behavior of the structure under the specific load type. These values are close to unity if the structure has a stretching-dominated mode, and are close to $2$ if the structure has a bending-dominated mode. Moreover, this behavior mode depends not only on the geometry but also on the thickness of the walls in the sheet-based TPMS \citep{lee17_stiffness,viet22_mechanical}. Hence, the value $k_m$ might be used to determine the separated region for each mode. However, this value might not coincide with all three moduli.

In this study, the value of $k_m$ is selected first based on the strong difference in deformation modes of two relative density regions subjected to uniaxial load. This allows the power $n_2$ to reach a higher value than the value of $n_1$ because of the shifting from the stretching-dominated mode to the bending-dominated mode. The value of $k_m$ is equal to $0.25$, $0.45$, and $0.35$ for types P, G, and IWP, respectively. For instance, in the Primitive case, the second region is found to be fully bending while stretching mode can be observed for both regions of the IWP structure. While the Gyroid structure walls, however, might be stated to be always bending while carrying load despite any value of relative density. Therefore, the second region of this TPMS can be split due to the value of $n_2$ which is higher than $2$. Furthermore, it is interesting to note that this $k_m$ value can be further employed in simple shear and hydrostatic compression responses with high accuracy in fitting results. 

As a result, the remaining coefficients are $C_1$, $n_1$, and $n_2$. While the power coefficients ($n_1$ and $n_2$) provide the deformation modes, the coefficient $C_1$ indicates the material alignment to the loading direction. Moreover, this $C1$ coefficient is also sensitive to fabrication imperfections, namely printing defects, solidification cracks or voids, relative density deviation, etc. Consequently, this coefficient might receive a smaller value in the experiment.

Although numerous studies on TPMS properties have been published, the mechanical behaviors of these structures have been shown to depend on the material scale and the number of units in the specimen. In fact, Simsek \textit{et al.} \citep{simsek20_modal} indicated that the $20\%$ relative density sheet-based TPMS blocks with at least four layers could provide the constant elastic modulus which was consistent with micro-mechanics simulation results. This affirmation could be inappropriate with different relative densities; however, this feature is not within the scope of this study. Therefore, in adopted resources, the effective moduli were investigated for one representative unit cell using micro-mechanics simulations involving the periodic boundary condition. By employing the fitting tool, the effective properties equation can be generated with high accuracy. For the purpose of relative density investigation, the data used in the fitting process are collected from two of the previous key studies; one provides data in the range of $0$ to $0.15$, and the other covers the range of 0.15 to 0.8. In addition, these studies employed the same parent material properties i.e. isotropic solid with the elastic modulus ($E_s$) and Poisson's ratio ($\nu_s = 0.3$). The publications data that have been used in the fitting process are given in Table \ref{tab.Fit_Data}.
\begin{table}[ht!]
    \centering
    \caption{Published data used in the present study}
    \label{tab.Fit_Data}
	{\renewcommand\arraystretch{1.3}
	{\tabcolsep = 2mm
	\begin{tabular}{lcccc}
        \hline \\[-4mm]
        \multirow{2}{*}{Property} & \multirow{2}{*}{TPMS} & \multicolumn{2}{c}{Adopted sources} & \multirow{2}{*}{\shortstack[c]{Comparison \\[1.2mm] sources}} \\[2mm]
        \cline{3-4} \\[-4mm]
         & & Light foam & Heavy foam & \\[2mm]
        \hline \\[-4mm]
        
		\multirow{3}{*}{Elastic modulus, $E$} & Primitive & \citep{Abu16} & \citep{viet22_mechanical} & \citep{lee17_stiffness, chen19_on-hybrid, abueidda17_mechanical, poltue21_design, Viet21} \\[2mm]
         & Gyroid & \citep{Abu16} & \citep{viet22_mechanical} & \citep{al-ketan19_multifunctional, chen19_on-hybrid, al-ketan18_mircoarchitected, poltue21_design, Viet21}\\[2mm]
         & IWP & \citep{Abu16} & \citep{viet22_mechanical} & \citep{rashid20_effective, abueidda17_mechanical, al-ketan18_the-effect, poltue21_design, Viet21} \\[2mm]
        \hline \\[-4mm]
        
		\multirow{3}{*}{Shear modulus, $G$} & Primitive & \citep{Abu16} & \citep{viet22_mechanical} & \citep{lee17_stiffness, chen19_on-hybrid} \\[2mm] 
         & Gyroid & \citep{Abu16} & \citep{viet22_mechanical} & \citep{al-ketan19_multifunctional, chen19_on-hybrid} \\[2mm]
         & IWP & \citep{Abu16} & \citep{viet22_mechanical} & \citep{rashid20_effective} \\[2mm]
        \hline \\[-4mm]
        
		\multirow{3}{*}{Bulk modulus, $K$} & Primitive & \citep{Abu16} & \citep{chen19_on-hybrid} & \citep{lee17_stiffness} \\[2mm]
         & Gyroid & \citep{al-ketan19_multifunctional} & \citep{chen19_on-hybrid} & \citep{Abu16} \\[2mm]
         & IWP & \citep{Abu16} & \citep{Abu16} & \citep{rashid20_effective} \\[2mm]
        \hline \\
	\end{tabular}}}
\end{table}

The final factors of the fitting curve for the effective elastic modulus, shear modulus, and bulk modulus are listed in Table \ref{tab.Fit_EGK}. These fitting curves are compared with other data from several previous investigations which were not used in the fitting data, and these studies are included in Table \ref{tab.Fit_Data}. As shown in Fig. \ref{fig:EGK_TPMS}, the homogenization expression in the present study shows accurate predictions of the effective moduli of sheet-based TPMS structures.

\begin{table}[ht!]
    \centering
    \caption{Fitting curves for the effective elastic modulus, shear modulus, and bulk modulus}
    \label{tab.Fit_EGK}
	{\renewcommand\arraystretch{1.3}
	{\tabcolsep = 1.3mm
	\begin{tabular}{lccccccc}
        \hline \\[-4mm]
        \multirow{1}{*}{\shortstack[l]{Effective \\[1.22mm] property}} & \multirow{1}{*}{TPMS} & \multirow{1}{*}{$C_1$} & \multirow{1}{*}{$n_1$} & \multirow{1}{*}{$n_2$} & \multirow{1}{*}{Expression} & \multirow{1}{*}{$R$-square} & \multirow{1}{*}{\shortstack[l]{Adjusted \\[1.2mm] $R$-square}}\\[4mm]
        \hline \\[-4mm]
		\multirow{3}{*}{\shortstack[l]{Elastic \\[1.2mm] modulus,\\[1.2mm] $E^*$}} & Primitive & 0.317 & 1.264 & 2.006 & $E^* = \begin{cases} 
            0.317 \rho^{1.264}, & \rho \leq 0.25 \\ 
            1.007 \rho^{2.006} - 0.007. & \rho > 0.25 
        \end{cases}$ & 0.9998 & 0.9998 \\[5mm]
         & Gyroid & 0.596 & 1.467 & 2.351 & $E^* = \begin{cases} 
    		0.596 \rho^{1.467}, & \rho \leq 0.45 \\
    		0.962 \rho^{2.351} + 0.038. & \rho > 0.45 \\
        \end{cases}$ & 0.9996 & 0.9995 \\[5mm]
         & IWP & 0.597 & 1.225 & 1.782 & $E^* = \begin{cases} 
    		0.597 \rho^{1.225}, & \rho \leq 0.35 \\
    		0.987 \rho^{1.782} + 0.013. & \rho > 0.35
        \end{cases}$ & 0.9968 & 0.9959 \\[5mm]
        \hline \\[-4mm]
		\multirow{3}{*}{\shortstack[l]{Shear \\[1.2mm] modulus,\\[1.2mm] $G^*$}} & Primitive & 0.705 & 1.189 & 1.715 & $G^* = \begin{cases} 
            0.705 \rho^{1.189}, & \rho \leq 0.25 \\ 
            0.953 \rho^{1.715} + 0.047. & \rho > 0.25 
        \end{cases}$ & 0.9996 & 0.9996 \\[5mm]
         & Gyroid & 0.777 & 1.544 & 1.982 & $G^* = \begin{cases} 
    		0.777 \rho^{1.544}, & \rho \leq 0.45 \\
    		0.973 \rho^{1.982} + 0.027. & \rho > 0.45 \\
        \end{cases}$ & 0.9999 & 0.9998 \\[5mm]
         & IWP & 0.529 & 1.287 & 2.188 & $G^* = \begin{cases} 
    		0.529 \rho^{1.287}, & \rho \leq 0.35 \\
    		0.960 \rho^{2.188} + 0.040. & \rho > 0.35
        \end{cases}$ & 0.9999 & 0.9999 \\[5mm]
        \hline \\[-4mm]
		\multirow{3}{*}{\shortstack[l]{Bulk \\[1.2mm] modulus,\\[1.2mm] $K^*$}} & Primitive & 0.487 & 1.127 & 2.380 & $K^* = \begin{cases} 
            0.487 \rho^{1.127}, & \rho \leq 0.25 \\ 
            0.932 \rho^{2.380} + 0.068. & \rho > 0.25 
        \end{cases}$ & 0.9993 & 0.9992 \\[5mm]
         & Gyroid & 0.536 & 1.240 & 3.336 & $K^* = \begin{cases} 
    		0.536 \rho^{1.240}, & \rho \leq 0.45 \\
    		0.861 \rho^{3.336} + 0.139. & \rho > 0.45 \\
        \end{cases}$ & 0.9988 & 0.9986 \\[5mm]
         & IWP & 0.439 & 1.064 &  & $K^* = 0.439 \rho^{1.064}, \rho \leq 0.35$ & 0.9990 & 0.9987 \\[5mm]
        \hline \\
	\end{tabular}}}
\end{table}

\begin{figure}[!ht]
    \centering
    \begin{subfigure}[c]{\textwidth}
        \centering
        \includegraphics[trim=1.3cm 6.3cm 2cm 6.3cm,clip=true,width=0.32\textwidth]{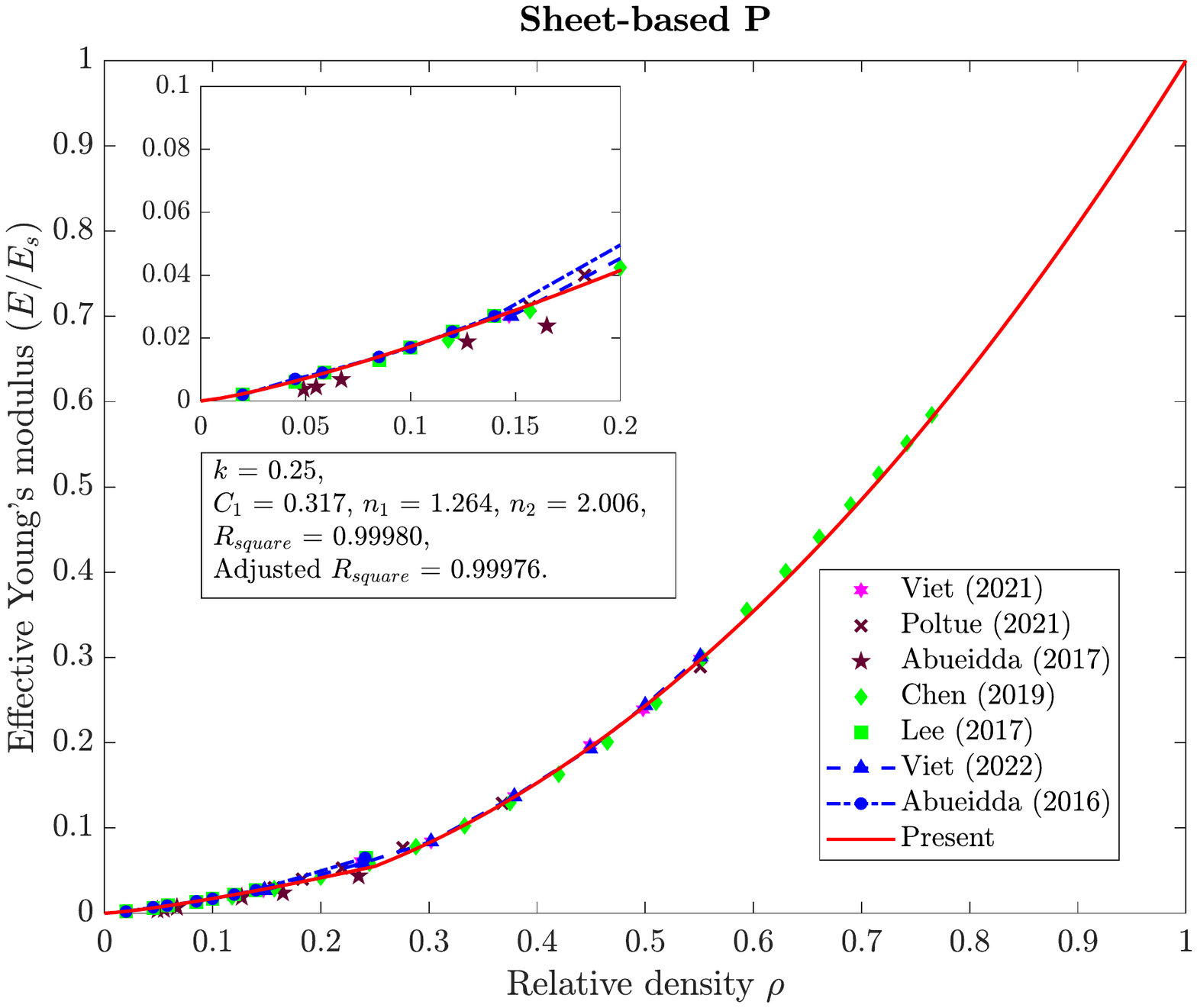}
        \hfill
        \includegraphics[trim=1.3cm 6.3cm 2cm 6.3cm,clip=true,width=0.32\textwidth]{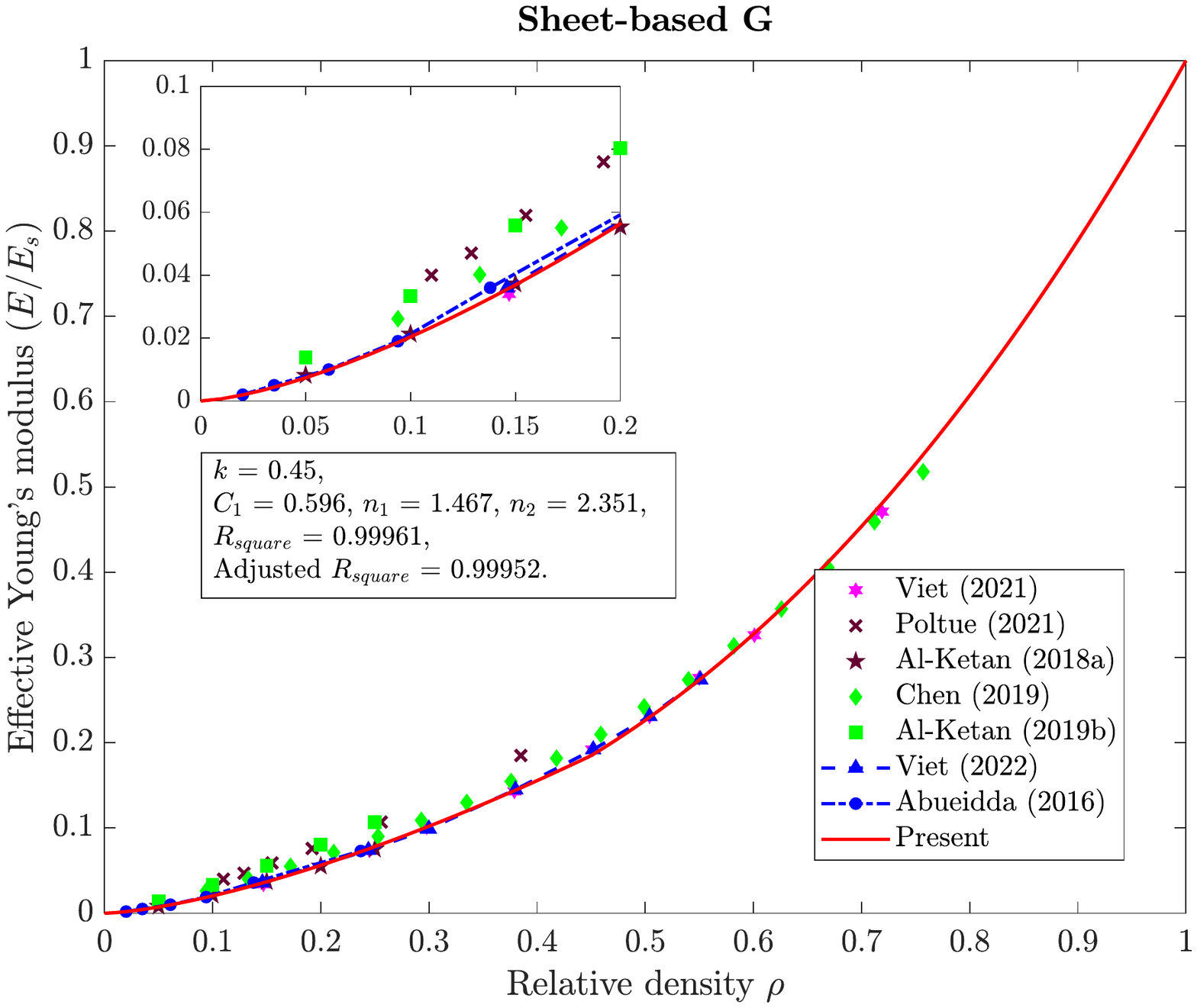}
        \hfill
        \includegraphics[trim=1.3cm 6.3cm 2cm 6.3cm,clip=true,width=0.32\textwidth]{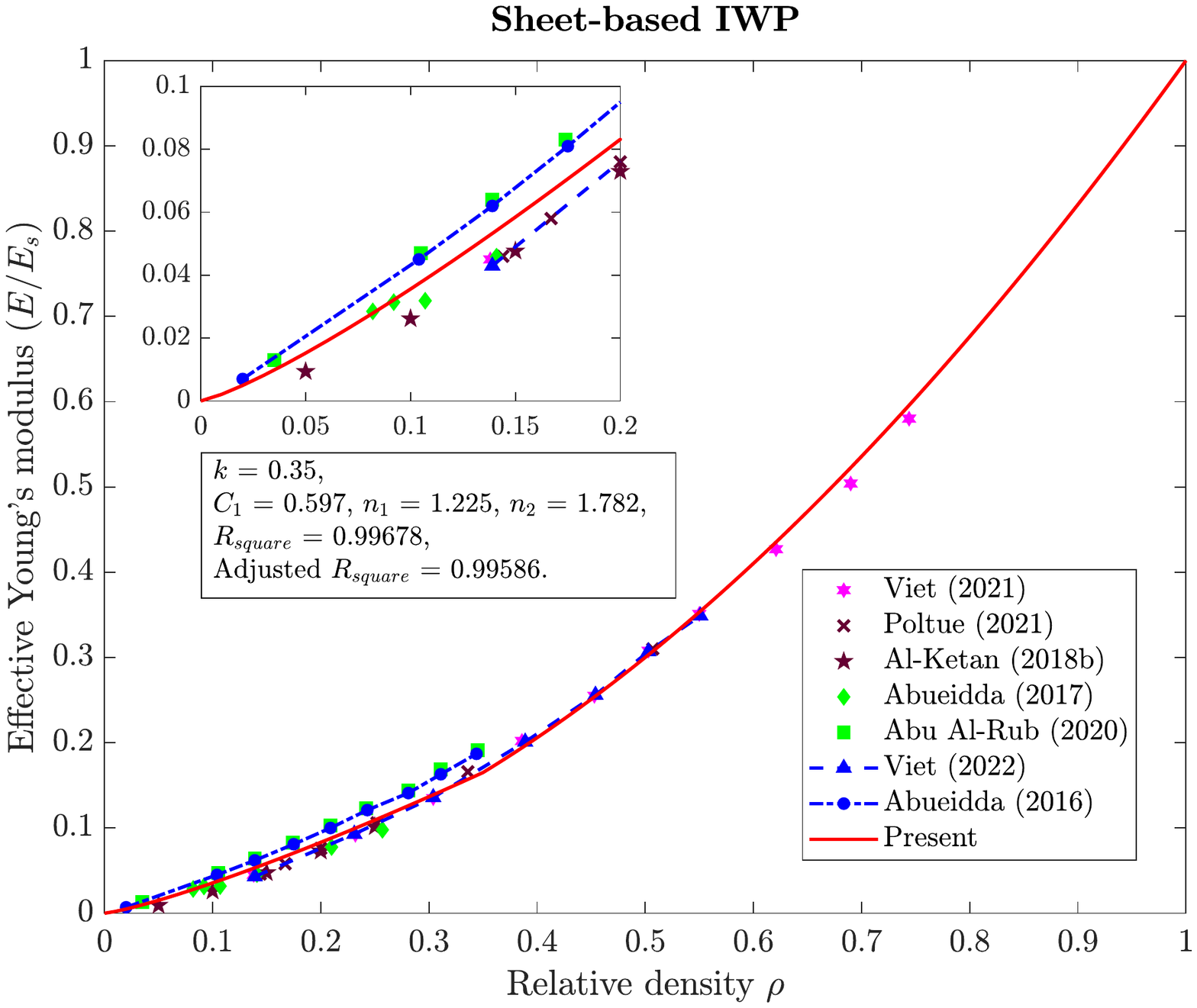}
        \caption {Effective elastic modulus, $E^*$}
    \end{subfigure}
    \medskip
    
    \begin{subfigure}[c]{\textwidth}
        \centering
        \includegraphics[trim=1.3cm 6.3cm 2cm 6.3cm,clip=true,width=0.32\textwidth]{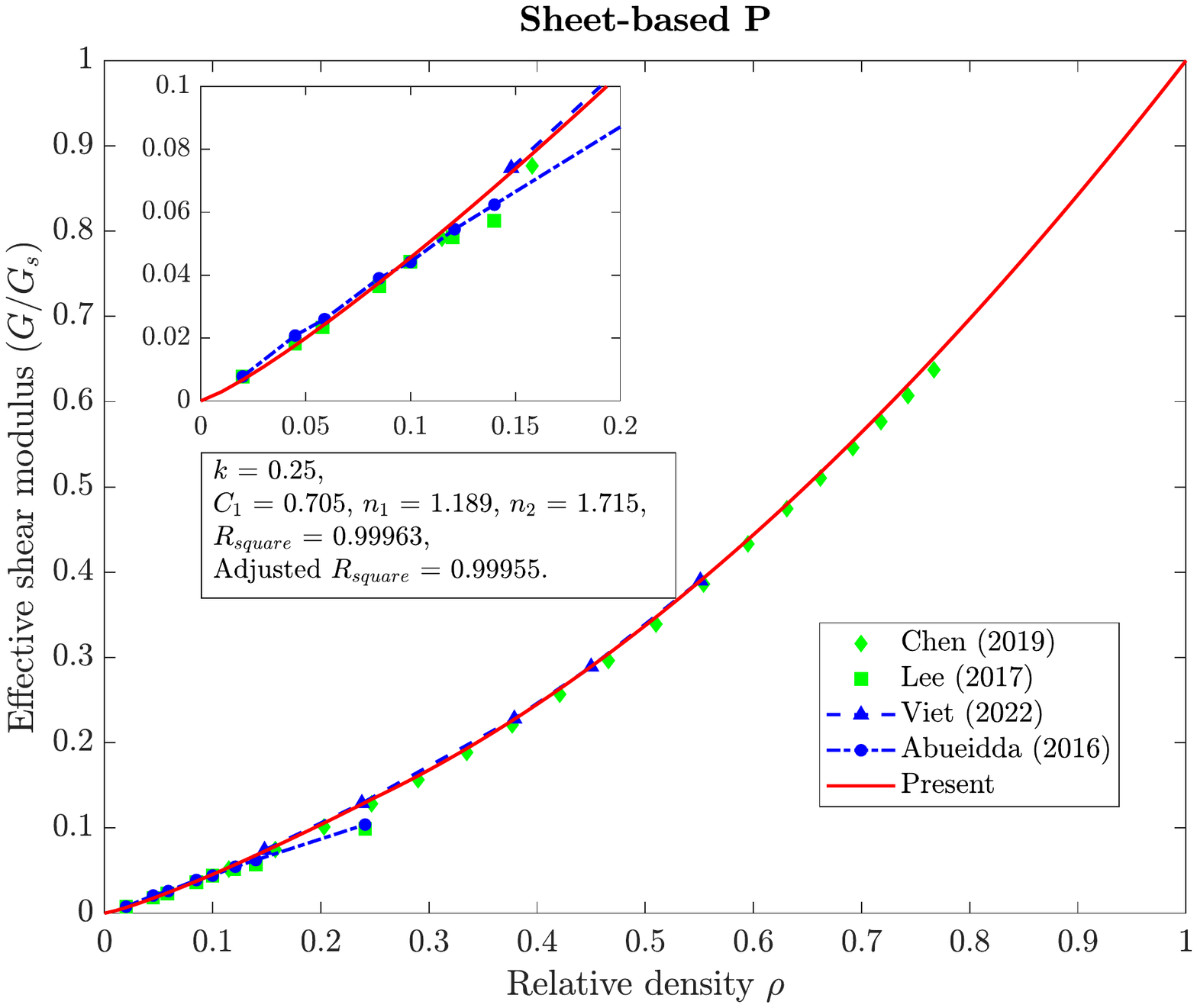}
        \hfill
        \includegraphics[trim=1.3cm 6.3cm 2cm 6.3cm,clip=true,width=0.32\textwidth]{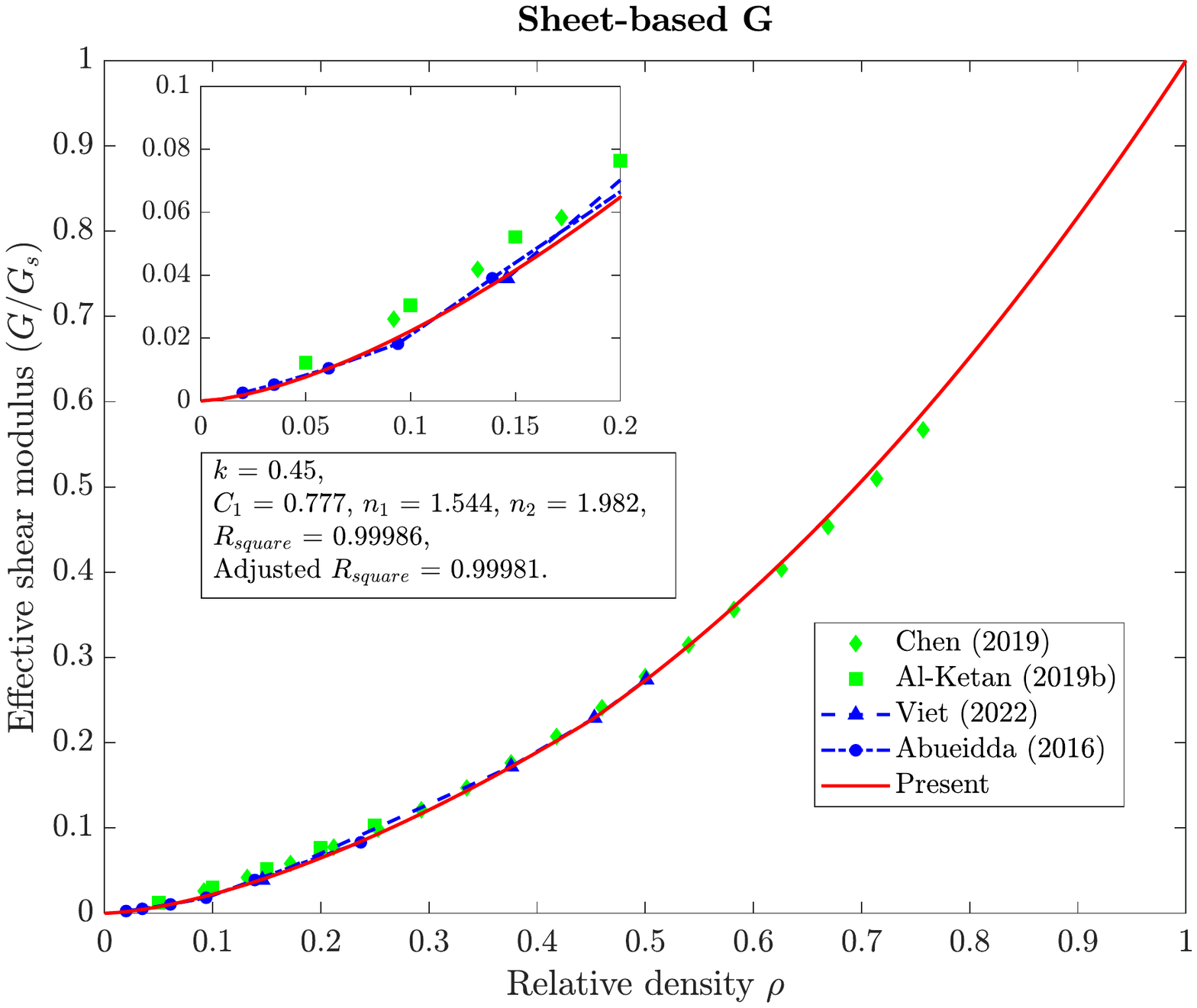}
        \hfill
        \includegraphics[trim=1.3cm 6.3cm 2cm 6.3cm,clip=true,width=0.32\textwidth]{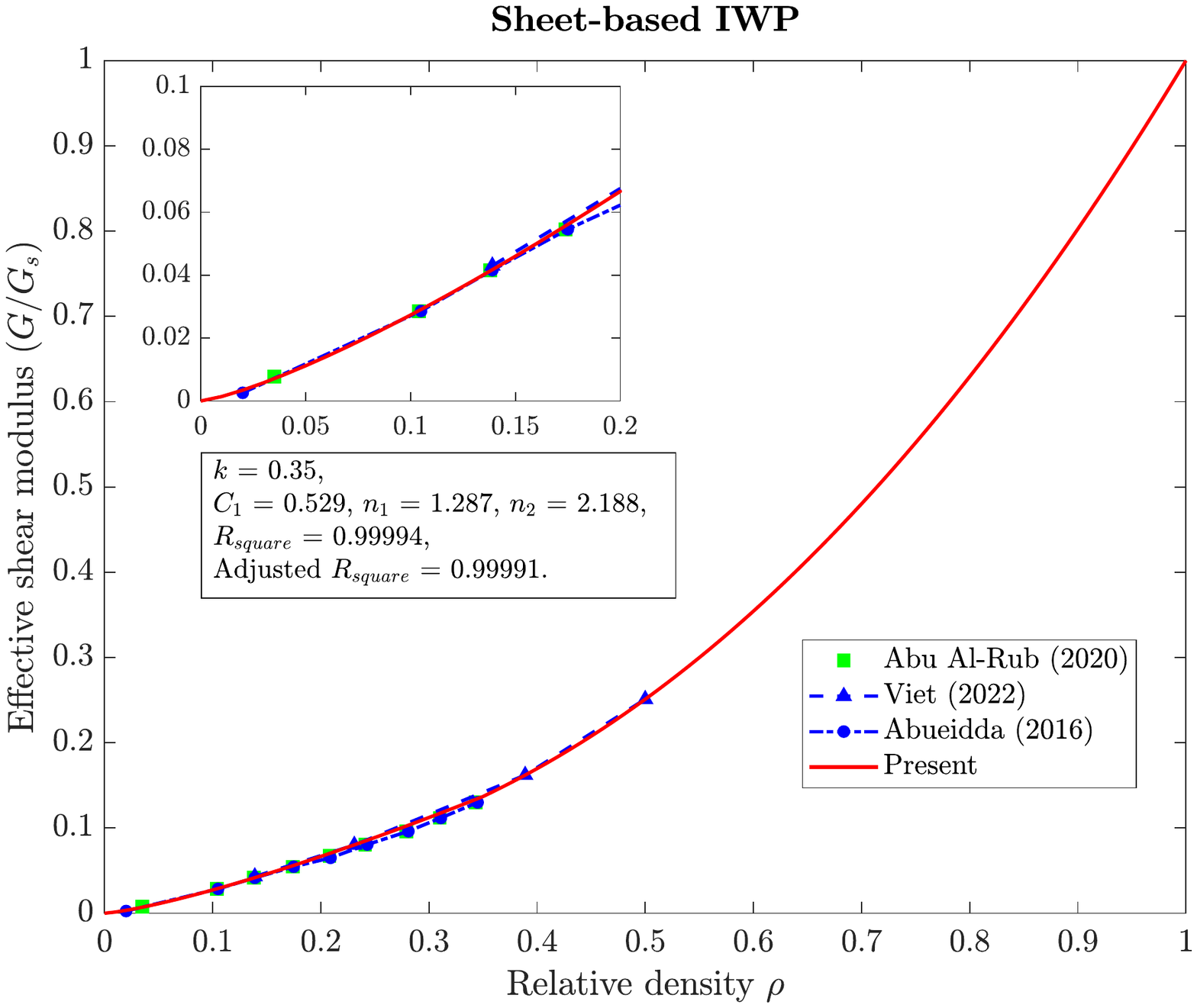}
        \caption {Effective shear modulus, $G^*$}
    \end{subfigure}
    \medskip
    
    \begin{subfigure}[c]{\textwidth}
        \centering
        \includegraphics[trim=1.3cm 6.3cm 2cm 6.3cm,clip=true,width=0.32\textwidth]{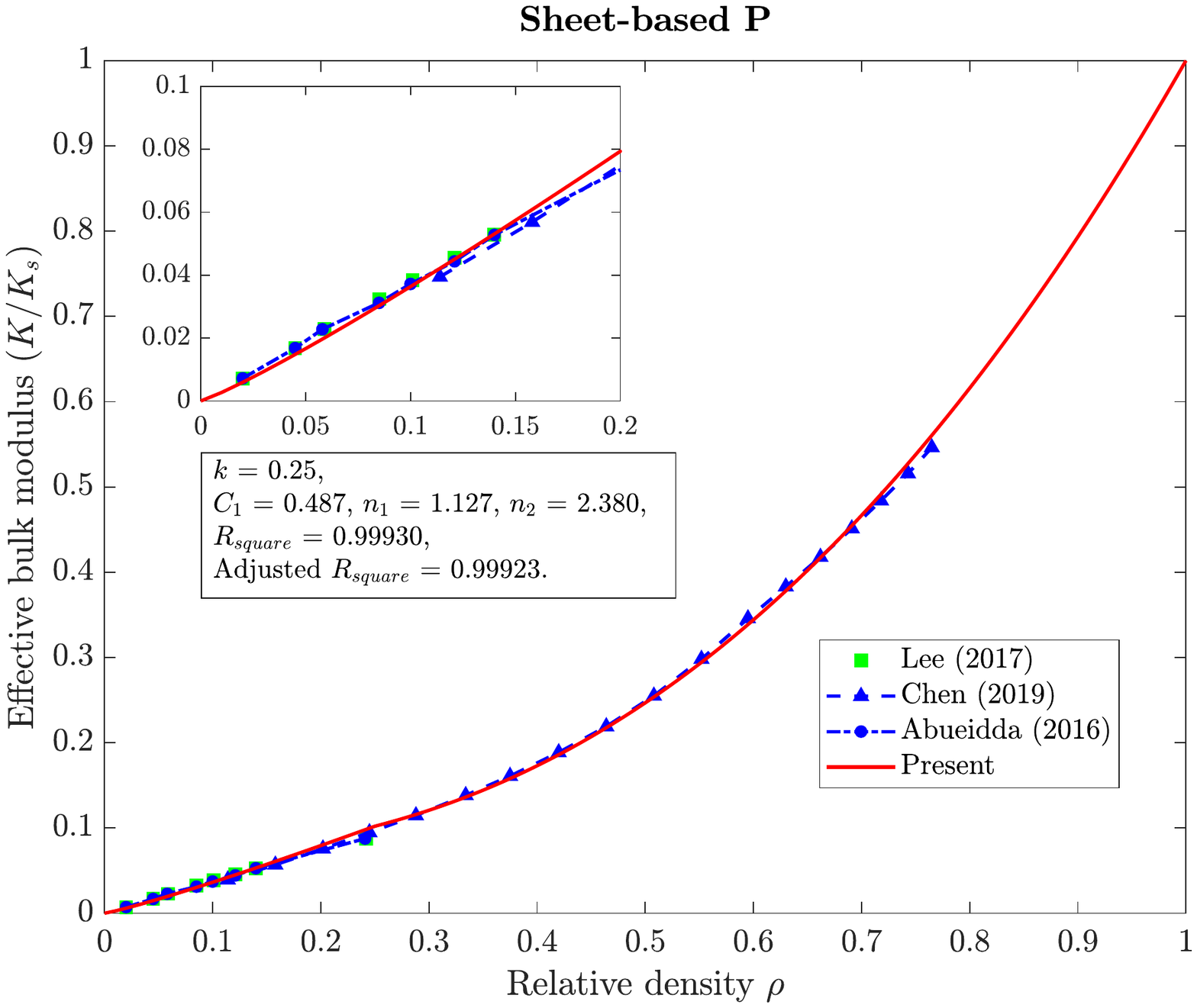}
        \hfill
        \includegraphics[trim=1.3cm 6.3cm 2cm 6.3cm,clip=true,width=0.32\textwidth]{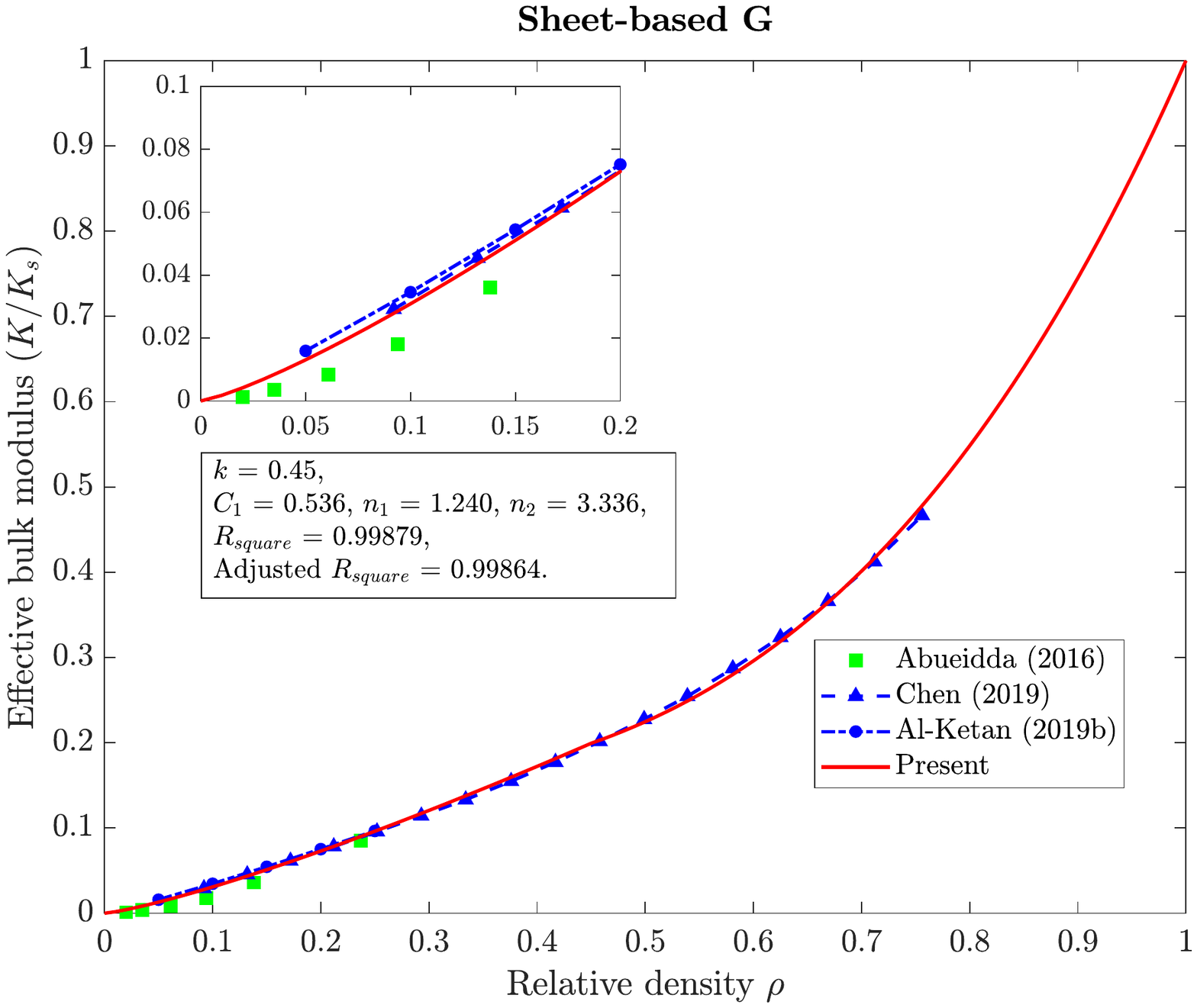}
        \hfill
        \includegraphics[trim=1.3cm 6.3cm 2cm 6.3cm,clip=true,width=0.32\textwidth]{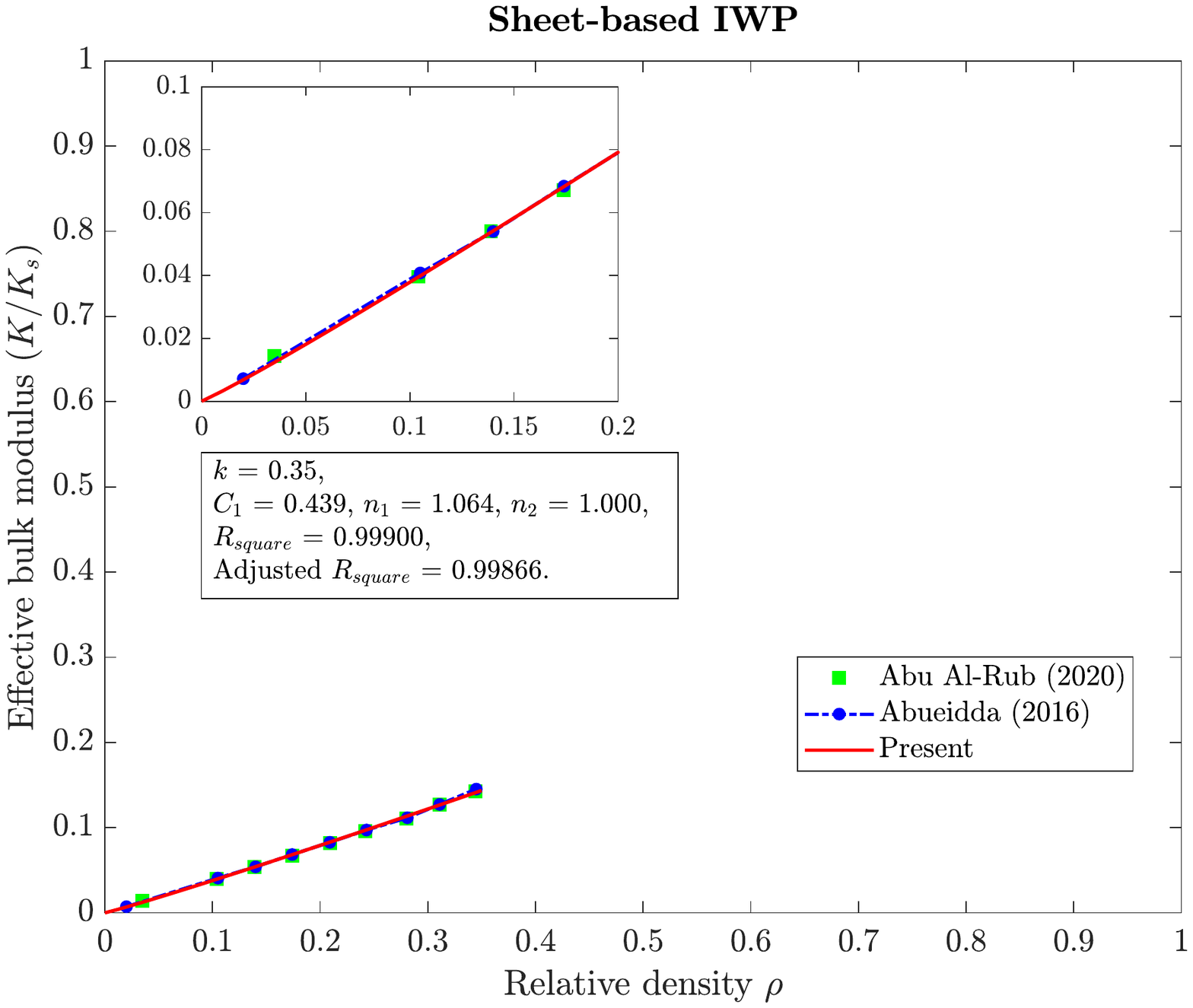}
        \caption {Effective bulk modulus, $K^*$}
    \end{subfigure}
    \caption {Effective mechanical moduli of Primitive, Gyroid, and IWP.}
    \label{fig:EGK_TPMS}
\end{figure}

For the uniaxial response, the power values of the first region ($n_1$) of all three structures are higher than unity, which means that both stretching and bending modes can be found during deformation. This corresponds to the new type of structure (shellular) \citep{han15_a-new}. The Primitive has the value of this coefficient being 1.264, and this result is close to the value of $1.17$ reported in \citep{lee17_stiffness}. A similar result is also obtained with IWP ($n_1 = 1.225$). In contrast, Gyroid reaches the highest value ($n_1 = 1.467$), which shows that the bending mode appears to be significant even under low relative density. The comparable values of these power coefficients could also be found in \citep{abueidda17_mechanical, rashid20_effective} and \citep{al-ketan18_mircoarchitected, abueidda19_mechanical} for IWP and Gyroid structures, respectively. Moreover, a large difference in the deformation behavior of these TPMSs is observed from the value of the second region power coefficient ($n_2$). For instance, as the relative density reaches $0.25$, Primitive provides mostly bending mode with this value being approximately $2$. In the case of IWP structure, stretching-dominated deformation can occur with any state of porosity ($n_2 = 1.782$). An exceptional value of this power coefficient belongs to the Gyroid structure. This value indicates the combination of bending and another deformation mode which results in a higher value of the power coefficient taking $2.351$.

Furthermore, the region-splitting values ($k_m$) for elastic modulus show high accuracy on both shear modulus and bulk modulus. Specifically, in the light foam region which has been mainly investigated in previous studies, the shear response of P-type TPMS is suggested as a stretching-dominated deformation ($n_1 = 1.189$) which was reported by Lee \textit{et al.} \citep{lee17_stiffness}. The power coefficient of $1.544$ is slightly greater than the value of $1.33$ achieved in \citep{Abu16}. However, both values indicate the combination of stretching and bending behaviors of the Gyroid structure. A similar agreement can be concluded with IWP according to the publication of Abu Al-Rub \textit{et al.} \citep{rashid20_effective}. In the case of hydrostatic load, the bulk modulus power coefficients ($n_1$) of $1.127$, $1.240$, and $1.064$ are equivalent to the values of $0.980$, $1.649$, and $1.072$ corresponding to P, G, and IWP structures listed in the comparison sources. However, due to the lack of data on the heavy foam region of IWP bulk modulus, we only provide the fitting results of the first region of this TPMS. In other words, these values might be varied when different splitting points $k_m$ of each structure are chosen. 

The Poisson's ratio ($\nu$) of the aforementioned TPMSs can be generated along with the fitting results of uniaxial and hydrostatic modulus given by Eq. (\ref{eq. nu-E-K_Relationship}). However, a similar fitting approach to other mechanical properties might simplify the above expression for application. A piece-wise formulation with an exponential function at the first region and a quadratic function at the other one as in Eq. (\ref{eq. fit.function_nu}) is investigated in this study. The major difference between the two proposed functions is that the two regions of the second formulation are separated by a value of relative density ($k_{\nu}$) which provides Poisson's ratio of the parent material (i.e. $\nu = \nu_s$). This assumption was concluded from the similar tendencies of considering TPMSs and could be considered a two-phase continuous condition. In addition, at fully dense material, the Poisson's ratio value needs to the consistent with the constituent material. The mathematical equations for these conditions are shown in Eq. (\ref{eq. fit_nu.BC}):
\begin{equation} \label{eq. fit.function_nu}
    \nu = 
    \begin{cases}
		a_1 e^{b_1 \rho} + d_1, & \rho \leq k_{\nu} \\
		a_2 \rho^2 + b_2 \rho + d_2, & \rho > k_{\nu}
	\end{cases}
\end{equation}
\begin{equation} \label{eq. fit_nu.BC}
    \begin{array}{*{20}{l}}
        \textrm{1}^{st} \textrm { Continuous Condition} &  a_1 e^{b_1 k_{\nu}} + d_1 = \nu_s\\
        \textrm{2}^{nd}\textrm { Continuous Condition} &  a_2 k_{\nu}^2 + b_2 k_{\nu} + d_2 = \nu_s\\
        \textrm {Boundary Condition} & a_2 + b_2 + d_2 = \nu_s.
	\end{array}
\end{equation}

In fact, the coefficient $k_{\nu}$ can be pre-defined according to the constituent material Poisson's ratio in this study ($\nu_s = 0.3$). This coefficient value can be considered the common value in metallic materials. Generally, three independent coefficients ($a_1$, $a_2$, and $b_1$) are chosen to create the homogenization function. The others could be generated by using the above boundary conditions as $d_1 = \nu_s - a_1 e^{b_1 k_{\nu}}$, $b_2 = -a_2 (k_{\nu}+1)$, and $d_2 = \nu_s - a_2 - b_2$. Poisson's ratio curve coefficients and graphs after the fitting process are shown in Table \ref{tab.H_nu} and Fig. \ref{fig:nu_TPMS}, respectively. The IWP structure could be fitted using this strategy despite the missing data, and Poisson's ratio value of this structure in the heavy foam region might be considered a prediction. Nonetheless, it is suggested that these fitting results should only be applied while Poisson's ratio of the constituent material coincides with the fitting data. TPMS structures fabricated by various parent materials might behave differently in this property due to the divergence in effective bulk modulus ($K^*$). In general, this fitting strategy can give a good estimation of the subordination of Poisson's ratio on the relative density of TPMS structures.

\begin{table}[ht!]
    \centering
    \caption{Fitting curve for Poisson's ratio}
    \label{tab.H_nu}
	{\renewcommand\arraystretch{1.3}
	{\tabcolsep = 1.5mm
	\begin{tabular}{lcccccc}
        \hline \\[-4mm]
        \multirow{1}{*}{TPMS} & \multirow{1}{*}{$a_1$} & \multirow{1}{*}{$b_1$} & \multirow{1}{*}{$a_2$} & \multirow{1}{*}{Expression} & \multirow{1}{*}{$R$-square} & \multirow{1}{*}{\shortstack[l]{Adjusted \\[1.2mm] $R$-square}}\\[4mm]
        \hline \\[-4mm]
		Primitive & 0.314 & $-$1.004 & 0.152 & $\nu = \begin{cases} 
    		0.314 e^{-1.004 \rho} + 0.119, & \rho \leq 0.55 \\
    		0.152 \rho^2 - 0.235 \rho + 0.383. & \rho > 0.55
        \end{cases}$ & 0.9862 & 0.9856 \\[5mm]
        \hline \\[-4mm]
		Gyroid & 0.192 & $-$1.349 & 0.402 & $\nu =\begin{cases} 
    		0.192 e^{-1.349 \rho} + 0.202, & \rho \leq 0.50 \\
    		0.402 \rho^2 - 0.603 \rho + 0.501. & \rho > 0.50
        \end{cases}$ & 0.9879 & 0.9874 \\[5mm]
        \hline \\[-4mm]
		IWP & 2.597 & $-$0.157 & 0.201 & $\nu = \begin{cases} 
    		2.597 e^{-0.157 \rho} - 2.244, & \rho \leq 0.13 \\
    		0.201 \rho^2 - 0.227 \rho + 0.326. & \rho > 0.13
        \end{cases}$ & 0.9932 & 0.9923 \\[5mm]
        \hline \\
	\end{tabular}}}
\end{table}

\begin{figure}[!ht]
    \centering
    \begin{subfigure}[c]{0.32\textwidth}
        \centering
        \includegraphics[trim=1.3cm 6.3cm 2cm 6.3cm,clip=true,width=\textwidth]{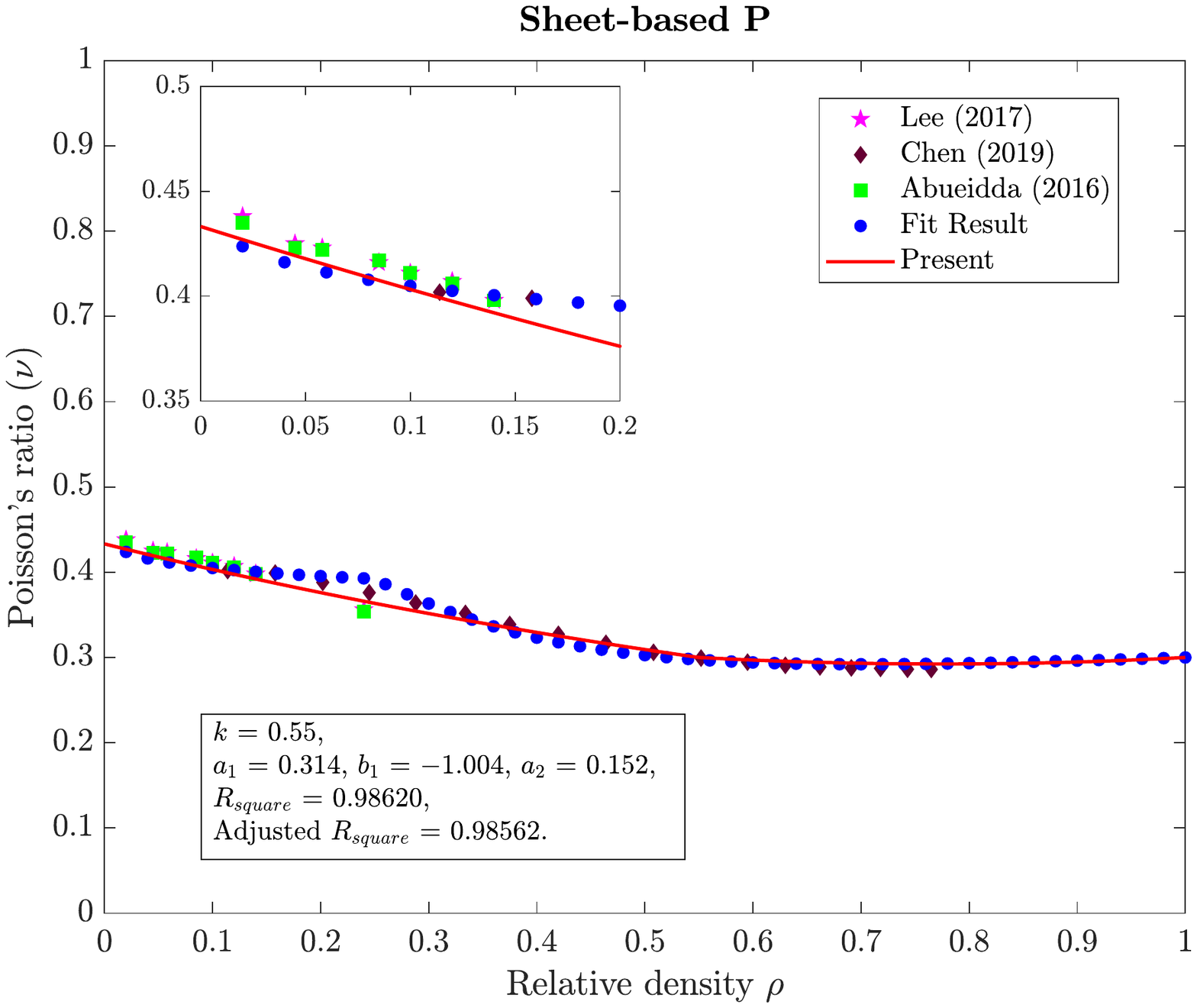}
        \caption {Primitive (P)}
    \end{subfigure}
    \hfill
    \begin{subfigure}[c]{0.32\textwidth}
        \centering
        \includegraphics[trim=1.3cm 6.3cm 2cm 6.3cm,clip=true,width=\textwidth]{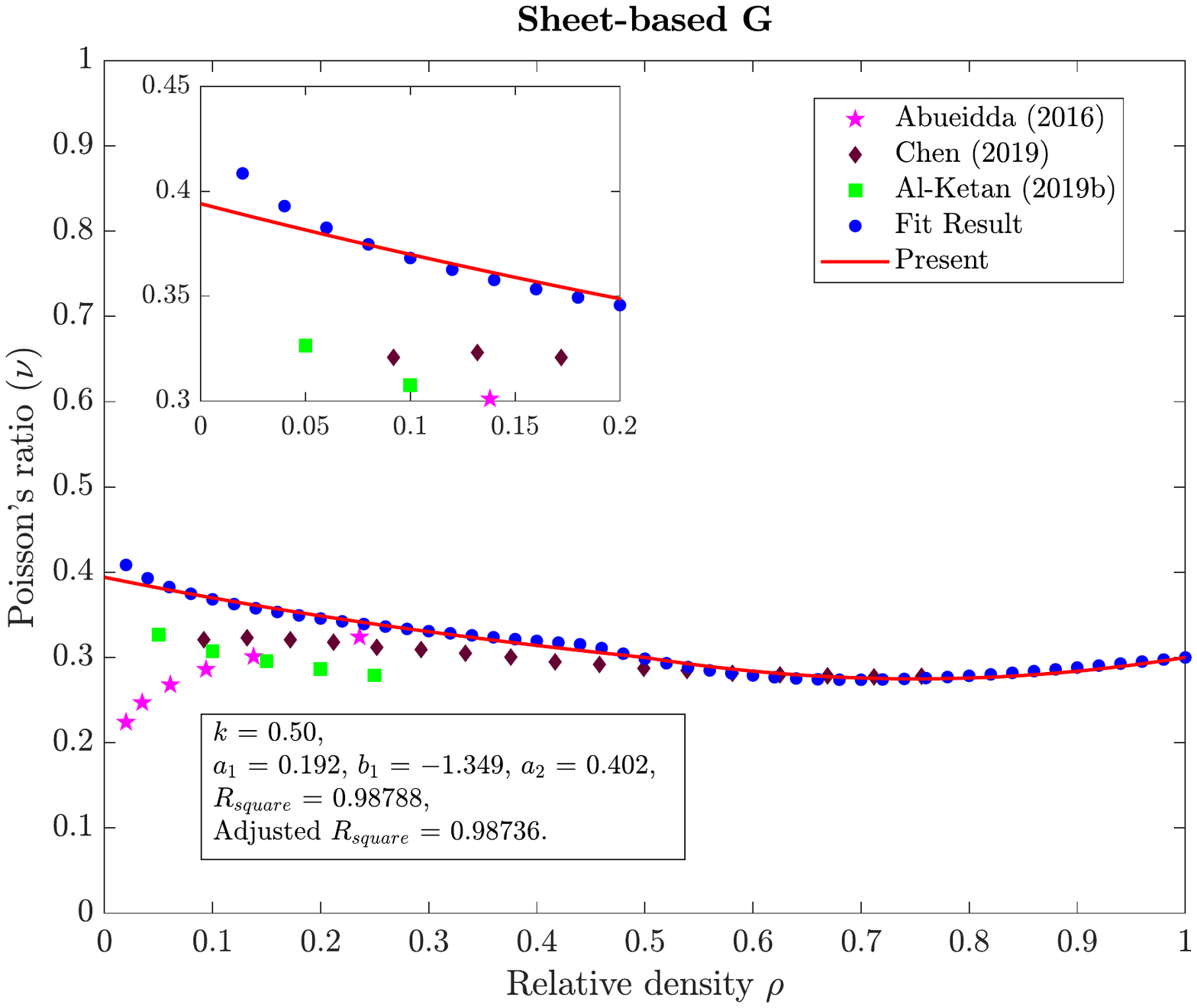}
        \caption {Gyroid (G)}
    \end{subfigure}
    \hfill
    \begin{subfigure}[c]{0.32\textwidth}
        \centering
        \includegraphics[trim=1.3cm 6.3cm 2cm 6.3cm,clip=true,width=\textwidth]{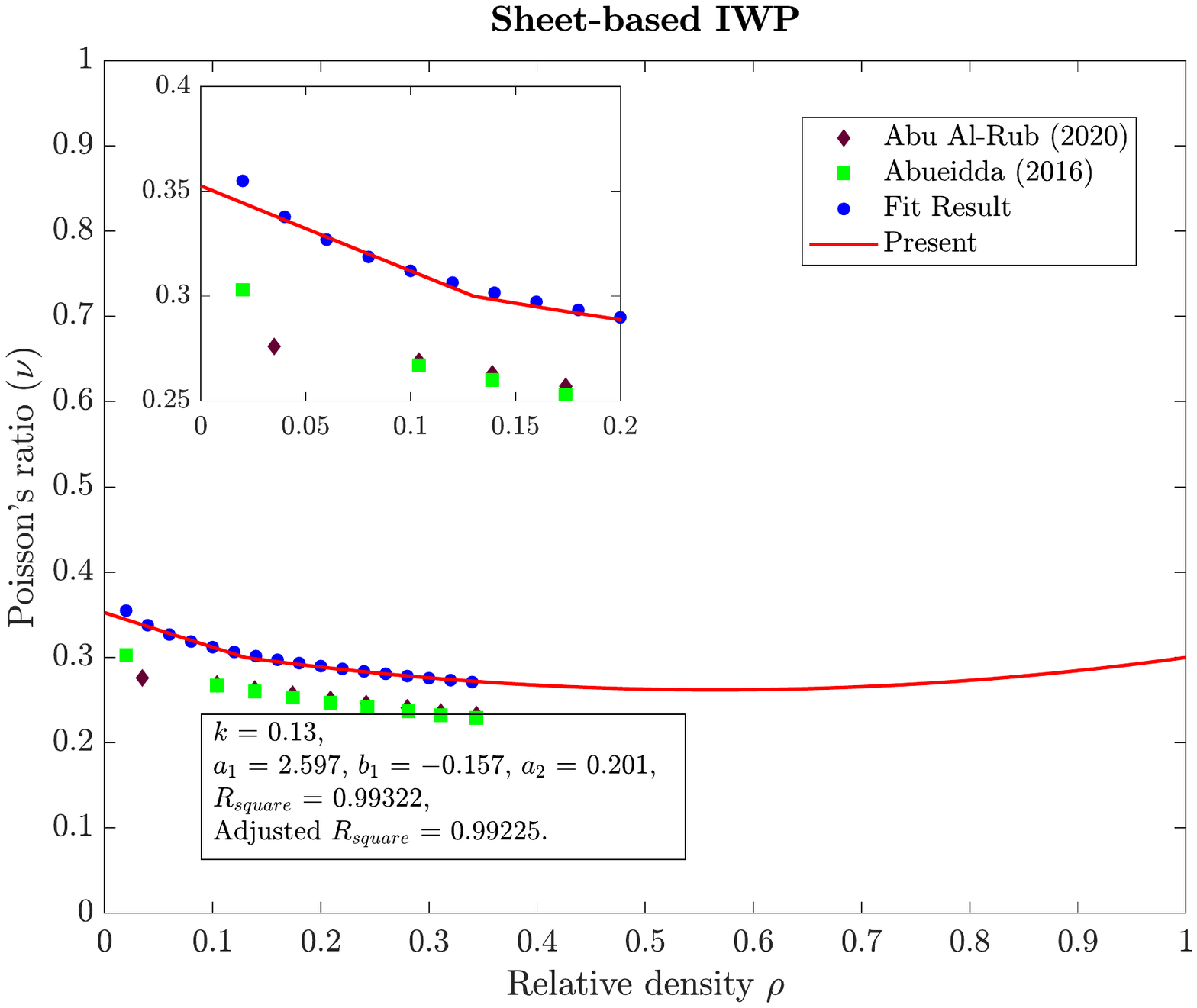}
        \caption {I wrapped package-graph (IWP)}
    \end{subfigure}
    \caption {Poisson's ratio of Primitive, Gyroid, and IWP.}
    \label{fig:nu_TPMS}
\end{figure}

\section{Isogeometric analysis of FG-TPMS plates}
\label{Section 3}
\subsection{\textit {Configuration of FG-TPMS plates}}
\label{Section 3.1}

In the current study, the FG-TPMS plate is described as a porous plate with an FG-TPMS core and two skin layers at both top and bottom of the plate. As presented in the previous section, the gradation approach is utilized for the TPMS shell thickness by changing the control parameter $t$ of the implicit function. We propose two main functional patterns for gradation that are given in Eq. (\ref{eq.Distribution}). The Cartesian coordinate system in physics is set as in Fig. \ref{fig:FGplateZoom}, where the plate mid-plane is adopted as the $x_{1}Ox_{2}$ plane and the $x_{3}$ direction demonstrates the plate thickness. The symbols $a$, $b$, and $h$ denote the dimension of the plate in the $x_1$, $x_2$, and $x_3$ axes, respectively. In addition, the thickness-to-length ratio ranging from $0.005$ to $0.2$ is applied in the numerical example to provide various views of both thin and thick plate behaviors. Furthermore, the skin layers of this plate might be fabricated simultaneously or separately from the inner core, using the same material. It is suggested that these skins are only used as a uniform load transmission and should not be included in the analysis. To reduce their influences on the comprehensive performance of the FG-TPMS plate, the maximum value of their thickness is set to be smaller than $0.001 h$. With this minor value, the contributions of these layers to all the in-plane, flexural, and shear stiffness can be negligible (e.g. less than $1\%$).
\begin{figure}[!ht]
    \centering
    \includegraphics[trim=0cm 0.5cm 14cm 3.5cm, clip=true, angle=270, width=0.9\textwidth]{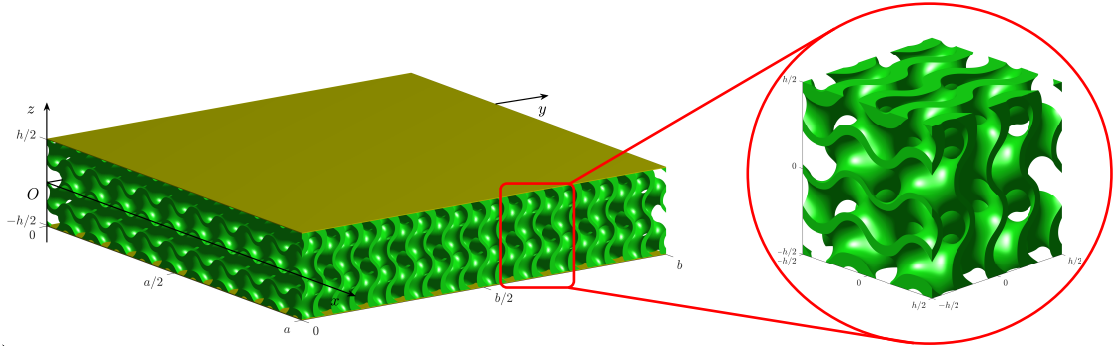}
    \caption {A detailed zoom of FGP-TPMS type G with density pattern B and the power $n=1$.}
    \label{fig:FGplateZoom}
\end{figure}

To yield accurate solutions, a five-variable SeSD plate model, which is coincided for both thick and thin plates, is utilized. The assumptions for this model can be listed as follows
\begin{itemize}
    \item Both bending stress and shell stress are taken into account, therefore, the displacement of all three directions are considered variables.
    \item As the impact of shear deformation on the flexural responses of the plate, the angles of a plate's line in $x_3$ direction after deformation with both in-plane axis are denoted as rotations variables.
    \item The shear stress and strain along the plate thickness can be expressed in a nonlinear function.
    \item The out-of-plane normal stress is negligible compared to others. This stress equals zero in the calculation. The bending stiffness of FG-TPMS could be generated by this presumption from the constitutive matrix in Eq. (\ref{eq: TPMS_constitutive}).
\end{itemize}

The mathematical description of the FG-TPMS plate can be provided with the following descriptions.
Let us consider a domain $\Omega$ bounded in $\mathbb{R}^2$ as a mid-plane surface of an FG-TPMS plate, cf. Fig. \ref{fig:FGplate}. Let $\uu_0=(u_{10},u_{20},u_{30})^T$ be the in-plane displacements and the transverse displacement, while  $\mathbf{\beta}=(\beta_1,\beta_2)^T$ are the rotations about the $x_1$ and $x_2$ axes. The plate made of FG-TPMS with elastic modulus $E$, shear modulus $G$, and Poisson's ratio $\nu$ is defined. A displacement field according to HSDT is widely used as
\begin{equation}
\begin{array}{*{20}{l}}
{u_1\left( {x_1,x_2,x_3} \right) = {u_{10}}\left( {x_1,x_2} \right) - x_3 u_{3,1} + f\left( x_3 \right){\beta _1}\left( {x_1,x_2} \right)}\\
{u_2\left( {x_1,x_2,x_3} \right) = {u_{20}}\left( {x_1,x_2} \right) - x_3 u_{3,2} + f\left( x_3 \right){\beta_2}\left( {x_1,x_2} \right)}\\
u_3\left( {x_1,x_2,x_3} \right) = {u_{30}}\left( {x_1,x_2} \right)
\end{array}
 \label{eq:equation1}
\end{equation}
where $f\left( x_3 \right)$ is a shear strain shape function controlling the nonlinear distribution of shear stresses and strains through the plate thickness. According to the previous work \citep{Tuan2016}, the shape functions available can be unified by polynomial forms. Therefore, a seventh-order shear deformation theory (SeSDT) is used for the present study to model and compute the FG-TPMS plate.
A different form of above equations can be written as
\begin{equation}
\uu = \uu_0 + x_3 \uu_1 + f(x_3)\uu_2
 \label{eq:equation2}
\end{equation}
in which
\begin{equation}
\uu_1=\{-u_{3,1},-u_{3,2}, 0 \}^T, \uu_2 = \{\beta_1, \beta_2, 0 \}^T
 \label{eq:equation2bis}
 \end{equation}
 
\subsection{\textit {Discrete governing equations}}
\label{Section 3.2}
From the definitions of the plate displacement field, the in-plane and transverse shear strains become
\begin{equation}
\bm{\varepsilon} = \bm{\varepsilon}_0 + x_3 \bm{\varepsilon}_1 + f(x_3)\bm{\varepsilon}_2 =[1 ~~x_3 ~~ f(x_3)]\bm{\varepsilon}_p, ~~ 
\gamma = f'(x_3)\bm{\varepsilon}_s 
\label{eq:strainfield}
\end{equation}
where \begin{equation}
\begin{array}{l}
{{\bm{\varepsilon }}_p} = \left[ {\begin{array}{*{20}{c}}
	{{{\bm{\varepsilon }}_0}}\\
	{{{\bm{\varepsilon }}_1}}\\
	{{{\bm{\varepsilon }}_2}}
	\end{array}} \right];{{\bm{\varepsilon }}_0} = \left[ {\begin{array}{*{20}{c}}
	u_{10,1}\\
	u_{20,2}\\
	u_{10,2} + u_{20,1}
	\end{array}} \right]\\
{{\bm{\varepsilon }}_1} =  - \left[ {\begin{array}{*{20}{c}}
	u_{3, 11}\\
	u_{3,22}\\
	2u_{3,12}
	\end{array}} \right];{{\bm{\varepsilon }}_2} = \left[ {\begin{array}{*{20}{c}}
	\beta _{1,1}\\
	\beta_{2,2}\\
	\beta _{1,2} + \beta _{2,1}
	\end{array}} \right];{{\bm{\varepsilon }}_s} = \left[ {\begin{array}{*{20}{c}}
	\beta _1\\
	\beta _2
	\end{array}} \right]
\end{array}
\label{eq:equation7}
\end{equation}

\label{eq:equation6}
\noindent Next we define the following material matrices 
\begin{equation}
{{\bf{D}}_p} = \left[ {\begin{array}{*{20}{c}}
	{\bf{A}}&{\bf{B}}&{\bf{E}}\\
	{\bf{B}}&{\bf{D}}&{\bf{F}}\\
	{\bf{E}}&{\bf{F}}&{\bf{H}}
	\end{array}} \right]
	\label{eq:equation8}
\end{equation}
\noindent where
\begin{equation}
\begin{array}{l}
\left[ A_{ij},{B_{ij}},{D_{ij}},{E_{ij}},{F_{ij}},H_{ij}\right]=
\int_{ - \frac{h}{2}}^{\frac{h}{2}} {\left[ 1,x_3,{x_3^2},f\left( x_3 \right),x_3f\left( x_3 \right),f{{\left( x_3 \right)}^2} \right]{{\overline C }_{ij}}\text{d}x_3} \\
D^s_{ij} = \int_{ - \frac{h}{2}}^{\frac{h}{2}} {f'{\left( x_3 \right)}^2 G_{ij} \text{d}x_3}
\end{array}
\label{eq:equation9}
\end{equation}
\noindent and $h$ is the plate thickness and $\bar{C}_{ij}, G_{ij}$ are the tensors derived from
\begin{equation}
\bar {\CC}  = \left[ {\begin{array}{*{20}c}
   \dfrac{E}
{1 - \nu^2 } & \dfrac{E\nu }
{1 - \nu^2 }& 0  \\
   \dfrac{E\nu}
{1 - \nu^2 }  & \dfrac{E}
{1 - \nu^2 } & 0  \\
   0 & 0 & G  \\
 \end{array} } \right];\bar {\GG}  = G\left[ {\begin{array}{*{20}c}
   1 & 0  \\
   0  & 1 
 \end{array} } \right]
\label{eq:bendingmodulus}
\end{equation}

\noindent Let $\cV$ and $\cV_0$ be denoted as
\begin{equation}
\cV = \{\tilde {\uu} =(\uu_0, \vbeta): \uu_0 \in H^1(\Omega)^2\times H^2(\Omega), \vbeta \in H^1(\Omega)^2 \}\cap
\mathcal{B} \label{eq:space}
\end{equation}
\begin{equation}
\cV_0=\{\tilde {\vv} = (\vv_0, \veta): \vv_0 \in H^1(\Omega)^2\times H^2(\Omega), \veta \in H^1(\Omega)^2:\vv_0=\mathbf{0}, \veta=\mathbf{0} \text{~on~} \partial\Omega \} \label{eq:space2}
\end{equation}
with $\mathcal{B}$ being a set of the essential boundary conditions.

A weak form of the static analysis is to
seek $\tilde {\uu} \in \cV$ such that
\begin{equation}
\forall \tilde {\vv} \in \cV_0, ~~a_p(\tilde {\vv},\tilde {\uu}) +  a_s (\veta, \vbeta) =(\tilde v_3, q)\label{eq:staticweakform}
\end{equation}
\noindent where $q$ is a transverse load; meanwhile the scalar and bilinear operators are expressed by 
\begin{equation}
a_p(\tilde {\vv},\tilde {\uu}) =\int_\Omega  { {\bm{\varepsilon }}_p(\tilde {\vv}):{\bf{D}}_p:{{\bm{\varepsilon }}_p(\tilde {\uu})}\text{d}\Omega },~
a_s (\veta, \vbeta) = \int_\Omega  { \bm{\varepsilon }_s(\veta):{\bf{D}}^s:{{\bm{\varepsilon }}_s}(\vbeta)\text{d}\Omega },~ (v_3,u_3)= \int_{\Omega} v_3 u_3 \text{d}\Omega
\label{eq:equation61}
\end{equation}

In the case of free vibration analysis, a weak form is to find $\omega \in \mathbb{R}^{+}$ and $\mathbf{0}\neq \tilde{\uu} \in \cV$ such
that
\begin{equation}
\forall \tilde {\vv} \in \cV_0, ~~a_p(\tilde {\vv},\tilde {\uu}) +  a_s (\veta, \vbeta) =\omega^2  a_m (\tilde {\vv},\tilde {\uu})\label{eq:dynamicweakform}
\end{equation}
where  
\begin{equation}
a_m (\tilde {\vv},\tilde {\uu}) = \int_{\Omega} \left[ {\begin{array}{*{20}{c}}
	\vv_0\\
	\vv_1\\
	\vv_2
	\end{array}} \right]^T\tilde \mm\left[ {\begin{array}{*{20}{c}}
	\uu_0\\
	\uu_1\\
	\uu_2
	\end{array}} \right]\text{d}\Omega, 
\label{eq:massI}
\end{equation}
\begin{equation}
	\tilde m_{ij} = 	\tilde m_{ji},~  \left[(\tilde m_{11}, \tilde m_{12}, \tilde m_{13}, \tilde m_{22}, \tilde m_{23}, \tilde m_{33}\right]  = \int_{ - \frac{h}{2}}^{\frac{h}{2}} \rho (x_3) \left[1,x_3,x_3^2, f(x_3), x_3f(x_3), f^2(x_3)\right]\text{d}x_3 
    \label{eq:mass_inertia}
\end{equation}
and $\rho$ and $\omega$ denote the mass density and natural frequency, respectively.

For in-plane buckling behavior with pre-buckling stresses $\hat \Sigma _0$, the weak form associated with the gradients of transverse deflection is to seek $\lambda_{cr} \in \mathbb{R}^{+}$ and $\mathbf{0}\neq \tilde{\uu} \in \cV$ s.t.
\begin{equation}
\forall \tilde {\vv} \in \cV_0, ~~a_p(\tilde {\vv},\tilde {\uu}) +  a_s (\veta, \vbeta)=\lambda _{cr} \{h~ a_g(\tilde v_3,u_3)\}
\label{eq:bucklingweakform}
\end{equation}
where
\begin{equation}
a_g(v_3, u_3)=\int_{\Omega}\nabla^T v_3 \hat \Sigma _0 \nabla u_3
\text{d}\Omega, ~~
 \hat \Sigma _0  = \left[ {\begin{array}{*{20}c}
   {\vsigma _1^0 } & {\vsigma _{12}^0 }  \\
   {\vsigma _{12}^0 } & {\vsigma _2^0 }  \\
 \end{array} } \right]
\label{eq:equation81}
\end{equation}
where $\nabla = [\partial/\partial_{x_1}~~\partial/\partial_{x_2}]^T$ indicates the gradient operator, $\sigma _1^0 = \sigma_{cr}, \sigma _2^0=\sigma _{12}^0 =0$ and  $\sigma_1^0 = \sigma _2^0=\sigma_{cr}, \sigma _{12}^0 =0$ stand for a uni-axial buckling analysis, and a bi-axial buckling analysis, respectively, where $\sigma_{cr}$ denotes critical buckling load.

It can be seen from the bilinear term $a_p(:,:)$ that the second derivatives in the displacement variable, $u_3$. Hence, approximate solutions require at least the $C^1$ continuity which causes difficulties using standard finite element methods. Alternatively, isogeometric analysis allows us to produce both exact geometry and arbitrarily high-order continuity. The next section describes numerical solutions using the NRUBS-based IGA.

\subsection{\textit {NURBS-based IGA approximations}}
\label{Section 3.3}
For the goal of this study, we mention only the basic background of isogeometric analysis. More details of the advanced techniques on IGA are provided in the textbook \citep{Cottrell2009}. A knot vector $\pmb{\Xi}=\{\xi_1,\xi_2,\xi_3,...\xi_{n+p+1}\}$, which increases in the interval of $\xi_1$ and $\xi_{n+p+1}$, is a primary component constituting NURBS basis functions. The B-spline basis functions provide non-negative values, $C^{p-1}$ continuous at a single knot, and $C^{\infty}$ continuous inside a knot span. A multiple knot means the same knot value repeated $k$ times for the continuity up to $C^{p-k}$. To ease in numerical computations, an open knot vector which has the first and last knots repeated ($p + 1$) times, is used.

By using Cox-de Boor algorithm, B-spline basis functions $N_{i,p}\left(\xi\right)$ of order $p$ $\ge 0$ in the one-dimensional (1D) space are compute by 

\begin{equation}\label{eq:Bspline1}
    N_{i,0}(\xi) = \left.
    \begin{cases}
    1 & if \quad \xi_i \le \xi < \xi_{i+1}\\
    0 & otherwise
\end{cases}
\right \}
\end{equation}
\begin{equation}\label{eq:Bspline2}
    N_{i,p}\left(\xi\right)=\frac{\xi-\xi_i}{\xi_{i+p}-\xi_i}N_{i,p-1}\left(\xi\right) +\frac{\xi_{i+p+1}-\xi}{\xi_{i+p+1}-\xi_{i+1}}N_{i+1,p-1}\left(\xi\right)
\end{equation}
Two-dimensional (2D) NURBS basis functions are then denoted as follows
\begin{equation}\label{eq:NURBS}
  R_I(\xi,\chi)=\frac{N_{i,p}(\xi)M_{j,q}(\chi)w_{i,j}}{\sum_{i=1}^{n}\sum_{j=1}^{m}N_{i,p}(\xi)M_{j,q}(\chi)w_{i,j}}
\end{equation}
where $N_{i,p}(\xi)$ and $M_{j,q}(\chi)$ are B-spline basis functions associated with two knot vectors $\pmb{\Xi}$ and $\pmb{X}$, respectively, and $w_{i,j}$ are the positive weights of $i,j$ control points.

\begin{figure}[!ht]
    \centering
    \begin{subfigure}[c]{0.48\textwidth}
        \centering
		\includegraphics[trim=1cm 0.5cm 1cm 0.5cm, clip=true, width=\textwidth]{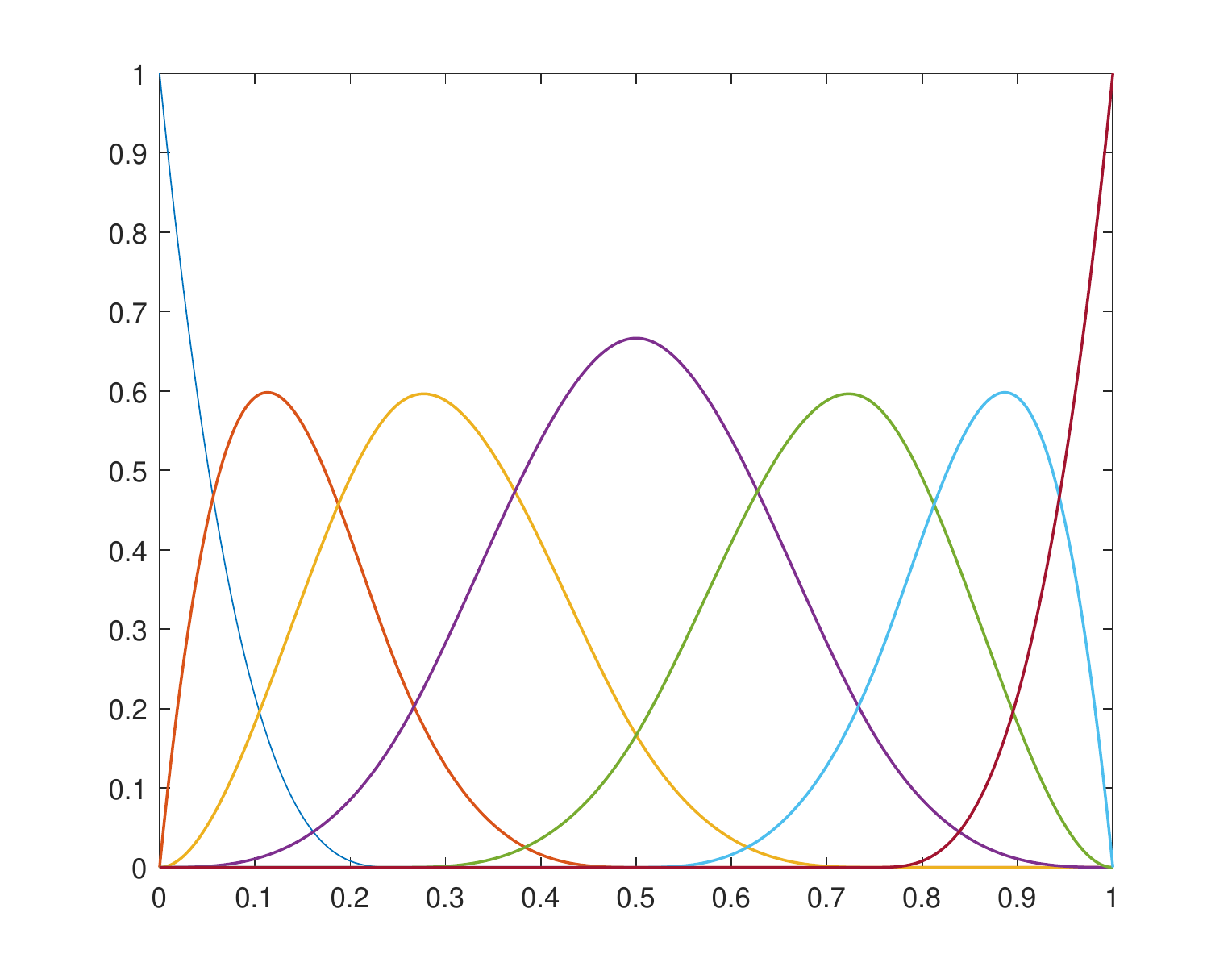}
		\caption{B-spline basis functions}
    \end{subfigure}
    \hfill
    \begin{subfigure}[c]{0.48\textwidth}
        \centering
		\scalebox{1}[1.05]{\includegraphics[trim=1cm 0.4cm 1cm 0.5cm, clip=true, width=\textwidth]{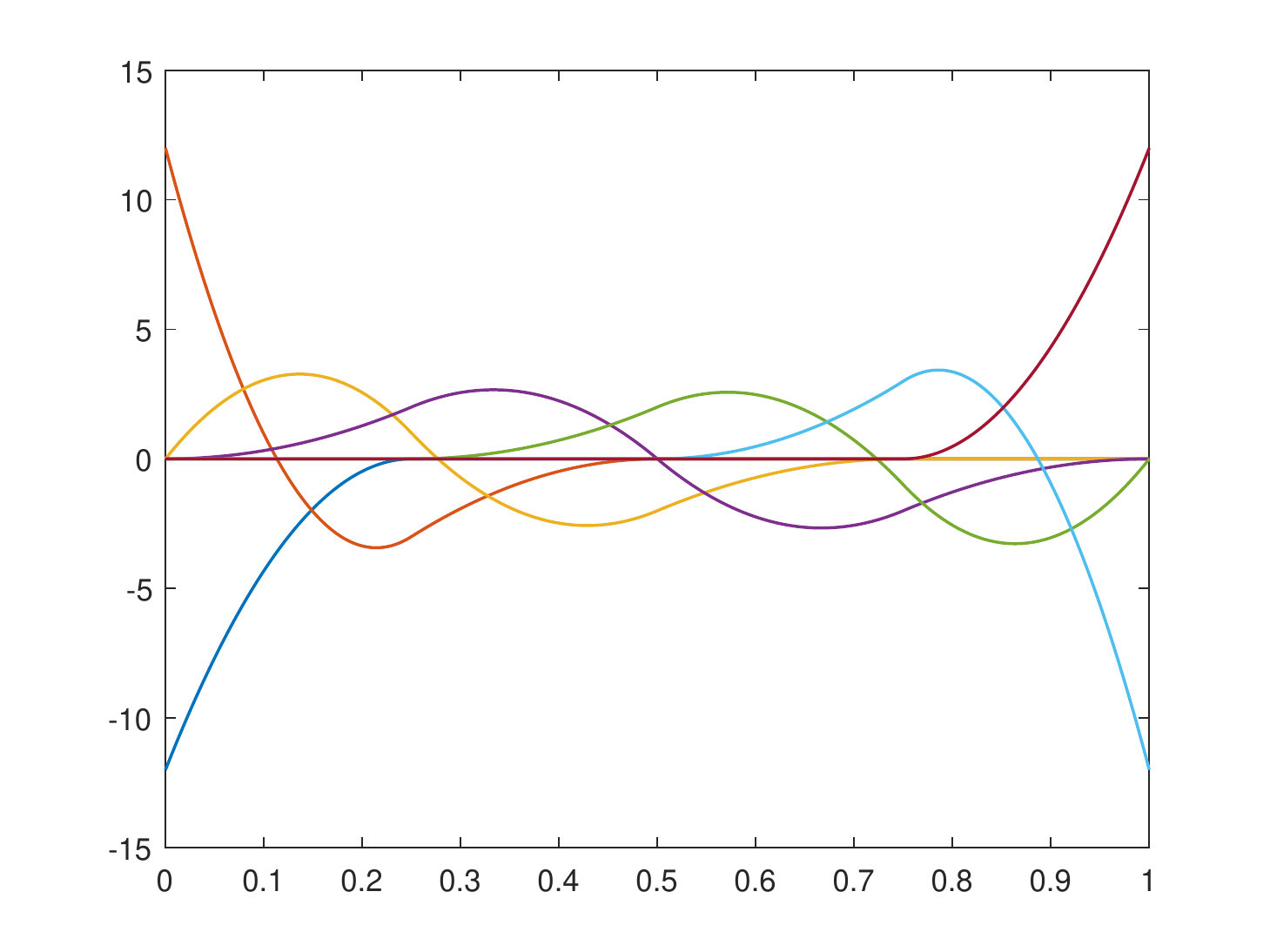}}
        \caption {Derivation}
    \end{subfigure}
	\caption{1D $3^{rd}$-order B-spline basis functions and its derivation, $\pmb{\Xi}=\{0,0,0,0, 1/4, 1/2, 3/4, 1, 1, 1, 1\}$.} \label{fig:1D_Bspline}
\end{figure}

\begin{figure}[!ht]
    \centering
    \includegraphics[trim=12cm 2.2cm 11cm 2.2cm,clip=true,width=0.6\textwidth]{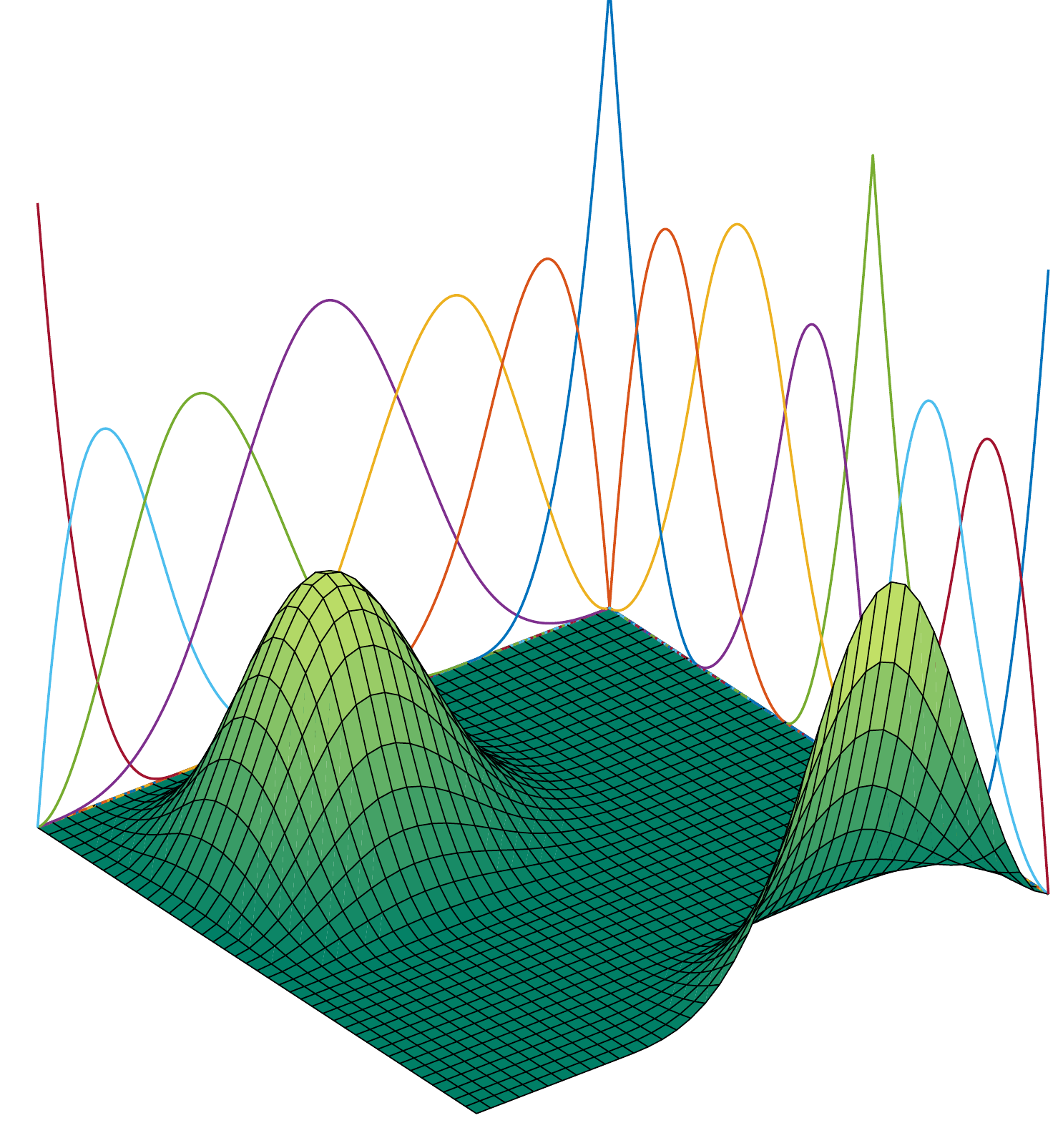}
    \caption{2D B-spline basis functions associated with two knot vectors $\pmb{\Xi}=\{0,0,0,0, 1/4, 1/2, 3/4, 1, 1, 1, 1\}$ and $\pmb{X}=\{0,0,0, 1/5, 2/5, 3/5, 3/5, 1, 1, 1\}$.}
    \label{fig:2D_Bspline}
\end{figure}

Then the approximate solution ${\bf {\tilde{u}}}^h$ can be expressed using NURBS basis functions as follows
\begin{equation}\label{eq:disNURBS}
{\bf \tilde{u}}^h(\xi,\chi) = \sum_{I}^{mxn}R_I(\xi,\chi)\mathbb{I}_{5\times 5}{\bf d}_I
\end{equation}
For computations, the bounded domain $\Omega$ is divided into $N_e$ patches such that
$\Omega  \approx  \bigcup _{e = 1}^{N_e} \Omega ^e$ and $\Omega ^i  \cap \Omega ^j  = \emptyset ~,~~i \ne j $. The approximate solution of the static problem using NURBS basis functions for the FG-TPMS plate is to seek $\tilde {\uu}^h \in \cV^h$ such that
\begin{equation}
\forall  \title \vv^h  \in \cV_0^h, ~~a_p(\tilde {\vv}^h ,\tilde {\uu}^h ) +  a_s (\veta^h,\vbeta^h)=(\tilde v_3^h,p)
\label{eq:staticFEM}
\end{equation}
\noindent where the NURBS patch spaces, $\cV^h$ and $\cV^h_0$, are specified by
\begin{equation}
\cV^h=\{\tilde {\uu}^h=(\uu_0^h,\vbeta^h) \in \cV, \uu_0^h|_{\Omega^e} \in P_k(\Omega^e)^3, \vbeta^h|_{\Omega^e} \in
P_k(\Omega^e)^2 \}\cap \mathcal{B}\\
\end{equation}
\begin{equation}
\cV^h_0=\{\tilde {\vv}^h\in \cV_0 : \tilde \vv^h=\mathbf{0} \text{~on~} \partial\Omega \} \label{eq:spaceFEM}
\end{equation}
where $P_k(\Omega^e)$ describes the set of NURBS bases of degree $k$ for each approximate variable.\\

The NURBS-based solution for the free vibration problem is to determine the natural frequency $\omega^h \in \mathbb{R}^{+}$ and $\mathbf{0}\neq \tilde{\uu}^h \in \cV^h$ 
such that
\begin{equation}
\forall \tilde\vv^h \in \cV_0^h, ~~a_p(\tilde\vv^h, \tilde\uu^h) +  a_s (\veta^h,\vbeta^h)
=(\omega^h)^2 a_m (\tilde {\vv}^h,\tilde {\uu}^h)
\label{eq:dynamicFEM}
\end{equation}

Thus, isogeometric analysis is also used to find the critical buckling load ($\lambda^h _{cr}$) of the buckling problem where $\lambda^h _{cr} \in \mathbb{R}^{+}$ and $\mathbf{0}\neq \tilde{\uu}^h \in \cV^h$
\begin{equation}
\forall \tilde\vv^h \in \cV_0^h, ~~a_p(\tilde\vv^h, \tilde\uu^h) +  a_s (\veta^h,\vbeta^h)
=\lambda^h_{cr} \{h~ a_g(\tilde v_3^h,u_3^h)\}
\label{eq:bucklingFEM}
\end{equation}
It is worth emphasizing that in solving the aforementioned problems, several well-known numerical methods \citep{Cottrell2009} are suitable candidates yet isogeometric analysis (IGA) is the most common choice to capture naturally high continuity in the weak form and achieve ultra-accuracy in solutions without any additional variables and computationally expensive requirements. 

\section{Numerical examples}
\label{Section 4}
In this section, the rectangular and circular plates with different architectures and boundary conditions are employed to validate the above IGA approach. Fig. \ref{fig:RecPlate_Input} a) and b) demonstrate the simply supported (SSSS) and fully clamped (CCCC) boundary of rectangular plates. While the corresponding samples of circular plates are plotted in Fig. \ref{fig:CirPlate_Input}. To compare with previous studies, free vibration and buckling behaviors of isotropic plates are investigated. For the buckling analysis, rectangular plates subjected to uniaxial and biaxial compression are shown in Fig. \ref{fig:RecPlate_Input} c) and d), respectively. Furthermore, to achieve good convergence results, the fine mesh with eleven refinements has been adopted for both rectangular and circular plates where an example of one refinement mesh is described in Fig. \ref{fig:Plate_Mesh}. As mentioned, the base materials adopted in this study are metallic materials that have the same Poisson's ratio of $0.3$. Furthermore, a plastic material is also listed, whose Poisson's ratio is $0.33$ instead, to emphasize the differences in this property. This dissimilarity between metallic and plastic TPMS structures is aimed at future investigations. The mechanical properties of these materials are tabulated in Table \ref{tab:mat-properties}.
\begin{figure}[!ht]
    \centering
    \includegraphics[trim=0cm 0cm 0cm 0cm,clip=true,width=\textwidth]{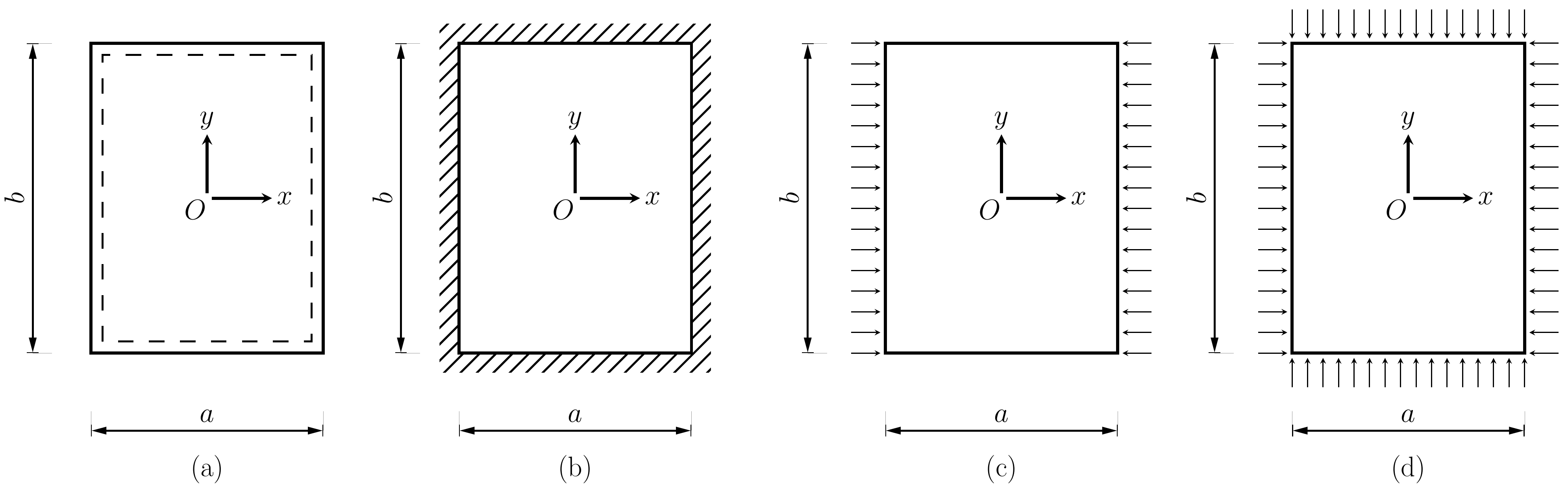}
    \caption {Rectangular plate: a) Simply supported plate, b) Fully clamped plate, c) Uniaxial compression along the $x$ direction, d) Biaxial compression.}
    \label{fig:RecPlate_Input}
\end{figure}
\begin{figure}[!ht]
    \centering
    \includegraphics[trim=0cm 0cm 0cm 0cm,clip=true,width=0.5\textwidth]{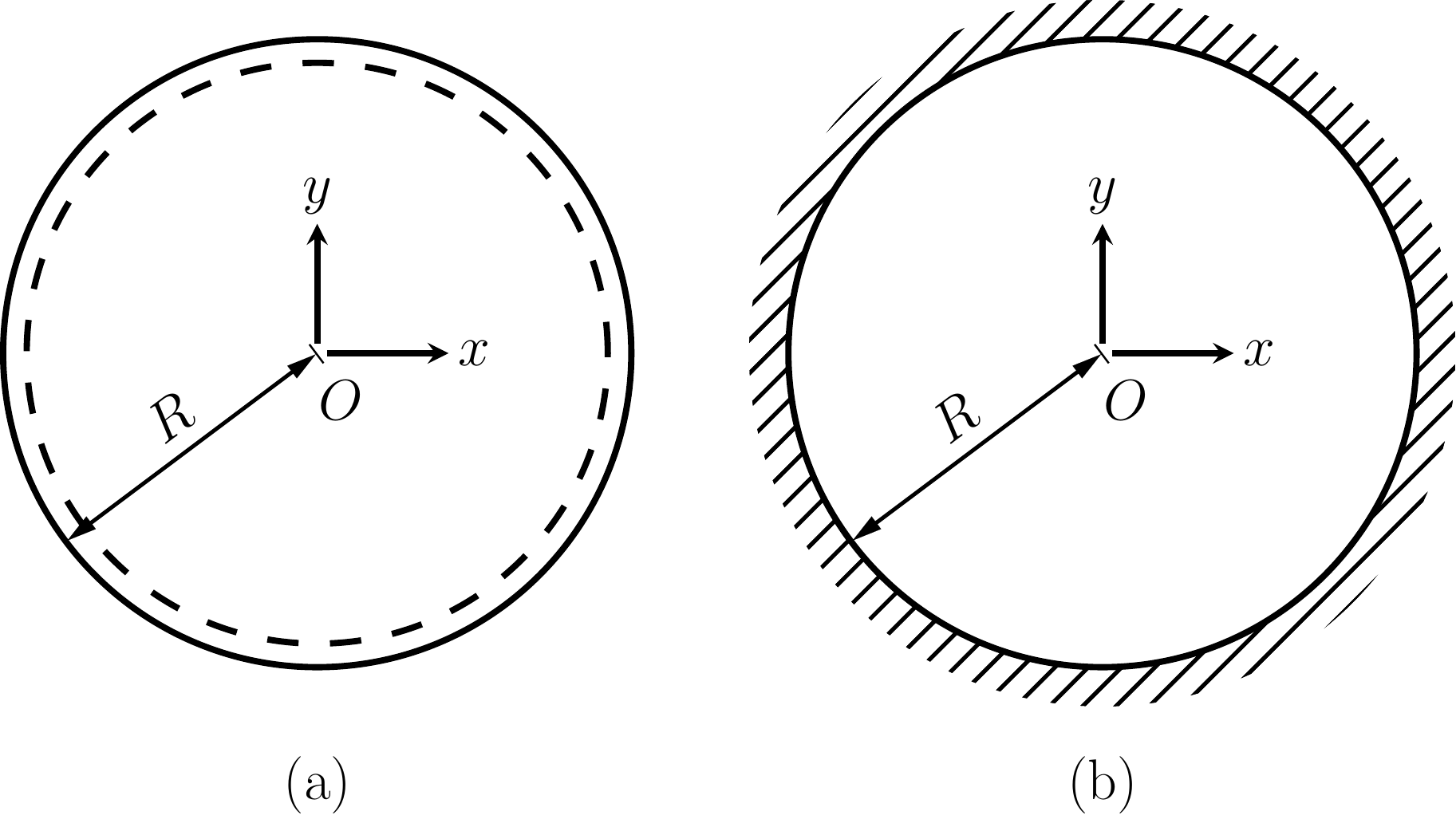}
    \caption {Circle plate: a) Simply supported plate, b) Fully clamped plate.}
    \label{fig:CirPlate_Input}
\end{figure}
\begin{figure}[!ht]
    \centering
    \begin{subfigure}[c]{0.42\textwidth}
        \centering
        \includegraphics[trim=3.7cm 7cm 3.7cm 5.8cm,clip=true,width=\textwidth]{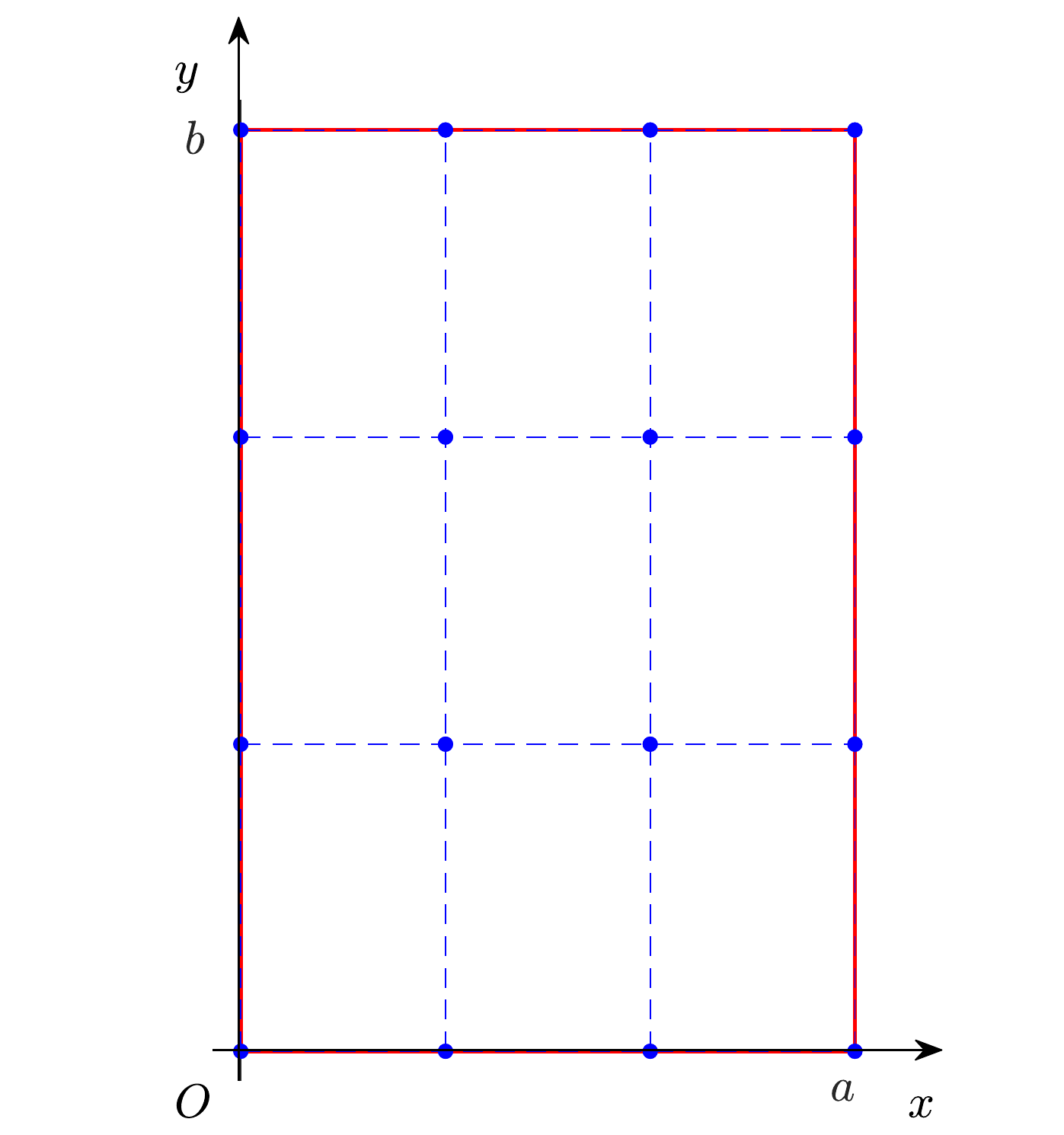}
        \caption {Rectangular plate}
    \end{subfigure}
    \begin{subfigure}[c]{0.45\textwidth}
        \centering
        \includegraphics[trim=4cm 7.3cm 3cm 6.2cm,clip=true,width=\textwidth]{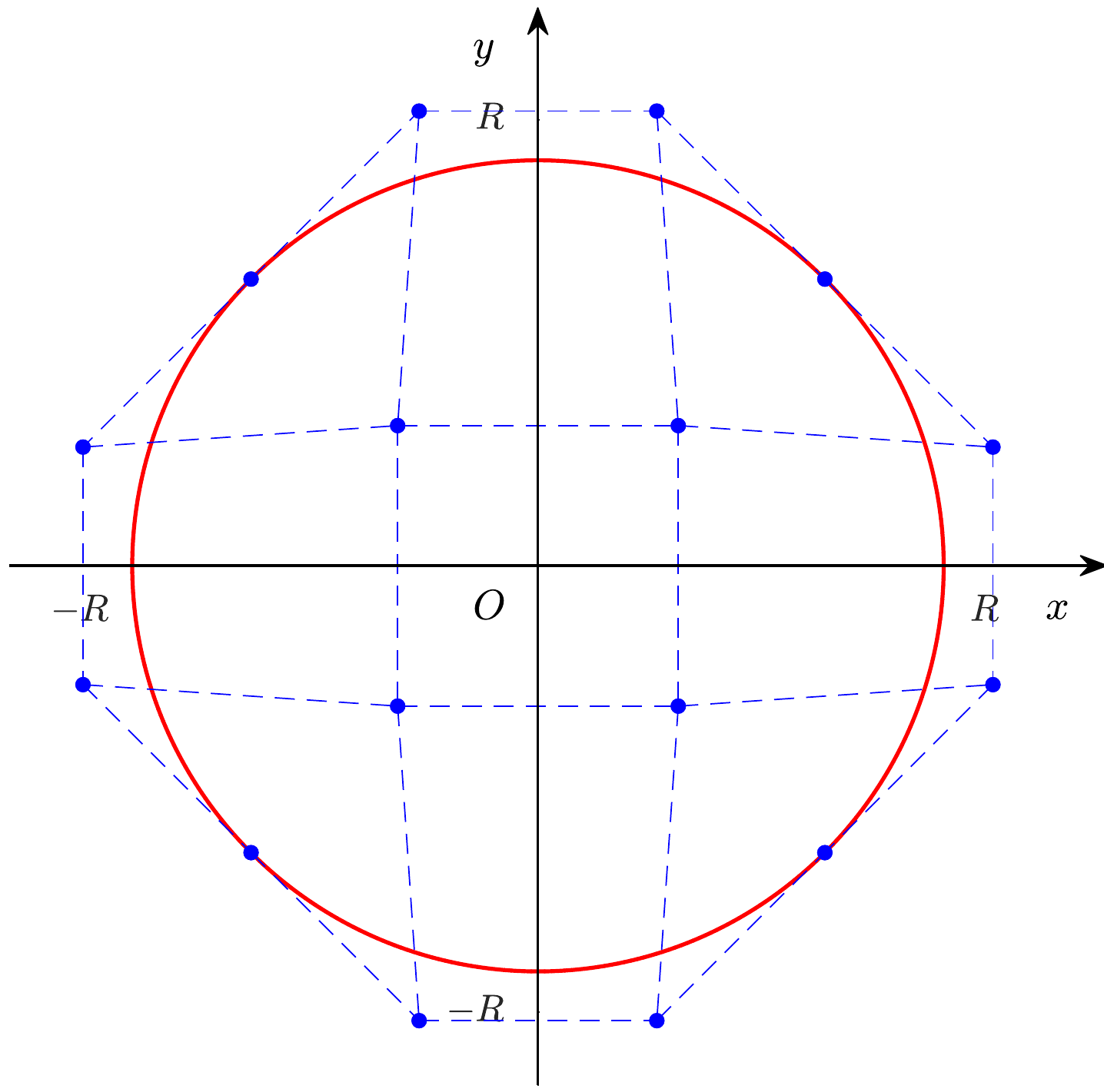}
        \caption {Circular plate}
    \end{subfigure}
    \caption {A coarse mesh with one NURBS element of rectangular and circular plates.}
    \label{fig:Plate_Mesh}
\end{figure}

\begin{table}[!ht]
    \centering
    \caption{Properties of parent materials}
    \label{tab:mat-properties}
    \begin{tabular}{@{}llll@{}} 
        \hline \\[-3mm]
        Material \hphantom{00000000} & $E_s$ (GPa) \hphantom{000000} & $\rho_s$ (kg/m$^3$) \hphantom{000000} & $\nu_s$ \\[1mm] 
        \hline \\[-3mm]
        Aluminum (Al)       &  70   & 2702  & 0.3   \\[1mm]
        Steel (SUS$304$)    &  200  & 8000  & 0.3   \\[1mm]
        Titanium (Ti)       &  106  & 4510  & 0.3   \\[1mm]
        Copper              &  117  & 8960  & 0.3   \\[1mm]
        Brass               &  90   & 8730  & 0.3   \\[1mm]
        Polylactic acid (PLA) & 3.15 & 1240 & 0.33  \\[1mm]
        \hline
    \end{tabular}
\end{table}

Next, six density distribution cases are utilized to illustrate the FG-TPMS plate behaviors. Their average relative densities are approximately $0.35$. Therefore, they might be considered to be foam materials. This value of mean density was adopted in previous studies with functionally graded beams \citep{Viet21}. The corresponding coefficients of porosity distribution patterns A and B are provided in Table \ref{tab:Cases_0.35}. Nevertheless, it is seen that various other scenarios of relative density distribution can be achieved from the changes in these coefficients.
\begin{table}[!ht]
    \centering
    \caption{Six density distribution cases with the same average value of $0.35$}
    \label{tab:Cases_0.35}
    {\renewcommand\arraystretch{1}
    {\tabcolsep = 4mm
    \begin{tabular}{lllll} 
        \hline \\[-3mm]
        Pattern & Case & $\rho_{min}$ & $\rho_{max}$ & $n$ \\[1mm]
        \hline \\[-4mm]
        \multirow{3}{*}{A} & A1 & 0.20 & 0.5 & 1.0 \\
         & A2 & 0.20 & 0.8 & 3.0 \\
         & A3 & 0.25 & 1.0 & 6.5 \\[1mm]
        \multirow{3}{*}{B} & B1 & 0.20 & 0.5 & 0.561 \\
         & B2 & 0.20 & 0.8 & 1.757 \\
         & B3 & 0.25 & 1.0 & 3.943 \\
        \hline
    \end{tabular}}}
\end{table}

Furthermore, Hashin-Shtrikman (HS) hypothesis can provide effective bulk and shear modulus boundaries for multiphase materials in terms of relative density \citep{hashin63_a-variational}. By assigning one of the materials with zero stiffness, the HS upper boundary for porous structure could be introduced which is defined in Eqs. (\ref{eq. HS_UBound_GK}). Due to the isotropic characteristic assumption for both base and result materials, the uniaxial elastic modulus and Poisson's ratio of these porous structures can be described by Eqs. (\ref{eq. HS_UBound_Enu}). A functionally graded porous plate (FGPP) with HS upper boundary mechanical properties is adopted to inspect the efficiency of TPMS geometries in functionally graded plate structures. 
\begin{equation} \label{eq. HS_UBound_GK}
    G_{HS} = G_s + \dfrac{1-\rho}{\dfrac{1}{-G_s} + \dfrac{6(K_s+2G_s)\rho}{5G_s(3K_s+4G_s)}}; \hphantom{000} K_{HS} = K_s + \dfrac{1-\rho}{\dfrac{1}{-K_s} + \dfrac{3\rho}{3K_s + 4G_s}}
\end{equation}
\begin{equation} \label{eq. HS_UBound_Enu}
    E_{HS} = \dfrac{9G_{HS} K_{HS}}{3K_{HS}+G_{HS}}; \hphantom{000} \nu_{HS} = \dfrac{3K_{HS}-2G_{HS}}{2(3K_{HS}+G_{HS})}
\end{equation}

\subsection{\textit{Static and buckling analysis}}
\label{Section 4.1}

At first, the central deflection under a uniform distribution load of both isotropic and porous plates is computed using the new isogeometric approach in this study. The results are given in Table. \ref{tab:Static_FG_0.35} where solutions from other numerical methods are adopted for comparisons, including finite element method FEM \citep{taylor93_linked}, and cell-based smoothed finite element method (CS-FEM) \citep{nguyen-xuan08_a-smoothed}. It can be seen that there is a remarkable agreement between the IGA approach and other methods with less than $1\%$ in deviations. This highest value belongs to the case of the fully clamped plate with a high thickness-to-length ratio ($h/a=0.1$). Next, the buckling analysis of these plates with different aspect ratios is observed by the above isogeometric approach in Table. \ref{tab:Buck_FG_0.35}. In addition to the analytical solutions by \citep{kitipornchai93_buckling}, an edge-based smoothed finite element method (ES-FEM) \citep{nguyen-xuan10_an-edge} has been adopted for verification. Consequently, the strong agreement between the isotropic results and these solutions can validate this method's efficacy. 
\begin{table}[!ht]
    \centering
    \caption{The normalized deflection of the center-point $\left( \bar{w_c}=w_{c} q a^4 / (100 D) \right)$ for square FG plates subjected to uniform load with various density distribution cases ($\rho_{average} = 0.35$), with $D = E_s h^3/(12(1-\nu_s^2))$}
    \label{tab:Static_FG_0.35}
    {\renewcommand\arraystretch{1.2}
    {\tabcolsep = 2mm
    \begin{tabular}{llllllllll} 
        \hline \\[-4mm]
        \multirow{2}{*}{$h/a$} & \multirow{2}{*}{Type} & \multicolumn{3}{c}{Pattern A} & \multicolumn{3}{c}{Pattern B} & \multicolumn{2}{c}{Isotropic plate}\\
        \cline{3-10} \\[-4mm]
         & & A1 & A2 & A3 & B1 & B2 & B3 & 35\% weight & 100\% weight\\
        \hline \\[-4mm]
        
         & & \multicolumn{8}{c}{SSSS} \\[1mm]
        \multirow{4}{*}{$0.01$} & P & 2.9084 & 2.6902 & 2.2201 & 1.8869 & 1.3170 & 1.3061 & \multirow{4}{*}{9.4755} & \multirow{4}{*}{\shortstack[l]{0.4064 (IGA) \\[1.2mm] 0.4064 \citep{nguyen-xuan08_a-smoothed} \\[1.2mm] 0.4064 \citep{taylor93_linked}}} \\
         & G & 3.1701 & 2.9055 & 2.2970 & 2.1466 & 1.4597 & 1.4045 & & \\
         & IWP & 2.7541 & 2.4877 & 2.0396 & 1.9172 & 1.3330 & 1.3078 & & \\
         & FGPP & 2.1041 & 1.9567 & 1.6354 & 1.5306 & 1.1738 & 1.1642 & & \\[2mm]
        \multirow{4}{*}{$0.1$} & P & 3.0073 & 2.7951 & 2.3323 & 2.0033 & 1.4472 & 1.4265 & \multirow{4}{*}{9.5347} & \multirow{4}{*}{\shortstack[l]{0.4272 (IGA) \\[1.2mm] 0.4273 \citep{nguyen-xuan08_a-smoothed} \\[1.2mm] 0.4273 \citep{taylor93_linked}}} \\
         & G & 3.3026 & 3.0475 & 2.4519 & 2.3086 & 1.6469 & 1.5763 & & \\
         & IWP & 2.8993 & 2.6396 & 2.2043 & 2.0906 & 1.5268 & 1.4880 & & \\
         & FGPP & 2.1974 & 2.0572 & 1.7418 & 1.6394 & 1.2962 & 1.2776 & & \\[2mm]
        \hline \\[-4mm]

         & & \multicolumn{8}{c}{CCCC} \\[1mm]
        \multirow{4}{*}{$0.01$} & P & 1.0462 & 0.9751 & 0.8022 & 0.6430 & 0.4341 & 0.4288 & \multirow{4}{*}{2.9516} & \multirow{4}{*}{\shortstack[l]{0.1268 (IGA) \\[1.2mm] 0.1268 \citep{nguyen-xuan08_a-smoothed} \\[1.2mm] 0.1267 \citep{taylor93_linked}}} \\
         & G & 1.0260 & 0.9326 & 0.7368 & 0.6972 & 0.4680 & 0.4476 & &\\
         & IWP & 0.8247 & 0.7482 & 0.6161 & 0.5747 & 0.4040 & 0.3987 & &\\
         & FGPP & 0.6561 & 0.6102 & 0.5102 & 0.4776 & 0.3666 & 0.3635 & &\\[2mm]
        \multirow{4}{*}{$0.1$} & P & 1.1603 & 1.0944 & 0.9280 & 0.7696 & 0.5692 & 0.5536 & \multirow{4}{*}{3.0176} & \multirow{4}{*}{\shortstack[l]{0.1485 (IGA) \\[1.2mm] 0.1504 \citep{nguyen-xuan08_a-smoothed} \\[1.2mm] 0.1499 \citep{taylor93_linked}}} \\
         & G & 1.1688 & 1.0833 & 0.8988 & 0.8674 & 0.6583 & 0.6222 & &\\
         & IWP & 0.9740 & 0.9038 & 0.7833 & 0.7497 & 0.5966 & 0.5787 & &\\
         & FGPP & 0.7544 & 0.7153 & 0.6204 & 0.5901 & 0.4912 & 0.4793 & &\\[2mm]
        \hline
    \end{tabular}}}
\end{table}

\begin{table}[!ht]
    \centering
    \caption{The non-dimensional buckling load $\left( \bar{P}=P_{cr} b^2 / (\pi^2 D) \right)$ along the $x$ direction for rectangular FG plates with various density distribution cases ($h/b = 0.05$, $\rho_{average} = 0.35$)}
    \label{tab:Buck_FG_0.35}
    {\renewcommand\arraystretch{1.2}
    {\tabcolsep = 2mm
    \begin{tabular}{llllllllll} 
        \hline \\[-4mm]
        \multirow{2}{*}{$a/b$} & \multirow{2}{*}{Type} & \multicolumn{3}{c}{Pattern A} & \multicolumn{3}{c}{Pattern B} & \multicolumn{2}{c}{Isotropic plate}\\
        \cline{3-10} \\[-4mm]
         & & A1 & A2 & A3 & B1 & B2 & B3 & 35\% weight & 100\% weight\\
        \hline \\[-4mm]
        
         & & \multicolumn{8}{c}{SSSS} \\[1mm]
        \multirow{4}{*}{$0.5$} & P & 0.7744	& 0.8270 & 0.9953 & 1.2189 & 1.7457 & 1.7721 & \multirow{4}{*}{0.2668} & \multirow{4}{*}{\shortstack[l]{6.0382 (IGA) \\[1.2mm] 5.9873 \citep{nguyen-xuan10_an-edge} \\[1.2mm] 6.0372 \citep{kitipornchai93_buckling}}} \\
         & G & 0.7598 & 0.8298 & 1.0380 & 1.0965 & 1.5749 & 1.6490 & & \\
         & IWP & 0.9147 & 1.004 & 1.2056 & 1.2813 & 1.7681 & 1.8036 & & \\
         & FGPP & 1.1715 & 1.2540 & 1.4870 & 1.5827 & 2.0204 & 2.0459 & & \\[2mm]
        \multirow{4}{*}{$1$} & P & 0.5512 & 0.5949 & 0.7187 & 0.8451 & 1.2005 & 1.2127 & \multirow{4}{*}{0.1712} & \multirow{4}{*}{\shortstack[l]{3.9447 (IGA) \\[1.2mm] 3.9412 \citep{nguyen-xuan10_an-edge} \\[1.2mm] 3.9444 \citep{kitipornchai93_buckling}}} \\
         & G & 0.5064 & 0.5517 & 0.6944 & 0.7413 & 1.0759 & 1.1201 & & \\
         & IWP & 0.5825 & 0.6435 & 0.7807 & 0.8287 & 1.1748 & 1.1996 & & \\
         & FGPP & 0.7633 & 0.8193 & 0.9768 & 1.0420 & 1.3471 & 1.3607 & & \\[2mm]
        \multirow{4}{*}{$2$} & P & 0.5512 & 0.5949 & 0.7187 & 0.8451 & 1.2005 & 1.2128 & \multirow{4}{*}{0.1712} & \multirow{4}{*}{\shortstack[l]{3.9448 (IGA) \\[1.2mm] 3.9811 \citep{nguyen-xuan10_an-edge} \\[1.2mm] 3.9444 \citep{kitipornchai93_buckling}}} \\
         & G & 0.5064 & 0.5517 & 0.6945 & 0.7413 & 1.0759 & 1.1201 & &\\
         & IWP & 0.5826 & 0.6435 & 0.7808 & 0.8287 & 1.1748 & 1.1996 & &\\
         & FGPP & 0.7634 & 0.8194 & 0.9768 & 1.0421 & 1.3472 & 1.3607 & &\\
        \hline \\[-4mm]

         & & \multicolumn{8}{c}{CCCC} \\[1mm]
        \multirow{4}{*}{$0.5$} & P & 2.0499 & 2.1697 & 2.5854 & 3.2421 & 4.5462 & 4.6641 & \multirow{4}{*}{0.8171} & \multirow{4}{*}{\shortstack[l]{17.2549 (IGA) \\[1.2mm] 17.2222 \citep{kitipornchai93_buckling}}} \\
         & G & 2.1472 & 2.3362 & 2.8527 & 2.9586 & 4.0257 & 4.2562 & &\\
         & IWP & 2.6716 & 2.8893 & 3.3670 & 3.5458 & 4.5486 & 4.6671 & &\\
         & FGPP & 3.3804 & 3.5822 & 4.1673 & 4.3984 & 5.3795 & 5.4940 & &\\[2mm]
        \multirow{4}{*}{$1$} & P & 1.1831 & 1.2640 & 1.5207 & 1.8760 & 2.6911 & 2.7401 & \multirow{4}{*}{0.4293} & \multirow{4}{*}{\shortstack[l]{9.5851 (IGA) \\[1.2mm] 9.5412 \citep{nguyen-xuan10_an-edge} \\[1.2mm] 9.5588 \citep{kitipornchai93_buckling}}} \\
         & G & 1.1955 & 1.3069 & 1.6271 & 1.7078 & 2.4310 & 2.5535 & &\\
         & IWP & 1.4690 & 1.6071 & 1.9173 & 2.0363 & 2.7629 & 2.8185 & &\\
         & FGPP & 1.8634 & 1.9908 & 2.3517 & 2.4984 & 3.1614 & 3.2070 & &\\[2mm]
         \multirow{4}{*}{$2$} & P & 0.9588 & 1.0276 & 1.2356 & 1.5015 & 2.1407 & 2.1769 & \multirow{4}{*}{0.3358} & \multirow{4}{*}{\shortstack[l]{7.5154 (IGA) \\[1.2mm] 7.487 \citep{kitipornchai93_buckling}}} \\
         & G & 0.9450 & 1.0322 & 1.2858 & 1.3527 & 1.9257 & 2.0195 & &\\
         & IWP & 1.1420 & 1.2520 & 1.4967 & 1.5870 & 2.1669 & 2.2119 & &\\
         & FGPP & 1.4607 & 1.5619 & 1.8463 & 1.9613 & 2.4871 & 2.5219 & &\\
        \hline
    \end{tabular}}}
\end{table}

Additionally, the porous plates in this study consist of combinations of six aforementioned density distribution cases and three investigating TPMSs, namely Primitive, Gyroid, and IWP along with the Hashin-Shtrikman-like structure. Various thickness-to-length ratios have been adopted to explore the influence of both flexural and shear stiffness of TPMS on functionally graded plates. From applying the fitting results in the previous section, the mechanical properties including effective elastic modulus, effective shear modulus, and Poisson's ratio of these TPMS structures with a relative density similar to that from other porous cases can be achieved (please see Table. \ref{tab:MechanicalProperties_0.35}). At the relative density of $35\%$, IWP and Primitive might produce the highest elastic modulus and shear modulus values, respectively. For this reason, P-type plates have the best performance when their thicknesses reach a noticeable value. On the other hand, IWP should be considered a superior structure for thin plate applications. In the case of the Gyroid, a medium value of shear modulus can be found, however, its elastic modulus is only slightly higher than the lowest one. As the result, Gyroid plates' efficiencies might not exceed other TPMS plates with the porous cases in this study. Furthermore, the insignificant differences in Poisson's ratio of all three TPMSs might be negligible due to the minor impact of this property on the bending stiffness of the FG-TPMS plates in Eq. (\ref{eq:bendingmodulus}). These conclusions can be obtained from the static investigation of uniform porous plates considering core design in Fig. \ref{fig:Static_NonFG}.
\begin{table}[!ht]
    \centering
    \caption{Mechanical properties of three considering TPMS structures with the relative density of $35\%$.}
    \label{tab:MechanicalProperties_0.35}
    {\renewcommand\arraystretch{1.15}
    {\tabcolsep = 4mm
    \begin{tabular}{llll} 
        \hline \\[-3mm]
        \multirow{2}{*}{Mechanical property} & \multicolumn{3}{c}{TPMS} \\[1mm]
        \cline{2-4} \\[-3mm]
         & Primitive & Gyroid & IWP \\
         \hline \\[-4mm]
        Effective elastic modulus, $E^{*}$ & 0.122 & 0.128 & 0.165 \\
        Effective shear modulus, $G^{*}$ & 0.204 & 0.154 & 0.137 \\
        Poisson's ratio, $\nu$ & 0.340 & 0.322 & 0.271 \\
        \hline
    \end{tabular}}}
\end{table}

\begin{figure}[!ht]
    \centering
    \begin{subfigure}[c]{\textwidth}
        \centering
        \includegraphics[trim=2.5cm 2.7cm 2.8cm 2.5cm,clip=true,width=0.49\textwidth]{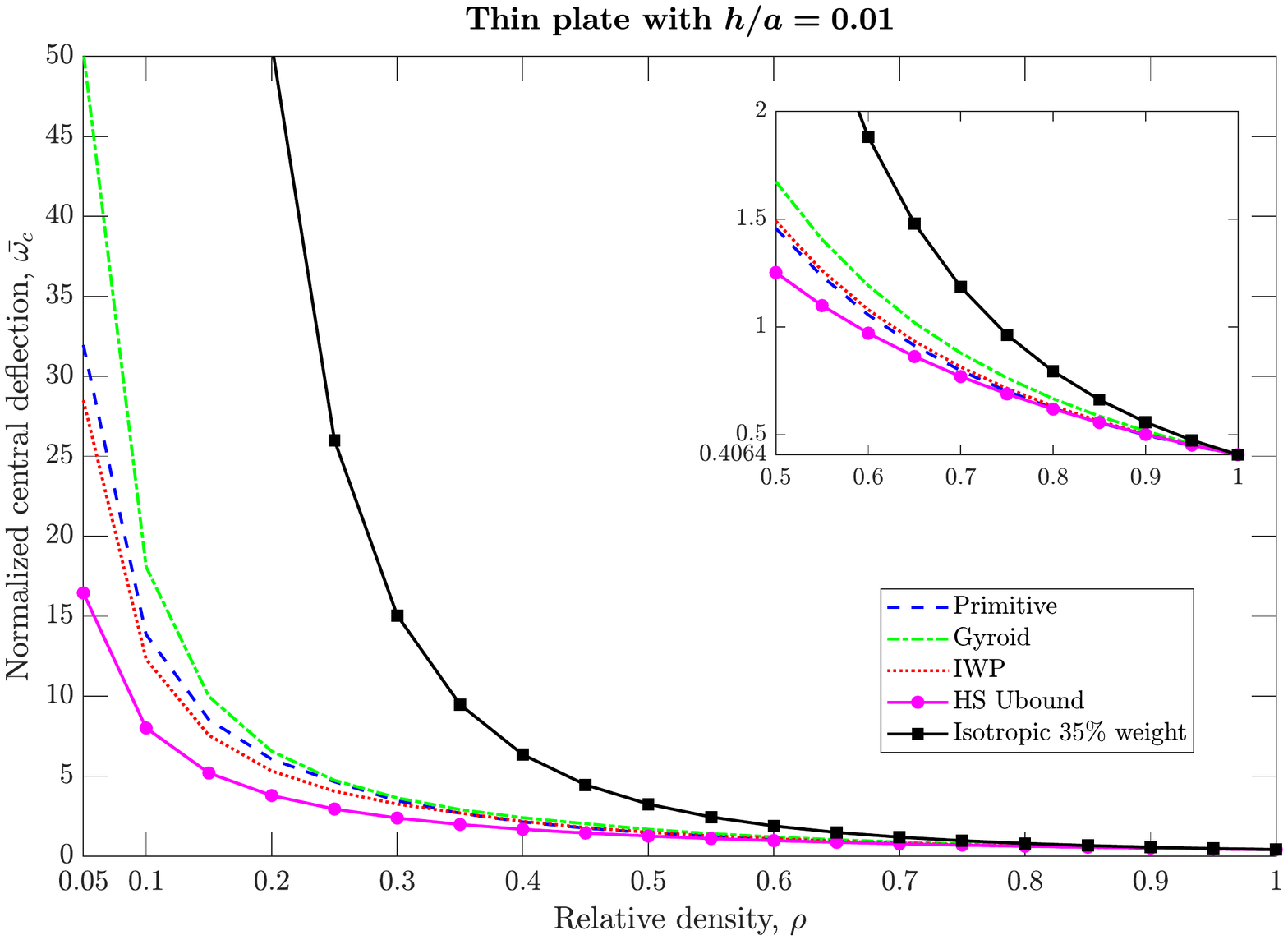}
        \hfill
        \includegraphics[trim=2.5cm 2.7cm 2.8cm 2.5cm,clip=true,width=0.49\textwidth]{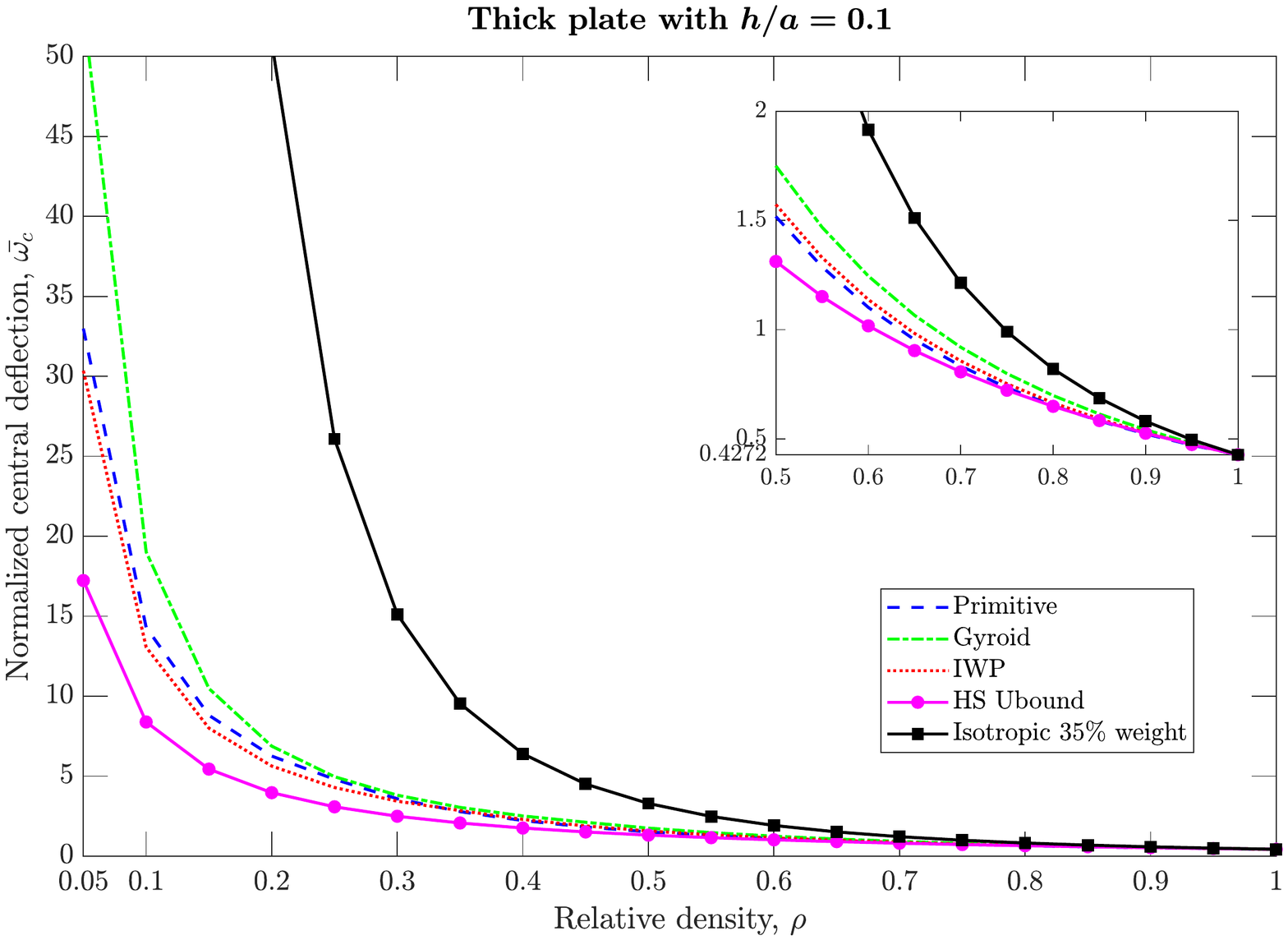}
        \caption {Simply supported boundary (SSSS)}
    \end{subfigure}
    \medskip
    
    \begin{subfigure}[c]{\textwidth}
        \centering
        \includegraphics[trim=2.5cm 2.7cm 2.8cm 2.5cm,clip=true,width=0.49\textwidth]{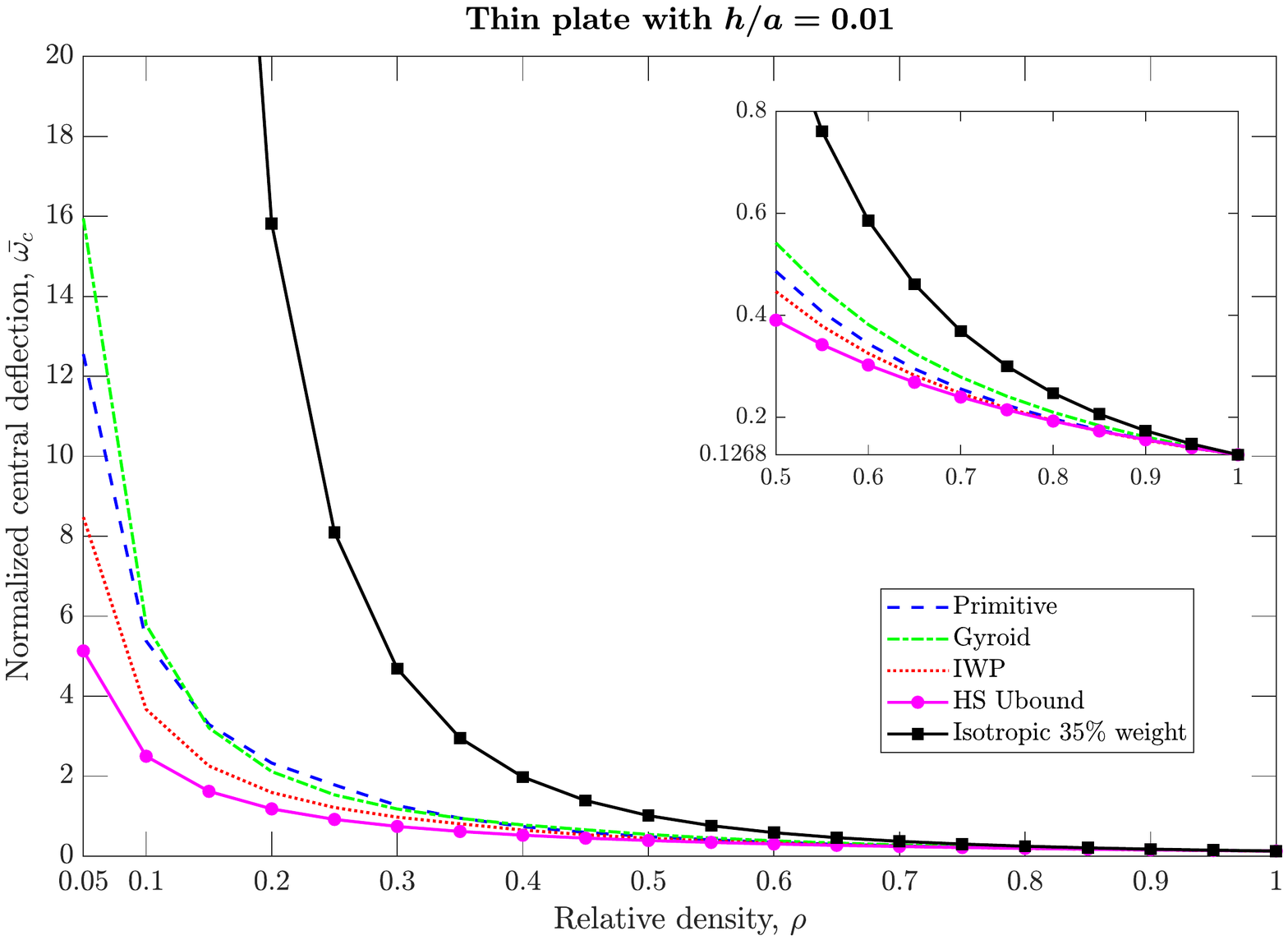}
        \hfill
        \includegraphics[trim=2.5cm 2.7cm 2.8cm 2.5cm,clip=true,width=0.49\textwidth]{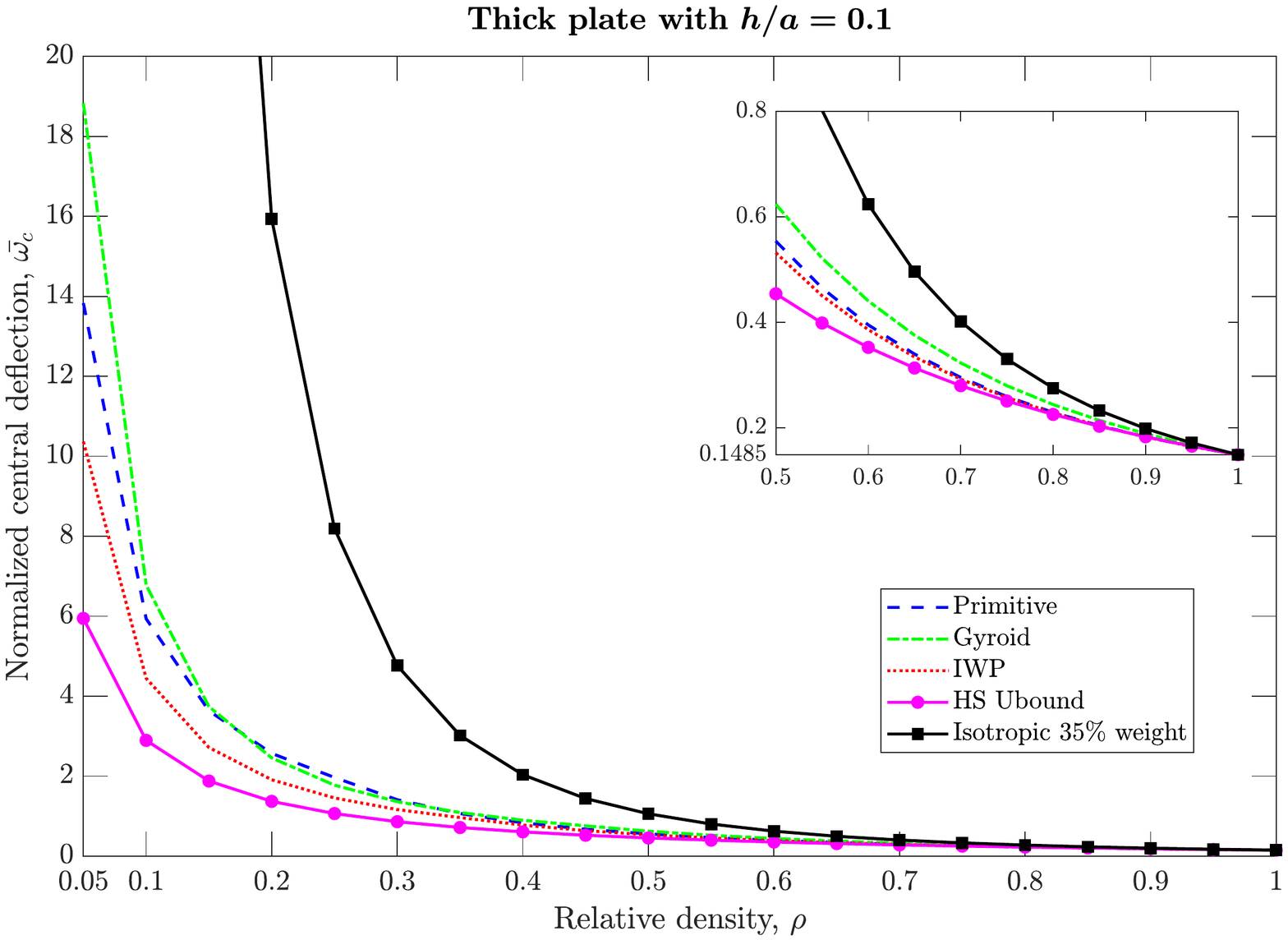}
        \caption {Fully clamped boundary (CCCC)}
    \end{subfigure}
    \caption {Normalized central deflection $\bar{w}_c$ of square plates with various core designs.}
    \label{fig:Static_NonFG}
\end{figure}

Nevertheless, these TPMS plates' performances can be modified by employing functional gradations. As a result, thicker zones of TPMS shells can be developed without increasing the average relative density. Therefore, the overall stiffness of the plates might increase. Generally, pattern B should be noted as an improved distribution because it provides two regions for higher relative density at the top and bottom surfaces of the plate, which cannot be found in pattern A. This confirmation can be obtained from both static and buckling analysis results.

With low relative density, the elastic modulus of the Gyroid is higher than that of the Primitive. This can be explained by the $C_1$ value in the fitting curve. However, as in previous statements, the shear response of the Primitive structure is more robust than other types. Consequently, the bending stiffness of P-type plates might outperform the G-type ones in almost all cases. A special case that can be indicated is clamped rectangular plate with distribution pattern A, where the Primitive plates have greater deflections and lower buckling load than the Gyroid ones. Similar results can be found in uniform TPMS plates' responses in Fig. \ref{fig:Static_NonFG} where the relative density ranges from $0.15$ to $0.35$. Nonetheless, with pattern B, where the lower porosity region of TPMS structures could be better exploited, the higher order deformation mode of Gyroid ($n_2 = 2.351 > 2$) may cause its flexural performance much lesser than others. This distinct behavior of the Gyroid structure should be noticed attentively when utilizing the TPMS structures in applications that require a certain value of porosity.

Although the IWP structure could provide the highest elastic modulus of all three TPMSs, its uniform plate deflection might be slightly larger than the P-based plate ones due to the influence of high shear modulus with considerable relative density, especially in simply supported boundary conditions. In fact, in high relative density ($\rho > 0.6$), Primitive's shear modulus can be greater than the shear modulus of the HS upper boundary \citep{chen19_on-hybrid}. Consequently, porosity distribution B can generate greater stiffness for Primitive FG-TPMS plates in SSSS schemes compared to IWP ones in both static and buckling analysis. In the case of clamped plates, the IWP structures have the best performance in both uniform and non-uniform cases. However, since the relative density reaches $0.75$ in the thin plate case and $0.65$ in the thick plate case, the static responses of Primitive and IWP applications might approach the same value (please see Fig. \ref{fig:Static_NonFG}). Combining the two comments above, for clamped boundary conditions, the P-core thick FG plates with pattern B distribution can have the minimum deflections instead of IWP core ones as the thin FG plates and uniform plates. These values can be seen in Tabe. \ref{tab:Static_FG_0.35}.

In buckling load analysis, the results of P-based and IWP-based plates are similar to the static results. For instance, pattern-A plates with IWP cores can produce higher buckling loads than that of the Primitive structures in both boundary conditions. To provide the view of the plate thickness influence, it is observed that for the cases $a/b=1$ and $a/b=2$, the minimum thickness-to-length ratio is constantly $0.05$ while the value of $0.1$ can be observed in the case $a/b=0.5$. Therefore, this case can be considered a thick-plate case while the rest of the cases are thin-plate cases. The agreement between the response of the thin plate structure in buckling load and that in static load might be achieved with the pattern-B application. However, thick plate responses might be strongly different, where the stiffness of P-core plates cannot surpass the IWP ones with both simply supported and clamped boundaries. This indicates that the influence of shear modulus in these cases has been reduced. This dissimilarity may come from the plate's aspect ratio and loading direction.

Interestingly, the P-type FG plate behaviors can closely reach the impractical HS porous plate behaviors in all cases. Distribution case B3 provides the lowest disparity among them, which is approximately $10.9\%$ and $12.2\%$ for static and buckling analysis, respectively. However, it is noted that all three TPMS structures investigated in this study could not exceed the HS upper boundary for elastic modulus \citep{chen19_on-hybrid}. Consequently, both two loading schemes of the FG-TPMS plates' responses in this section might be poorer than the boundary despite any relative density distribution. Furthermore, both boundary conditions and thickness of the plate can influence the correlation of FG-TPMS and isotropic plate behaviors. Evidently, the clamped thick plates of Primitive-based structure could only produce $87\%$ and $85\%$ of the HS-based plates' stiffness under static and buckling load, respectively.

To demonstrate the superiority of the FG-TPMS plate, a similar-weight isotropic plate with  $35\%$ weight has been investigated by reducing the applied thickness ($h_{applied} = 0.35 h$). The plate performances are normalized by the same value as other cases (with the same value of $h$). The results from Table. \ref{tab:Static_FG_0.35} and Table. \ref{tab:Buck_FG_0.35} show that the 35\% weight isotropic plate could only provide about $1/22$ the stiffness of full-weight isotropic plates for both static and buckling responses. This factor might derive from the bending stiffness ($D$), which is a cubic function of the thickness ($h$), whereas $(1/0.35)^{3} = 1/23.3 \approx 1/22$. In addition, the minimum deflection of all considering six cases is about $3.33$ times the value of isotropic plates. This indicates that the flexural behaviors of these plates are approximately $30\%$ the isotropic one. A similar value could also be found in non-dimensional buckling load. In other words, the FG-TPMS plate can generate an approximately seven-times-higher flexural stiffness than that of the weight-reduced isotropic plates. A detailed comparison of the maximum stiffness of all FG-TPMS plates is given in Table. \ref{tab:Stiff_Comparison}. However, it should be noted that the smallest correlation values were obtained from the clamped thick plate cases i.e. $5.45$ and $5.71$ for static and buckling responses, respectively.
\begin{table}[!ht]
    \centering
    \caption{Average performance comparisons of FG-TPMS plates with isotropic plates in various analysis types.}
    \label{tab:Stiff_Comparison}
    {\renewcommand\arraystretch{1.2}
    {\tabcolsep = 3mm
    \begin{tabular}{lllll} 
        \hline \\[-4mm]
        \multirow{3}{*}{Comparison case} & \multicolumn{4}{c}{Analysis type} \\[1mm]
        \cline{2-5} \\[-4mm]
         & \multicolumn{3}{c}{Rectangular plate} & Circular plate\\
        \cline{2-5} \\[-4mm]
         & Static & Uniaxial buckling & Free vibration & Free vibration \\
         & stiffness & load & frequency & frequency\\
         \hline \\[-4mm]
        \vspace{3mm}
        \shortstack[l]{FG-TPMS to \\[1.2mm] 35\% weight isotropic} & 6.70 & 6.63 & 1.60 & 2.53 \\
        \vspace{3mm}
        \shortstack[l]{FG-TPMS to \\[1.2mm] 100\% weight isotropic} & 0.30 & 0.30 & 0.96 & 0.92 \\
        \shortstack[l]{100\% weight isotropic to \\[1.2mm] 35\% weight isotropic} & 22.31 & 22.42 & 1.67 & 2.74 \\
        \hline
    \end{tabular}}}
\end{table}

Furthermore, there might be an outstanding distribution case that can allow the FG-TPMS plate behaviors to exceed the HS upper boundary regarding stiffness. On the other hand, there were only three types of TPMS structures that have been reviewed in the present study. Numerous other types, that might or might not be micro-mechanically reviewed, can be valuable choices to apply in this porous plate structure. Additionally, the hybridization approach should be considered another dominant solution due to the fact that each TPMS might have a specific strength and weakness. For example, the IWP plates have a remarkable bending modulus while the Primitive can generate the highest shear modulus at high relative density. However, the finding of these cases was not covered due to the scope of this study, yet could be considered for future research in this field.

\subsection{\textit{Free vibration analysis}}
\label{Section 4.2}

In this subsection, the six lowest frequencies of square and circular isotropic plates have been generated by the isogeometric method. Table \ref{tab:Free_compare_Rec} shows that the results of square plates show a superior agreement with analytical solutions in \citep{abbassian87_free} and \citep{blevins80_formulas} for both SSSS and CCCC boundaries, respectively. Moreover, the solutions from the ES-FEM approach have also been adopted for comparison. For circular plates, the solutions of Irie \textit{et al.} \citep{irie80_natural} are included in Fig. \ref{fig:Free_compare_Cir} to specify the robustness of the present method. It can be observed that the IGA deviations are much smaller than the ES-FEM ones in most cases. Similar to the previous analysis, one exceptional case is for the fully clamped rectangular plate with a high thickness-to-length ratio ($h/a=0.1$). However, it is suggested that the second and third frequencies provided by the IGA method are similar to each other, which is consistent with analytical results. Furthermore, the agreement between analytical and IGA results can also be seen in circular plates with various combinations of thickness-to-radius ratios and boundary conditions.
\begin{table}[!ht]
    \centering
    \caption{The first six natural frequencies $\left( \bar{\omega}= (\omega^2 \rho_s a^4 h/D ) ^{1/4}\right)$ for a square isotropic plate}
    \label{tab:Free_compare_Rec}
    {\renewcommand\arraystretch{1.2}
    {\tabcolsep = 2mm
    \begin{tabular}{llllllll} 
        \hline \\[-4mm]
        \multirow{2}{*}{$h/a$} & \multirow{2}{*}{Method} & \multicolumn{6}{c}{Mode} \\
        \cline{3-8} \\[-4mm]
         & & 1 & 2 & 3 & 4 & 5 & 6 \\
        \hline \\[-4mm]
        
         & & \multicolumn{6}{c}{SSSS} \\[1mm]
        \multirow{3}{*}{$0.005$} & IGA & 4.4427 & 7.0242 & 7.0242 & 8.8843 & 9.9341 & 9.9341 \\
         & ES-FEM \citep{nguyen-xuan10_an-edge} & 4.4537 & 7.0565 & 7.0729 & 8.9731 & 10.0410 & 10.0422 \\
         & Ref \citep{abbassian87_free} & 4.443 & 7.025 & 7.025 & 8.886 & 9.935 &  9.935 \\[2mm]
        \multirow{3}{*}{$0.1$} & IGA & 4.3666 & 6.7454 & 6.7454 & 8.3568 & 9.2269 & 9.2269 \\
         & ES-FEM \citep{nguyen-xuan10_an-edge} & 4.3759 & 6.7692 & 6.7834 & 8.4173 & 9.2968 & 9.2976 \\
         & Ref \citep{abbassian87_free} & 4.37 & 6.74 & 6.74 & 8.35 & 9.22 & 9.22 \\
         \hline \\[-4mm]
        
         & & \multicolumn{6}{c}{CCCC} \\[1mm]
        \multirow{3}{*}{$0.005$} & IGA & 5.9981 & 8.5657 & 8.5657 & 10.4005 & 11.4706 & 11.4978 \\
         & ES-FEM \citep{nguyen-xuan10_an-edge} & 6.0158 & 8.6075 & 8.6353 & 10.5252 & 11.6032 & 11.6293 \\
         & Ref \citep{blevins80_formulas} & 5.999 & 8.568 & 8.568 & 10.407 & 11.472 & 11.498 \\[2mm]
        \multirow{3}{*}{$0.1$} & IGA & 5.7198 & 7.9181 & 7.9181 & 9.3911 & 10.2067 & 10.2494 \\
         & ES-FEM \citep{nguyen-xuan10_an-edge} & 5.7141 & 7.8990 & 7.9206 & 9.3896 & 10.1935 & 10.2411 \\
         & Ref \citep{blevins80_formulas} & 5.71 & 7.88 & 7.88 & 9.33 & 10.13 & 10.18 \\
        \hline
    \end{tabular}}}
\end{table}
\begin{figure}[!ht]
    \centering
    \begin{subfigure}[c]{0.48\textwidth}
        \centering
        \includegraphics[trim=0.8cm 6.4cm 1.2cm 7cm,clip=true,width=\textwidth]{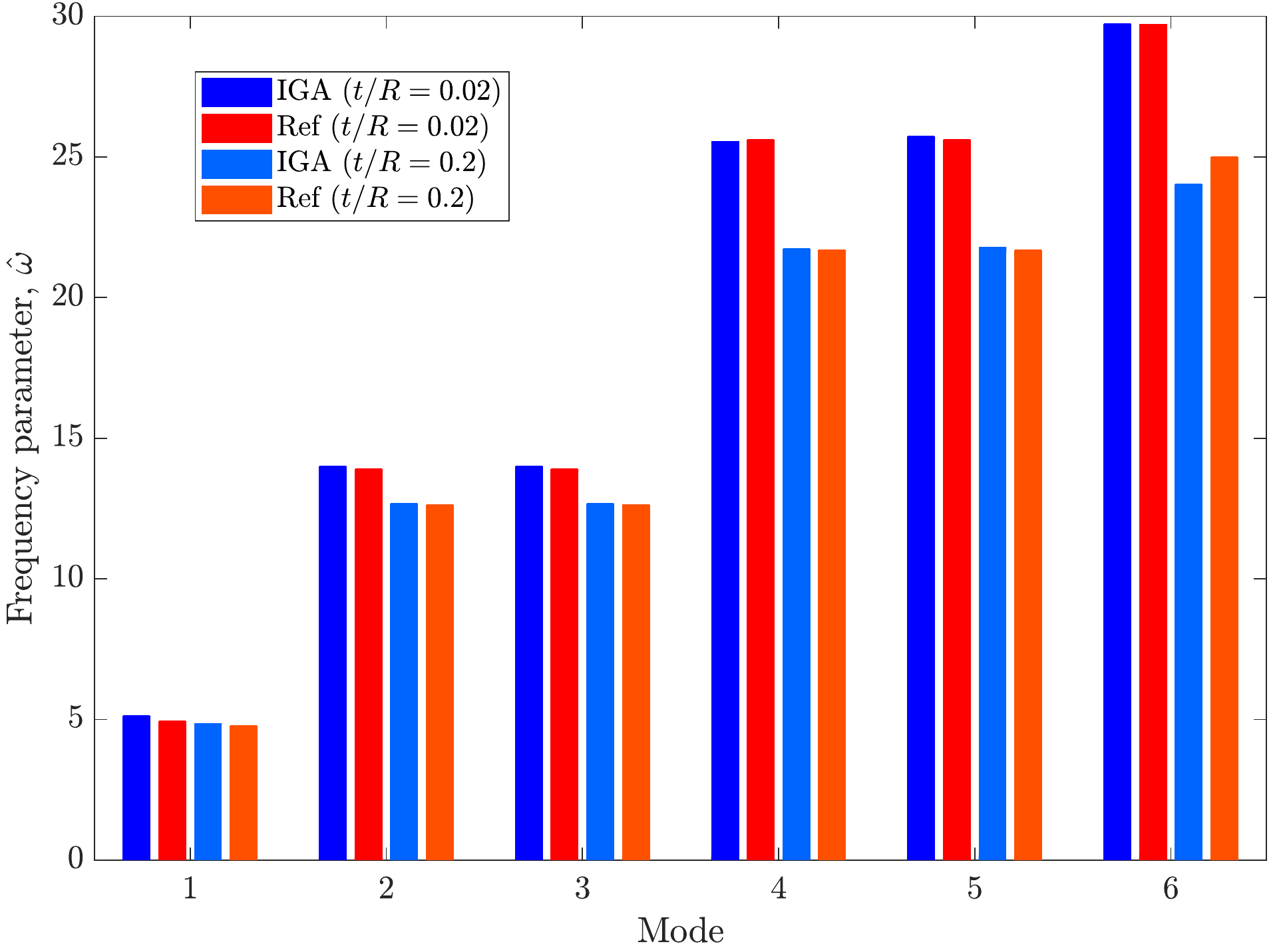}
        \caption {SSSS}
    \end{subfigure}
    \hfill
    \begin{subfigure}[c]{0.48\textwidth}
        \centering
        \includegraphics[trim=0.8cm 6.4cm 1.2cm 7cm,clip=true,width=\textwidth]{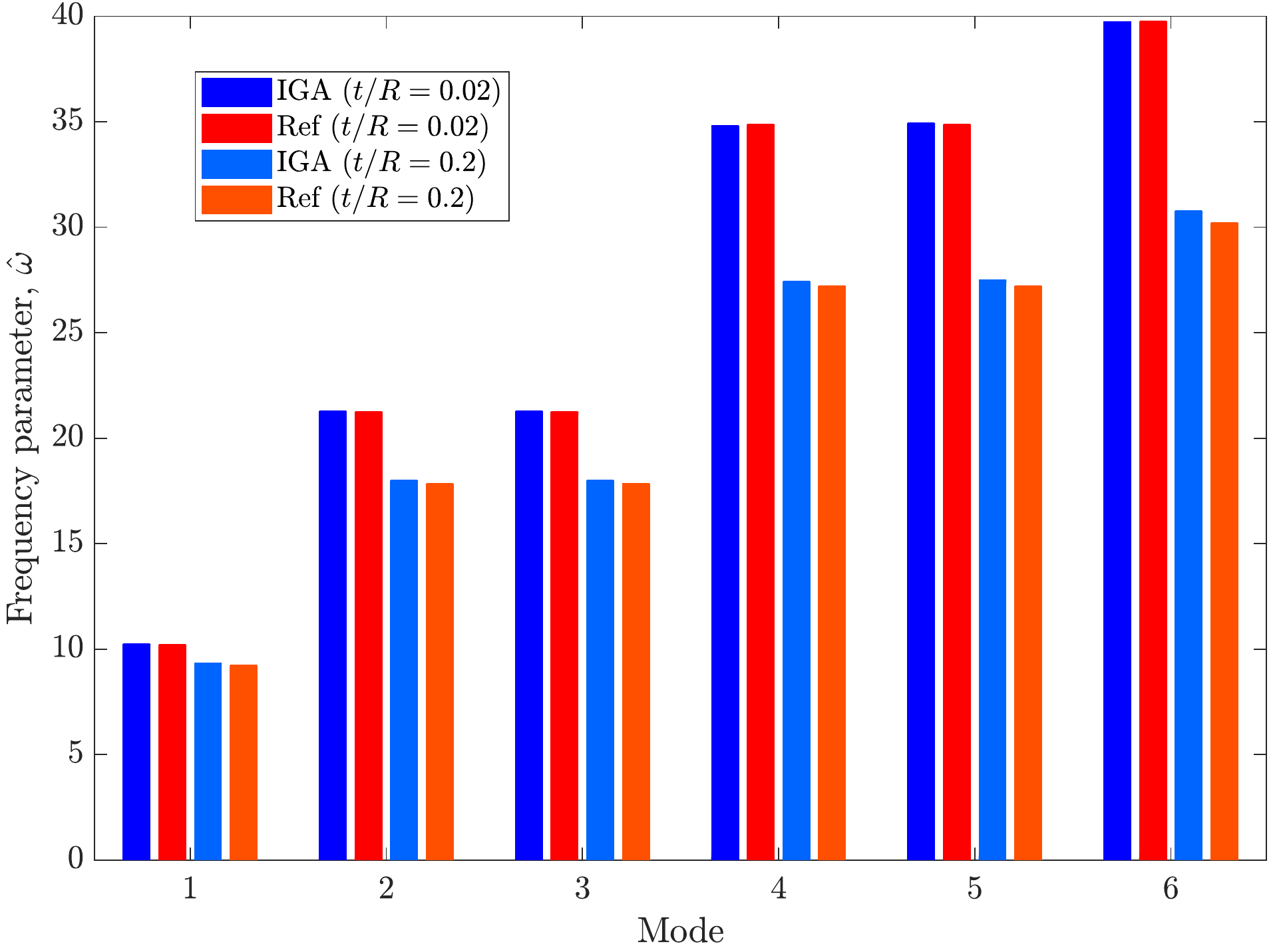}
        \caption {CCCC}
    \end{subfigure}
    \caption{The first six natural frequencies $\left( \hat{\omega}=\omega R^2 \sqrt{\rho_s h/D} \right)$ of the circular isotropic plate, with $D = E_s h^3/(12(1-\nu_s^2))$}
    \label{fig:Free_compare_Cir}
\end{figure}

Furthermore, the non-dimensional frequencies of constituent functionally graded porous plates are provided in Table. \ref{tab:Free_RecFG_0.35} and Table. \ref{tab:Free_CirFG_0.35} for square shape and circle shape plates, respectively. 
\begin{table}[!ht]
    \centering
    \caption{The fundamental frequency parameter $\left( \bar{\omega}=(\omega^2 \rho_s a^4 h/D)^{1/4} \right)$ for square FG plates with various density distribution cases ($\rho_{average} = 0.35$)}
    \label{tab:Free_RecFG_0.35}
    {\renewcommand\arraystretch{1.2}
    {\tabcolsep = 2mm
    \begin{tabular}{llllllllc} 
        \hline \\[-4mm]
        \multirow{2}{*}{$h/a$} & \multirow{2}{*}{Type} & \multicolumn{3}{c}{Pattern A} & \multicolumn{3}{c}{Pattern B} & \multirow{2}{*}{\shortstack[c]{Isotropic plate \\[1.2mm] with 35\% weight}}\\
        \cline{3-8} \\[-4mm]
         & & A1 & A2 & A3 & B1 & B2 & B3 & \\
        \hline \\[-4mm]
        
         & & \multicolumn{7}{c}{SSSS} \\[1mm]
        \multirow{4}{*}{$0.005$} & P & 3.5274 & 3.5965 & 3.7735 & 3.9322 & 4.3050 & 4.3148 & \multirow{4}{*}{2.6284} \\
         & G & 3.4553 & 3.5317 & 3.7455 & 3.8090 & 4.1968 & 4.2384 & \\
         & IWP & 3.5811 & 3.6733 & 3.8603 & 3.9206 & 4.2951 & 4.3161 & \\
         & FGPP & 3.8292 & 3.8994 & 4.0783 & 4.1463 & 4.4325 & 4.4424 & \\[2mm]
        \multirow{4}{*}{$0.1$} & P & 3.4826 & 3.5452 & 3.7065 & 3.8517 & 4.1759 & 4.1928 & \multirow{4}{*}{2.6226} \\
         & G & 3.4047 & 3.4726 & 3.6638 & 3.7183 & 4.0429 & 4.0891 & \\
         & IWP & 3.5188 & 3.6009 & 3.7639 & 3.8131 & 4.1210 & 4.1490 & \\
         & FGPP & 3.7708 & 3.8325 & 3.9926 & 4.0519 & 4.2939 & 4.3110 & \\
        \hline \\[-4mm]
                
         & & \multicolumn{7}{c}{CCCC} \\[1mm]
        \multirow{4}{*}{$0.005$} & P & 4.5914 & 4.6724 & 4.9064 & 5.1894 & 5.7319 & 5.7507 & \multirow{4}{*}{3.5490} \\
         & G & 4.6208 & 4.7332 & 5.0212 & 5.0902 & 5.6285 & 5.6928 & \\
         & IWP & 4.8852 & 5.0055 & 5.2551 & 5.3478 & 5.8442 & 5.8634 & \\
         & FGPP & 5.1699 & 5.2647 & 5.5060 & 5.5977 & 5.9838 & 5.9971 & \\[2mm]
        \multirow{4}{*}{$0.1$} & P & 4.4454 & 4.5076 & 4.6924 & 4.9193 & 5.2988 & 5.3386 & \multirow{4}{*}{3.5253}  \\
         & G & 4.4413 & 4.5241 & 4.7343 & 4.7741 & 5.1064 & 5.1814 & \\
         & IWP & 4.6490 & 4.7339 & 4.8999 & 4.9511 & 5.2322 & 5.2746 & \\
         & FGPP & 4.9562 & 5.0202 & 5.1965 & 5.2588 & 5.4982 & 5.5347 & \\
        \hline
    \end{tabular}}}
\end{table}
\begin{table}[!ht]
    \centering
    \caption{The fundamental frequency parameter $\left( \hat{\omega}=\omega R^2 \sqrt{\rho_s h/D} \right)$ for circular FG plates with various density distribution cases ($\rho_{average} = 0.35$)}
    \label{tab:Free_CirFG_0.35}
    {\renewcommand\arraystretch{1.2}
    {\tabcolsep = 2mm
    \begin{tabular}{llllllllc} 
        \hline \\[-4mm]
        \multirow{2}{*}{$h/R$} & \multirow{2}{*}{Type} & \multicolumn{3}{c}{Pattern A} & \multicolumn{3}{c}{Pattern B} & \multirow{2}{*}{\shortstack[c]{Isotropic plate \\[1.2mm] with 35\% weight}}\\
        \cline{3-8} \\[-4mm]
         & & A1 & A2 & A3 & B1 & B2 & B3 & \\
        \hline \\[-4mm]
        
         & & \multicolumn{7}{c}{SSSS} \\[1mm]
        \multirow{4}{*}{$0.02$} & P & 2.9576 & 3.0848 & 3.3705 & 3.7597 & 4.5789 & 4.6192 & \multirow{4}{*}{1.8762} \\
         & G & 3.0467 & 3.2259 & 3.6042 & 3.6432 & 4.4228 & 4.5387 &\\
         & IWP & 3.4057 & 3.5989 & 3.9439 & 4.0402 & 4.7981 & 4.8363 & \\
         & FGPP & 3.7645 & 3.9220 & 4.2748 & 4.3846 & 5.0041 & 5.0384 & \\[2mm]
        \multirow{4}{*}{$0.2$} & P & 2.7985 & 2.9012 & 3.1623 & 3.5466 & 4.2655 & 4.3139 & \multirow{4}{*}{1.7599} \\
         & G & 2.8952 & 3.0474 & 3.3840 & 3.4249 & 4.0770 & 4.1928 & \\
         & IWP & 3.2299 & 3.3913 & 3.6881 & 3.7749 & 4.3849 & 4.4348 & \\
         & FGPP & 3.5773 & 3.7109 & 4.0258 & 4.1277 & 4.6491 & 4.6935 & \\
        \hline \\[-4mm]
                
         & & \multicolumn{7}{c}{CCCC} \\[1mm]
        \multirow{4}{*}{$0.02$} & P & 6.0679 & 6.2887 & 6.9333 & 7.7165 & 9.3878 & 9.4479 & \multirow{4}{*}{3.5917} \\
         & G & 6.0993 & 6.3954 & 7.1948 & 7.3990 & 9.0321 & 9.2369 & \\
         & IWP & 6.7852 & 7.1244 & 7.8522 & 8.1265 & 9.7022 & 9.7709 & \\
         & FGPP & 7.6174 & 7.8984 & 8.6375 & 8.9267 & 10.1956 & 10.2422 & \\[2mm]
        \multirow{4}{*}{$0.2$} & P & 5.7100 & 5.8759 & 6.3677 & 6.9584 & 8.0363 & 8.1569 & \multirow{4}{*}{3.5461}  \\
         & G & 5.6468 & 5.8537 & 6.4066 & 6.5218 & 7.4325 & 7.6508 & \\
         & IWP & 6.1469 & 6.3731 & 6.8237 & 6.9595 & 7.7543 & 7.8888 & \\
         & FGPP & 7.0098 & 7.1906 & 7.7011 & 7.8853 & 8.6064 & 8.7243 & \\
        \hline
    \end{tabular}}}
\end{table}

As a result, the free vibration responses of square plates coincide with the static ones. For instance, the Gyroid plates might have the lowest vibration stiffness among others in almost all cases except for clamped pattern-A structures. Both Primitive and IWP plates provide similar stiffness tendencies to the previous section \ref{Section 4.1}. However, an unusual case should be noted for the simply supported thin plate with pattern B3. This distribution case with a higher value of power $n$ might produce a larger zone with low relative density in the plate thickness direction. From Fig. \ref{fig.por.dis}, the minimum density value was adopted from $-h/4$ to $h/4$ when $n=3 \div 5$, while a half smaller range was found when $n=2$. In addition, the thickness-to-length ratio of thin rectangular plates in the previous section is twice the ratio from this section. By combining these contributions, the enhancement of the Primitive shear modulus to plate stiffness might be slightly reduced. However, the deviation between the P-based and IWP-based plates in this application is only about $0.03\%$ and therefore can be negligible. Due to this inconsistency, the best performance density distribution might be influenced by the plate thickness and the analysis types. In other words, the most valuable functional gradation design should be evaluated meticulously.

Moreover, not only the aspect ratio, the plate geometry might give a strong impact on the FG-TPMS behaviors. With an even greater thickness and high relative density zone, the Primitive plates cannot exceed the IWP one in simply supported circular schemes. Besides, the value of disparity between the best-performance FG-TPMS plates and FGPP plates in case B3 is $2.8\%$ typically for square plates while a value of $5.2\%$ is obtained for circular plates. On the other hand, these deviations can indicate that the porous TPMS plates might provide robust imitation to the Hashin-Shtrikman in free vibration responses.

Similar to the previous section, a $35\%$ weight isotropic plate is adopted with the same normalized value as other plates for comparison. The results show that these plates can only produce about $1/1.60 = 62.4\%$ and $1/2.53 = 39.5\%$ the maximum stiffness of TPMS rectangular and circular plates in all distribution cases, respectively. In other words, a higher stiffness can be produced by the FG-TPMS plate with a similar weight. This can be the result of the increment in plate thickness. These thicknesses are $1/0.35 = 2.857$ times larger than the one of $35\%$ weight isotropic plate. Therefore, the greater flexural and shear stiffness could contribute to the overall performance. Furthermore, the highest frequency value of all six cases, which belongs to case B3, can closely reach the value of the $100\%$ weight isotropic plate's one with $96\%$ and $92\%$, respectively, for the rectangular and circular plates. Consequently, the FG-TPMS plates have proven their superior robustness in vibration application compared to static and buckling cases. 

Although the above robustness can also be found in the HS plates with slightly higher values in all cases, there might be only a few realistic materials that give the behavior of the HS upper boundary. Therefore, the HS plate is considered an impractical structure. In contrast, FG-TPMS structures could be fabricated simply by the advances in additive manufacturing (AM) despite any porosity distribution. Numerous studies have verified the mechanical behaviors of metallic TPMS structures that were manufactured by the laser powder bed fusion (LPBF) process. The experimental specimens showed that 3D printing technology could provide high precision for the products fabricated on the digital models, even with complicated functionally graded and hybridization methods \citep{ejeh22_flexural}.

Furthermore, plastic printing is suggested as an arising research area, which is more economical than a metal one. In fact, numerous plastic printing technologies have been used to manufacture complex bio-inspired structures namely material extrusion, material jetting, and VAT polymerization \citep{siddique22_lessons}. Fused Deposition modelling (FDM), an AM method of material extrusion technology, can be considered a common method to create TPMS structures with thermal plastic material. Due to the limitation of the printing equipment, the 3D prototypes of the proposed FG-TPMS plates were fabricated with PLA plastic instead of metallic materials. These products can be found in Fig. \ref{fig:Plastic_FGplate} where the skin layers could be attached afterward. Also, the metallic ones are in progress. However, it is noteworthy that the bulk modulus and the shear modulus of these plastic TPMS might be slightly different from the results of this study due to the inconsistency in the Poisson's ratio of the base materials. Nonetheless, the proposed fitting approach is universally applicable in investigating TPMS structures despite any type of material because of the coincidence of the mechanical response tendencies \citep{al-ketan19_multifunctional}.

\begin{figure}[!ht]
    \centering
    \begin{subfigure}[c]{0.28\textwidth}
        \centering
        \includegraphics[trim=0cm 1.2cm 1.5cm 5cm,clip=true,width=\textwidth]{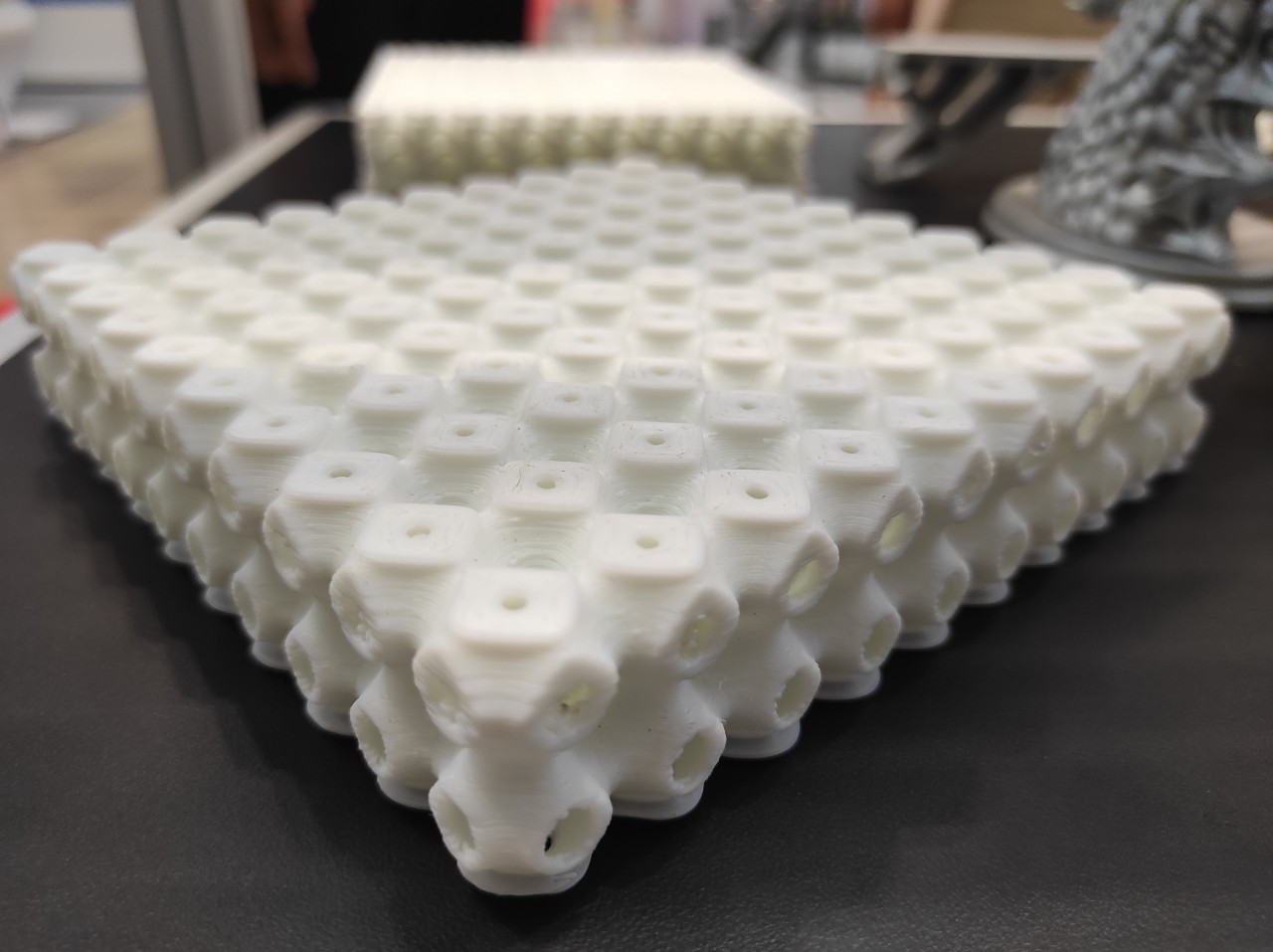}
        \caption {Primitive (P)}
    \end{subfigure}
    \begin{subfigure}[c]{0.345\textwidth}
        \centering
        \includegraphics[trim=8.5cm 0cm 2.5cm 3cm,clip=true,width=\textwidth]{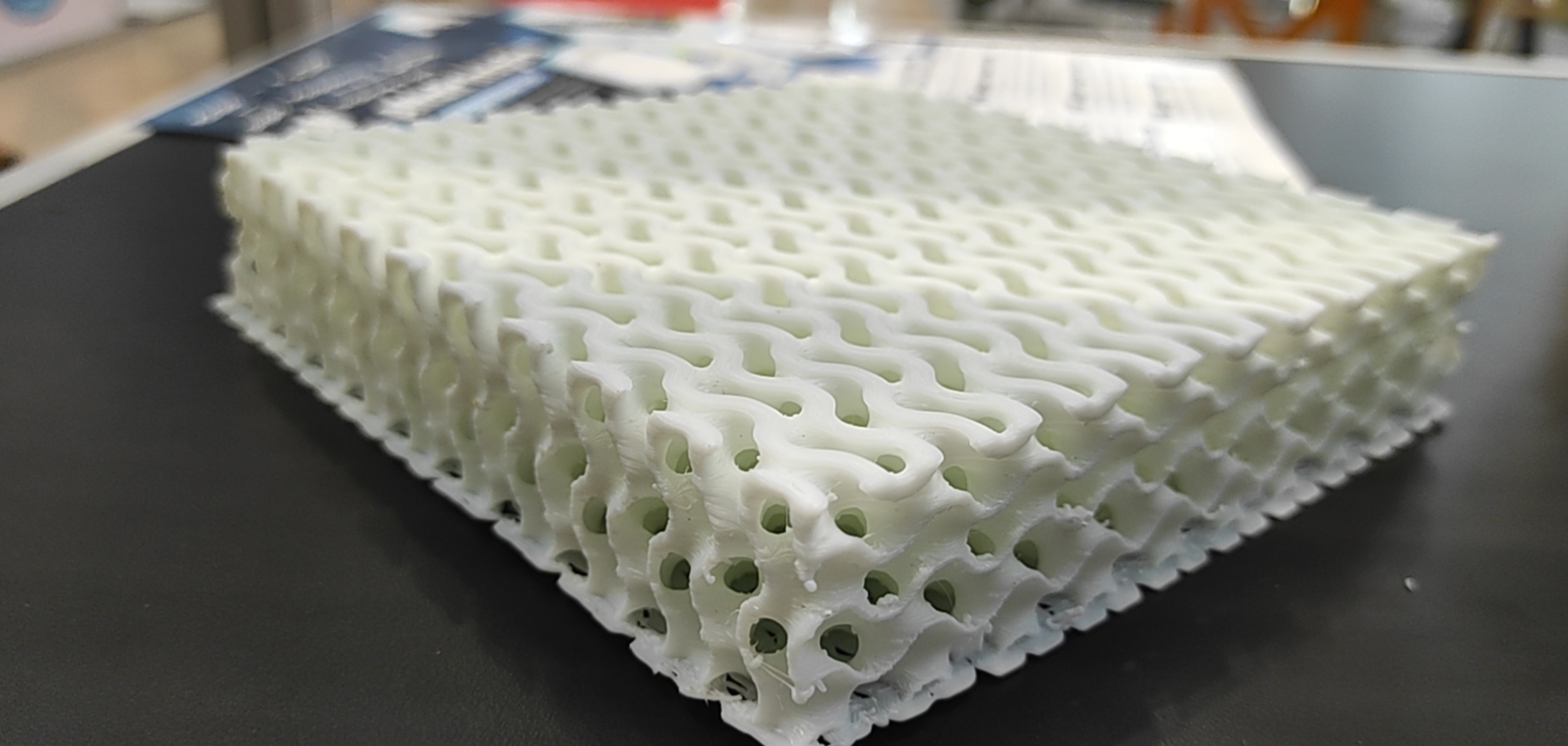}
        \caption {Gyroid (G)}
    \end{subfigure}
    \begin{subfigure}[c]{0.325\textwidth}
        \centering
        \includegraphics[trim=4cm 5cm 3cm 8cm,clip=true,width=\textwidth]{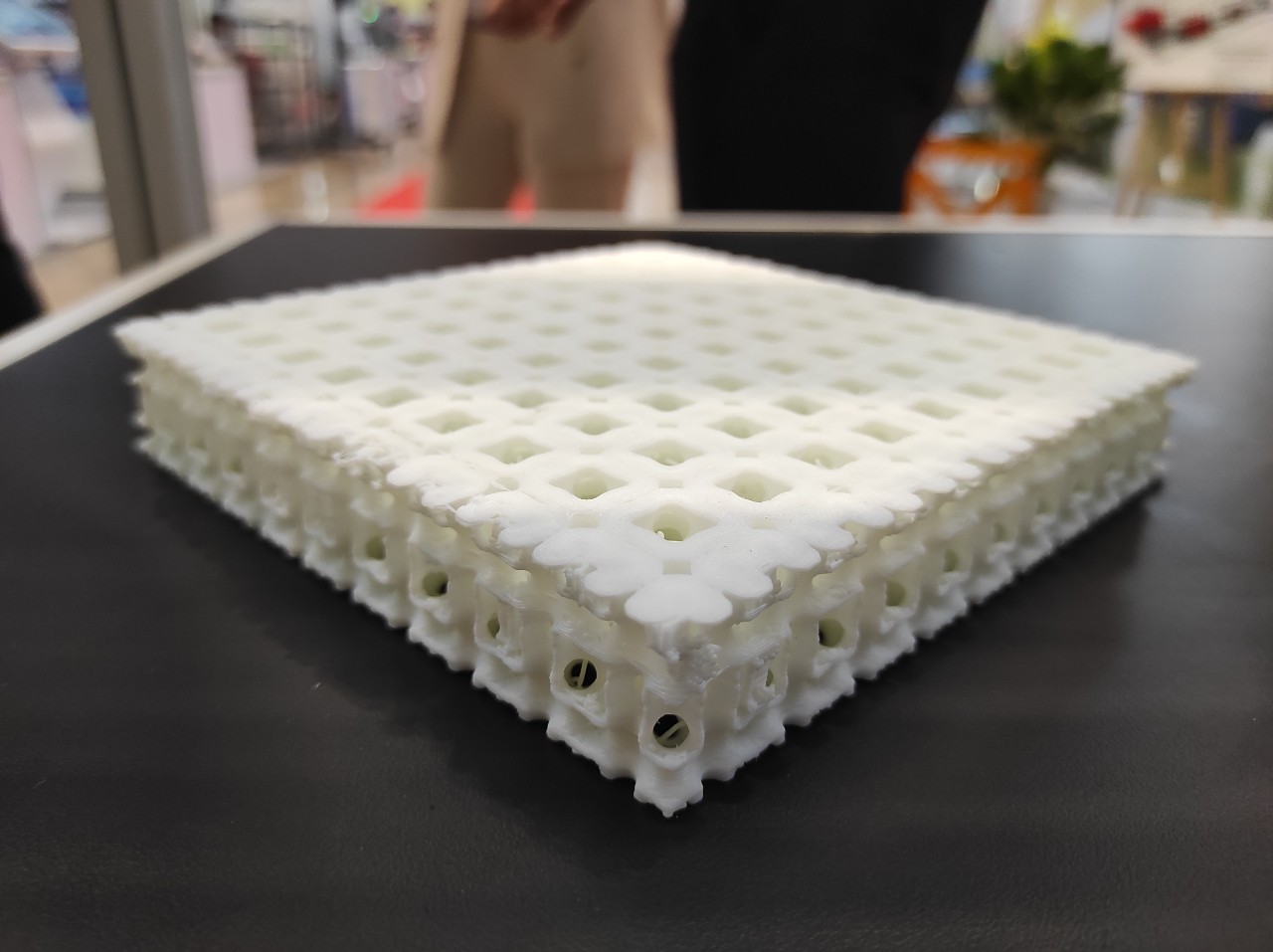}
        \caption {I wrapped package-graph (IWP)}
    \end{subfigure}
    \caption {The plastic 3D printed versions of FG-TPMS plates with density pattern A and the power $n=3$.}
    \label{fig:Plastic_FGplate}
\end{figure}

Therefore, these printed FG-TPMS plates have validated the robustness of additive manufacturing while other traditional fabrication methods might not feasibly create the complicated geometry of TPMSs. Although the presented FG-TPMS plates in Fig. \ref{fig:Plastic_FGplate} are made of PLA plastic with the FDM printing method, the metallic printing process of these structures is straightforward with the LPBF technique. With high-tech equipment in printing metal, complex geometry structures such as TPMSs can be fabricated more accurately than plastic ones. For the functional gradation through the thickness of the FG-TPMS plate, plastic printing might not provide sufficient granularity to the product, where a high-order relative density distribution function can be adopted (high value of the power $n$). In contrast, with the advances in the LPBF method, the resolution of the printing product can be as small as a few dozen of micrometers. However, each method might produce different beneficial and undesirable features for the products. For instance, the filament direction might create a large impact on the performance of specimens in FDM printing. The LPBF printing, however, might consist of multiple micro-porosity which can be created from the chemical composition of the binder. To eliminate these deflects in the printing product, the post-processing should be included, namely the heat treatment process for Selective Laser Sintering (SLS) \citep{al-ketan19_on-mechanical}.

\section{Conclusions}
\label{Section 5}
In this study, a new functionally graded TPMS plate model was proposed. Three TPMS architectures including Primitive (P), Gyroid (G), and wrapped package-graph (IWP) with different graded functions were considered. Their effective properties were derived from a new fitting technique based on a two-phase function in terms of relative density. This technique is inspired by the fact that sheet-based TPMS structures' responses strongly depend on their relative density or their thickness. Also, in smaller values of the relative density, these complex geometries have been well-known as stretching-dominated structures. On the other hand, a combining mode of stretching and bending, or even higher-order behaviors, should be noted in heavy foam TPMS structures. By applying the fitting process for the effective moduli, the above conclusion could be observed explicitly in the function parameters. The findings were obtained and further utilized to estimate Poisson's ratios for each TPMS type. Then, a seventh-order shear deformation theory (SeSDT) was investigated to describe the generalized displacement field of the FG-TPMS plate. A modification in the constitutive matrix has been conducted to capture the cubic symmetrical behavior of TPMS structures. NURBS-based isogeometric analysis (IGA) was employed to fulfill the $C^1$ continuity in approximations. Static, buckling, and free vibration analyses of rectangular and circular FG-TPMS plates were illustrated. Some concluding remarks can be coined as
\begin{itemize}
\item A two-phase fitting function in terms of the relative density produced ultra-accurate results in comparison with the previously published results.
\item The obtained solutions from the present approach show good agreement with the results of analytical theories and other numerical methods. 
\item The FG-TPMS plate model exhibited an excellent alternative for the isotropic plate. With only $35\%$ weight, the FG-TPMS plate is still able to mimic the fundamental frequency of the $100\%$ weight isotropic one. 
\item It was observed that the constituent structures provide dominant responses compared with similar-weight isotropic plates ($0.35$-thickness plates). The approximately three times higher and seven times higher values could be indicated from the previous section's results for fundamental frequency and both non-dimensional buckling and static load, respectively. 
\item Another major aspect of the FG-TPMS plates is their ability to be fabricated with multiple scales from macroscale to nanoscale with various manufacturing approaches. Among all, 3D printing technology might be the most feasible one with rapid developments in both macroscale and microscale printing.
\end{itemize}

As demonstrated in the previous section, additive manufacturing should be indicated as a suitable approach to fabricating sophisticated TPMS structures. Thanks to the advances in 3D printing technology, various types of materials may be adopted for the TPMS porous plates taken into account metal, plastic, ceramic, fibre-reinforced concrete, and other existing ones. Although this manufacturing technology might remain several limitations such as nanoscale printing ability, mirco-deflections, etc. Numerous solutions have been introduced and are continuously investigated to improve the technology. In sum, by adopting AM process, these FG-TPMS structures could expand numerous new potential application features in the future.

\section*{Acknowledgements}
This research used resources of the high-performance computer cluster (HPCC) at the Advanced Institute of Materials Science, Ton Duc Thang University (TDTU).

\bibliographystyle{unsrtnat}
\bibliography{Ref}

\end{document}